\documentclass[12pt]{article}
\pdfoutput=1
\usepackage{jheppub}

\usepackage{graphicx}
\usepackage{color}
\usepackage{verbatim}
\usepackage{amsfonts}
\usepackage{slashed}
\usepackage{epsfig}
\usepackage{latexsym,amssymb,amsmath,makeidx,mathrsfs,multirow}

\allowdisplaybreaks[4]
\usepackage{hyperref}

\newcommand{\fr}[2]{\mbox{$\frac{\,{#1}\,}{#2}$}}

\renewcommand{\rm}{\mathrm}
\def\bge{\begin{equation}}
\def\ede{\end{equation}}
\def\bga{\begin{aligned}}
\def\eda{\end{aligned}}
\newcommand{\beq}{\begin{equation}}
\newcommand{\eeq}{\end{equation}}
\newcommand{\bq}{\begin{equation}}
\newcommand{\eq}{\end{equation}}
\newcommand{\ba}{\begin{array}}
\newcommand{\ea}{\end{array}}
\newcommand{\beqa}{\begin{eqnarray}}
\newcommand{\eeqa}{\end{eqnarray}}
\newcommand{\beqs}{\begin{subequations}}
\newcommand{\eeqs}{\end{subequations}}

\def\nn{\nonumber}

\def\dis{\displaystyle}

\def\({\left(}
\def\){\right)}
\def\[{\left[}
\def\]{\right]}

\def\deg{\circ}
\def\End{\end{document}}

\def\over{\overline}

\def\bd{\boldsymbol}

\def\al{\alpha}
\def\be{\beta}

\def\lam{\lambda}

\def\ga{\gamma}

\def\ep{\epsilon}
\def\lam{\lambda}

\def\si{\sigma}

\def\di{\mathrm{d}}

\def\to{\rightarrow}

\def\ii{\mathrm{i}}

\def\mm{\text{m}}
\def\sm{\text{sm}}

\def\cut{\Lambda}

\newcommand{\mL}{\mathcal{L}}
\newcommand{\mO}{\mathcal{O}}

\newcommand{\TeV}{\textrm{TeV}}
\newcommand{\GeV}{\textrm{GeV}}

\def\LL{\mathcal{L}}
\def\abinv{\text{ab}^{-1}}

\def\vvb{\nu\bar{\nu}}

\def\SZZ{\mathcal{Z}}

\def\phic{\phi_{\text{c}}}
\def\phicc{\phi_{\text{c}}^{}}
\def\phicm{\phi_{\text{c}}^{\text{m}}}
\def\vvb{\nu\bar{\nu}}

\def\End{\end{document}}
\title{\boldmath Probing the Scale of New Physics\\
       in the $ZZ\gamma$ Coupling at $e^+e^-$ Colliders}

\author[a,b]{John Ellis,}
\author[b]{~Shao-Feng Ge,}
\author[b,c]{~Hong-Jian He,}
\author[b]{~Rui-Qing Xiao}

\affiliation[a]{Department of Physics, King's College London, Strand, London WC2R 2LS, UK;\\
Theoretical Physics Department, CERN, CH-1211 Geneva 23, Switzerland;\\
NICPB, R\"{a}vala 10, 10143 Tallinn, Estonia.}
\affiliation[b]{T.\ D.\ Lee Institute $\&$ School of Physics and Astronomy, \\
Shanghai Jiao Tong University, Shanghai 200240, China.}
\affiliation[c]{Institute of Modern Physics and Department of Physics, \\
                Tsinghua University, Beijing 100084, China;\\
Center for High Energy Physics, Peking University, Beijing 100871, China.}

\emailAdd{john.ellis@cern.ch}
\emailAdd{gesf@sjtu.edu.cn}
\emailAdd{hjhe@sjtu.edu.cn}
\emailAdd{xiaoruiqing@sjtu.edu.cn}

\abstract{\\[1mm]
The $ZZ\gamma$ triple neutral gauge couplings are absent
in the Standard Model (SM) at the tree level. They receive no contributions
from dimension-6 effective operators, but can arise from effective operators of dimension-8.
We study the scale of new physics associated with such dimension-8 operators
that can be probed by measuring the reaction $\,e^+e^-\!\!\to\! Z\gamma$\,,\,
followed by $\,Z\!\!\to\!\ell\bar\ell,\vvb$\, decays,
at future $e^+e^-$ colliders including the CEPC, FCC-ee, ILC and CLIC.
We demonstrate how angular distributions of the final state mono-photon and leptons
can play a key r\^ole in suppressing SM backgrounds.
We further show that using electron/positron beam polarizations
can significantly improve the signal sensitivities.
We find that the dimension-8 new physics scale can be probed up to
the multi-TeV region at such lepton colliders.
\\[2cm]
KCL-PH-TH/2019-11, CERN-TH/2019-008
\\[2mm]
e-Print: arXiv:1902.06631.v3 
}

\begin{document}
\maketitle
\flushbottom

\renewcommand{\thefootnote}{\arabic{footnote}}
\setcounter{page}{2}

\section{\hspace{-6mm}.~Introduction}
\label{sec:1}

At the time of writing, there is no confirmed evidence for phenomena in accelerator experiments
that require new physics beyond the Standard Model (SM)\,\cite{John},
pending clarifications of the apparent discrepancy between the SM prediction
and the experimental value of the anomalous magnetic moment of the muon,
and of the apparent anomalies in $b$-hadron decays into strange and charmed particles.
It is therefore plausible to assume that the SM particles have the same dimension-4
interactions as in the SM, and seek to characterize possible deviations from SM predictions
in terms of higher-dimensional effective operators constructed out of SM fields,
whose contributions are suppressed by some power of an underlying new physics scale
$\Lambda \gg 100$\,GeV\,\cite{dim6}.

\vspace*{1mm}	

This Standard Model Effective Field Theory (SMEFT) approach
has mainly been applied with the
assumption that only dimension-6 SMEFT operators\,\cite{dim6x}
contribute to the experimental observables
under study~\cite{reviews}. With this restriction, global SMEFT analyses\,\cite{SMEFT}
have been made of the available data
from the LHC and other accelerators, and the sensitivities of experiments
at possible future accelerators to the scales of new physics in dimension-6 operators
have also been estimated\,\cite{SMEFT,CLIC-dim6,eepp}.
However, there are some instances in which dimension-6 contributions are absent,
and the first SMEFT operators to which experimental measurements
are sensitive are those of higher dimensions\,\cite{above6}.
For instance,
the neutral triple gauge couplings (nTGCs) $ZZ\gamma$ and $Z\gamma\gamma$
provide a promising way of probing directly the relevant dimension-8
operators\,\cite{nTGC1,Degrande:2013kka}.

\vspace*{1mm}	

These neutral triple gauge couplings are absent in the SM and
receive no dimension-6 contributions\,\cite{dim6,dim6x}.
Within the SMEFT approach,
the first contributions arise from effective operators of dimension 8\,.\,
These operators involve the Higgs doublet $H$ \cite{Degrande:2013kka}
and the induced nTGCs vanish as the vacuum expectation value
$\left<H\right>\!\to 0$\,.\,
Hence, {\it the origin of these dimension-8 operators is linked to
spontaneous electroweak symmetry breaking.}
This means that probing these neutral TGCs also opens up a new window to
the physics of the Higgs boson and electroweak symmetry breaking.
We study here how these dimension-8 operators can be probed via
the reaction $\,e^+e^-\!\!\to\! Z\gamma$\,
(with $Z\!\to \ell^+\ell^-\!,\nu\bar{\nu}$\, decays)
at future $e^+e^-$ colliders including
CEPC~\cite{CEPC}, FCC-ee~\cite{FCCee}, ILC~\cite{ILC}, and CLIC~\cite{CLIC},
offering one of the rare direct windows on new physics at dimension-8.
The test of nTGCs at the FCC-hh via future $pp$(100TeV) collisions
was also considered recently\,\cite{pp-nTGC}.
Other examples where dimension-8 operators dominate include light-by-light scattering\,\cite{gaga-gaga},
$\gamma \gamma \to \gamma \gamma$, which has recently
been measured for the first time in heavy-ion collisions at the LHC~\cite{Aaboud:2017bwk},
and $gg \!\to\! \gamma \gamma$ scattering\,\cite{gg-gaga},
which is constrained by ATLAS measurements of
events with isolated diphotons in $pp$ collisions at the LHC\,\cite{Aaboud:2017yyg}.
The effect of dimension-8 operators on Higgs observables
was discussed in \cite{Sanz8}.

\vspace*{1mm}	

Our analysis framework is described in Section\,\ref{sec:2}.
We first discuss in Section\,\ref{sec:2.1} how the neutral triple-gauge couplings ${ZV\gamma}$
($V = Z,\gamma$) can be generated by effective dimension-8 operators,
and then present cross sections for $\,e^+e^-\!\!\to\! Z\gamma$\, production in the
different $Z$ polarization states $Z_{T,L}^{}$ in Section\,\ref{sec:2.2}.
Since the SM produces $Z_T^{} \gamma$ final states copiously, with the vector
bosons emerging preferentially in the forward and backward directions, we can make use of
angular distributions in the $e^+ e^-$ centre-of-mass frame and $Z$ decay frame
to separate the SM contribution to $Z \gamma$ final states and distinguish
$Z_L^{}$ from $Z_T^{}$ via their decays into dileptons $\ell^+\ell^-$.\,
We study angular observables in Section\,\ref{sec:3},
where the angular distributions are presented in Section\,\ref{sec:3.1}
and their uses for isolating and analyzing new physics contributions
are discussed in Section\,\ref{sec:3.2}, with the focus on ${\cal O}(\Lambda^{-4})$
contributions in Section\,\ref{sec:3.2.1} and on ${\cal O}(\Lambda^{-8})$
contributions in Section\,\ref{sec:3.2.2}. A systematical analysis of the sensitivities
to $\Lambda$ by measurements at different $e^+ e^-$ collider energies $\sqrt{s\,}$\,
from 250\,GeV to 5\,TeV is presented in Section\,\ref{sec:3.2.3}.
We present a refined analysis in Section\,\ref{sec:3.3} by including additional non-resonant
SM backgrounds. In Section\,\ref{sec:44}, we analyze the probe of new physics scale
via the invisible decay channel $\,Z\!\to\nu\bar{\nu}$,\, which we then combine with the
sensitivity of the dilepton channels $\,Z\!\!\to\!\ell^+\ell^-$.\,
Furthermore, we study the improved sensitivity in Section\,\ref{sec:55}
obtainable by using the $e^\mp$ beam polarizations.
Finally, we summarize our conclusions in Section\,\ref{sec:66}.
The $5\sigma$ sensitivity to $\Lambda$ may reach into the multi-TeV range,
depending on the $e^+ e^-$ collision energy, even after taking into account the fact
that in many new physics scenarios the SMEFT approach may be valid only
when $\Lambda \!\gtrsim\!\! \sqrt{s\,}$ or $\Lambda \!\gtrsim\!\!\sqrt{s}/2$\,.\,
Hence, the reaction $e^+e^-\!\!\to\! Z\gamma$\, will provide a unique and
interesting probe of new physics in $e^+ e^-$ collisions.

\vspace*{1.5mm}	
\section{\hspace{-6mm}.~Neutral Triple-Gauge Couplings and
\boldmath{${e^+ e^-\!\!\to\! Z\gamma}$} Production}
\label{sec:2}

In this Section, we first discuss the neutral triple-gauge couplings ${ZV\gamma}$ ($V\!= Z,\gamma$),
and the corresponding dimension-8 effective operators as their unique lowest-order gauge-invariant
formulations in the SMEFT.
We then analyze the scattering amplitudes for $e^+e^-\!\!\to\! Z \gamma$\,,\,
considering separately the transverse and longitudinal polarizations of the final-state $Z$ bosons.

\vspace*{1mm}
\subsection{\hspace{-5.1mm}.\,\boldmath{${ZV\gamma}$} Coupling from Dimension-8 Operator}
\label{sec:2.1}

The neutral triple gauge couplings (nTGCs) ${ZV\gamma}$ ($V\!=Z,\gamma$) vanish at tree level
in the SM and do not receive contributions from any dimension-6 effective operators.
However, at the dimension-8 level there are four CP-conserving effective operators
that include Higgs doublets and can contribute to the nTGCs,
\vspace*{-2mm}
\beqa
\Delta\mathcal{L}(\text{dim-8})
\,=\, \sum_{j=1}^4 \frac{c_j}{\,\tilde{\cut}^4\,}\mathcal{O}_j^{}
\,=\, \sum_{j=1}^4 \frac{\,\text{sign}(c_j^{})\,}{\,\cut_j^4\,}\mathcal{O}_j^{} \,,
\label{cj}
\eeqa
where the dimensionless coefficients $\,c_j^{}$ may be ${\cal O}(1)$,
with signs $\,\text{sign}(c_j^{})=\pm$, and
the corresponding ultraviolet (UV) cutoff scales are
$\,\cut_j^{} \equiv \tilde{\cut}/|c_j^{}|^{1/4}\,$.\,
The four dimension-8 CP-even effective operators
$\mathcal{O}_j^{}$ contributing to the  nTGCs
may be written as\,\cite{Degrande:2013kka},
\beqs
\label{eq:dim8H}
\begin{eqnarray}
\mathcal{O}_{\widetilde{B}W} &=&
\ii\, H^\dagger  \widetilde{B}_{\mu\nu}W^{\mu\rho}
\!\left\{D_{\!\rho}^{},D^\nu\right\}\! H+\text{h.c.},
\label{eq:obtw}
\\
\mathcal{O}_{B\widetilde{W}} &=&
\ii\, H^\dagger  B_{\mu\nu}\widetilde{W}^{\mu\rho}
\!\left\{D_{\!\!\rho}^{},D^\nu\right\}\! H+\text{h.c.},
\label{eq:obwt}\\
\mathcal{O}_{\widetilde{W}W} &=&
\ii\, H^\dagger  \widetilde{W}_{\mu\nu}W^{\mu\rho}
\!\left\{D_{\!\!\rho}^{},D^\nu\right\}\! H+\text{h.c.},
\label{eq:owwt}\\
\mathcal{O}_{\widetilde{B}B} &=&
\ii\, H^\dagger\widetilde{B}_{\mu\nu}B^{\mu\rho}
\!\left\{D_{\!\!\rho}^{},D^\nu\right\}\! H+\text{h.c.},
\label{eq:obbt}
\end{eqnarray}
\eeqs
where $H$ denotes the SM Higgs doublet.
The above operators are Hermitian and we take their coefficients $c_j^{}$
to be real for the present study, as we assume CP conservation.
We define the dual field strengths
$\,\widetilde{B}_{\mu\nu}^{} \!= \ep_{\mu\nu\al\be}B^{\al\be}$\, and
$\,\widetilde{W}_{\mu\nu}^{} \!= \ep_{\mu\nu\al\be}W^{\al\be}$,\,
where $\,W_{\mu\nu}^{}\!= W_{\mu\nu}^a\sigma^a\!/2\,$ and $\,\sigma^a$ 
denotes Pauli matrices.
Among the above operators,
one can use the equations of motion (EOM) and integration by parts to show
that $\mathcal{O}_{B\widetilde{W}}^{}$ is equivalent to
$\mathcal{O}_{\widetilde{B}W}^{}$
up to operators with more currents, or more field-strength tensors, or
with quartic gauge boson couplings.
Moreover, the operators
$\mathcal{O}_{\widetilde{W}W}^{}$ and $\mathcal{O}_{\widetilde{B}B}^{}$
do not contribute to $ZV\gamma$ coupling for on-shell $Z$ and $\gamma$.
Thus, there is only one independent CP-conserving dimension-8 operator
to be considered in our nTGC study.
We choose $\mathcal{O}_{\widetilde{B}W}^{}$ for our analysis, and
denote the corresponding cutoff scale
$\,\cut_{\widetilde{B}W}^{}\!\!=\cut\,$,
for simplicity.
	
\vspace*{1mm}

We note that all the dimension-8 operators in Eq.\eqref{eq:dim8H}
involve Higgs doublets and the induced nTGCs
vanish as the Higgs vacuum expectation value (VEV)
$\left<H\right>\!\!\to\!0$\,,\, so their origin is connected to
the spontaneous electroweak symmetry breaking (EWSB).
Hence, the nTGCs given by dimension-8 operators \eqref{eq:dim8H}
{\it provide a probe of new physics connected to the spontaneous EWSB.}
One could write down dimension-8 operators with three gauge-field-strength tensors
and two covariant derivatives (but {without} Higgs doublet)
that contribute to the nTGC. For instance, the following pure gauge operator
can contribute to nTGC:
\beqa
\label{eq:BWWDD}
g\mathcal{O}_{\widetilde{B}WW}^{} &=&
\widetilde{B}_{\!\mu\nu}^{}W^{a\mu\rho}\!\( D_{\!\!\rho}^{} D_\lambda^{} W^{a\nu\lambda}\!
+D^\nu D^\lambda W^{a}_{\!\rho\lambda}\).
\eeqa
But, the equations of motion (EOM) can be used
to convert such operators into operators with two Higgs doublets
[cf.\ Eq.\eqref{eq:dim8H}] plus extra operators involving the gauge current of left-handed
fermions\,\cite{Degrande:2013kka}.
In this connection, we note that the EOM of the gauge field $W^{a\mu}$ is given by
\beqa
\label{eq:EOM-W}
D^\nu W_{\!\mu\nu}^a \,=\,
\ii g\!\left[ H^\dag T^aD_{\!\mu}^{}H-(D_{\!\mu}^{}H)^\dag T^aH\right]
+g\,\overline{\psi_L^{}}T^a\ga_{\!\mu}^{}\psi_L^{} \,,
\eeqa
where $\,T^a\!=\tau^a/2$\, and $\,\psi_L^{}\,$ denotes the left-handed weak doublet fermions
(leptons or quarks). The summation over the fermion flavor indices is implied
in the last term of Eq.\eqref{eq:EOM-W},
Thus, for the pure gauge operator \eqref{eq:BWWDD},
we can make use of the EOM \eqref{eq:EOM-W} and
re-express the new dimension-8 operator \eqref{eq:BWWDD} as follows:
\beqa
\mathcal{O}_{\!\widetilde{B}WW}^{} &\,=\,&
\mathcal{O}_{\!\widetilde{B}W}^{}\! +
\widetilde{B}_{\!\mu\nu}^{}W^{a\mu\rho}\!
\left[D_{\!\!\rho}^{}(\overline{\psi_{\!L}^{}}T^a\!\ga^\nu\!\psi_{\!L}^{}\!)
      +D^\nu(\overline{\psi_{\!L}^{}}T^a\!\ga_{\!\rho}^{}\psi_{\!L}^{})
\right]\! ,
\label{eq:BBW-BW-eeZA}
\eeqa
where $\mathcal{O}_{\!\widetilde{B}W}^{}$ on the right-hand-side (RHS)
is just the original dimension-8 operator \eqref{eq:obtw}.
We have explicitly verified that for the reaction
$e^-e^+\!\!\!\to\! Z\ga$\, with on-shell final states,
the contribution from the above dimension-8
pure gauge operator $\,\mathcal{O}_{\!\widetilde{B}WW}^{}$
still vanishes in the limit $\left<H\right>\!\!\to\!0$\,,\,
because the extra fermionic contact contribution is proportional to
$\,M_Z^2\propto\!\left<H\right>^2$\,.\,
Hence, the key point is that the nTGC, as they are absent in the SM and at the
level of dimension-6 operators, can originate from the new physics generating the
dimension-8 operators, whose contributions vanish in the limit
$\left<H\right>\!\to\!0$\, and thus {\it the existence of these nTGCs hinges upon
the spontaneous EWSB.}
This shows that testing the nTGC via Eq.\eqref{eq:dim8H}
can provide a new window for probing
the new physics connected to the spontaneous EWSB.

\vspace*{1mm}

We note that the reaction $e^+e^-\!\!\to\! Z \gamma$\, may also contain
possible new physics contribution from a contact vertex $e^+e^-Z\gamma$ generated
by the dimension-8 fermionic operator in Eq.\eqref{eq:BBW-BW-eeZA}.
The three types of dimension-8 operators are constrained by Eq.\eqref{eq:BBW-BW-eeZA},
so only two of them are independent.
We can choose $\mathcal{O}_{\!\widetilde{B}W}^{}$
and the fermionic contact operator (contributing to $e^+e^-Z\gamma$)
as two independent operators.
For the current analysis of testing nTGC at future colliders,
we adopt the conventional approach of
{\it one operator at a time,} as widely used in the
literature\,\cite{reviews}-\cite{gaga-gaga}\cite{gg-gaga}\cite{nTGC1}\cite{Degrande:2013kka}.
Hence, we can focus on the dimension-8 operator \eqref{eq:obtw} for this nTGC study,
because Eq.\eqref{eq:obtw} contributes directly to the nTGC, while
the fermionic contact operator does not.
Furthermore, we work in the SMEFT formulation for the current study,
and follow the common practice in this approach of
not considering possible correlations among different dimension-8 operators
that may be given by certain specific underlying UV models and are highly model-dependent.

\vspace*{1mm}

Finally, we can expand Eq.\eqref{eq:obtw} and derive
the following effective $Z\gamma Z^*$ coupling from the
dimension-8 operator $\mathcal{O}_{\widetilde{B}W}^{}$
in momentum space:
\begin{eqnarray}
\label{eq:ZAZ*-vertex}
\ii\, \Gamma^{\mu\nu\al}_{Z\gamma Z^*}
({q}_1^{}, {q}_2^{}, {q}_3^{})
&\,=\,&
\text{sign}(c_j^{})
\frac{\,v M_Z^{} ({q}_3^2\!-\!M_Z^2)\,}{\,\Lambda^4\,}
\epsilon^{\mu\nu\al\be} q_{2\be}^{} \, ,
\end{eqnarray}
where $\,v/\!\sqrt{2}=\!\left<H\right>\,$ is the Higgs vacuum expectation value.
However, $\mathcal{O}_{\widetilde{B}W}$ does not contribute to the $Z\gamma\gamma^*$ coupling
for on-shell gauge bosons $Z$ and $\gamma$\,.
Moreover, there is no $\ga\ga\ga^*$ triple photon coupling with two photons on-shell.
This fact is consistent with our observation that the existence of nTGCs has to rely on
the Higgs VEV and thus the spontaneous EWSB, while the residual electromagnetic gauge
group $U(1)_{\text{em}}^{}$ holds unbroken.

\vspace*{1mm}
\subsection{\hspace{-5.1mm}.\,\boldmath{$Z\gamma$} Production at $e^+e^-$ Colliders}
\label{sec:2.2}	
\vspace*{1mm}

The SM contributes to the production process $e^-(p_1^{})e^+(p_2^{})\to Z(q_1^{})\gamma(q_2^{})$,\,
via $t$- and $u$-channel exchange diagrams at tree level.
In general, the final-state $Z$ boson may have either longitudinal or transverse
polarizations.

\vspace*{1mm}
	
Working in the centre of mass (c.m.) frame of the $e^+e^-$ collider
and neglecting the electron mass, we denote the
momenta of the initial- and final-state particles as follows:
\beqs
\vspace*{-5mm}
\label{eq:p1p2q1q2}
\begin{eqnarray}
&&
p_1^{} = E_1^{}(1,0,0,1),~~~~
p_2^{} = E_1^{}(1,0,0,-1),
\label{eq:p1p2}
\\[1.5mm]
&&
q_1^{} = \!\(E_Z^{},\, q\sin\!\theta,\, 0,\, q\cos\!\theta\)\!,~~~~
q_2^{} = q(1,\, -\sin\!\theta,\, 0,\, -\cos\!\theta),~~~~~~
\label{eq:q1q2}
\end{eqnarray}	
\eeqs	
where the electron (positron) energy
$\,E_1^{}\!=\frac{1}{2}\sqrt{s\,}\,$, the momentum
$\,q=\frac{1}{\,2\sqrt{s\,}\,}(s\!-\!M_Z^2)$,\,
and the $Z$ boson energy $E_Z^{}\!=\!\!\sqrt{q^2\!+\!M_Z^2\,}$.\,
The squared scattering amplitudes for the SM contributions to final states with
the different $Z$ polarizations take the following forms:
\beqs
\vspace*{-2mm}
\begin{eqnarray}
\label{eq:SM-A-ZL}
\overline{\left|\mathcal{T}_{\text{sm}}^{}\right|^2}[ Z_L^{}\ga_T^{}]
&=&
e^4\!\(8 s_W^4\!-4 s_W^2\!+\!1\)\!\frac{ M_Z^2\,s}{\,c_W^2 s_W^2
\!\(s\!-\!M_Z^2\)^2\,}\,,
\\[2mm]
\overline{\left|\mathcal{T}_{\text{sm}}^{}\right|^2}[Z_T^{}\ga_T^{}]
&=&  e^4\!\(8 s_W^4\!-4 s_W^2\!+\!1\)
\frac{\(1\!+\!\cos^2\theta\) \(s^2\!+\!M_Z^4\)}
{\,2s_W^2c_W^2\sin^2\theta\(s\!-\!M_Z^2\)^2\,}\,,
\hspace*{8mm}
\label{eq:SM-A-ZT}
\end{eqnarray}
\label{eq:SMT}
\eeqs
where we have averaged over the initial-state spins, and used the notations
$(s_W^{},\,c_W^{}) = (\sin\!\theta_W^{},\,\cos\!\theta_W^{})$
with $\theta_W^{}\!$ being the weak mixing angle.
We have verified that the above formulae agree with
the previous results in the literature\,\cite{Degrande:2013kka}.

\vspace*{1mm}

We see from the above equations that the squared amplitude for a final-state
longitudinal weak boson $Z_L^{}$ is suppressed by $\,1/s\,$
in the high-energy region $\,s\gg M_Z^2$.\, This behaviour can be understood via
the equivalence theorem \cite{ET}, which connects the
longitudinal scattering amplitude to the corresponding Goldstone boson amplitude
at high energies,
\beqa
\vspace*{-2mm}
\mathcal{T}[Z_L^{}\ga_T^{}] \,=\,
\mathcal{T}[\pi^0\ga_T^{}] + O(M_{\!Z}^{}/\!\sqrt{s\,})\,,
\eeqa
where $\pi^0$ is the would-be Goldstone boson absorbed
by the longitudinally-polarized $Z$ via the Higgs mechanism of the SM.
Since the SM does not contain any tree-level $ZV\ga$ and $\pi^0V\ga$ ($V=Z,\ga$)
triple couplings, at tree level the
production processes $e^+e^-\!\!\to Z_L^{}\ga_T^{}$ and
$e^+e^-\!\!\to \pi^0\ga_T^{}$ must proceed through the $t$-channel electron-exchange process.
Since the electron Yukawa coupling $\,y_e^{}=\sqrt{2}m_e^{}/v=O(10^{-6})\,$
is very small and can be neglected for practical purposes, we have for the SM contributions
\beqa
\label{eq:SM-ZL-ET}
\mathcal{T}_{\text{sm}}^{}[\pi^0\ga_T^{}]\simeq 0\,,
&~~~&
\left|\mathcal{T}_{\text{sm}}^{}[Z_L^{}\ga_T^{}] \right|^2 = O(M_Z^2/s)\,.
~~~~~
\eeqa
This explains the high-energy behavior of Eq.\eqref{eq:SM-A-ZL}.

\vspace*{1mm}

We note also that, for the final state with a transverse weak boson $Z_T^{}$,
Eq.\eqref{eq:SM-A-ZT} exhibits a collinear divergence at $\,\theta=0,\pi\,$
due to our neglect of the electron mass $m_e^{}\!\simeq 0$\,.\,
In the following analysis we implement a lower cut on the transverse momentum
of the final state photon: $\,P_T^{\gamma}=q\sin\!\theta >P_{T0}^{\gamma}\,$
to remove the collinear divergence,
corresponding to a lower cut on the scattering angle,
$\,\sin\theta>\sin\delta =P_{T0}^{\gamma}/q$\,.\,
For $\,\theta\neq 0,\pi$,\, Eq.\eqref{eq:SM-A-ZT} gives the asymptotic behavior,
$\,\mathcal{T}_{\text{sm}}^{}[Z_T^{}\ga_T^{}]= O(s^0)$,\,
in the high-energy regime $\,s\gg M_Z^2$\,,\, as expected.
This completes the explanation why production of the transversely polarized
final state $Z_T^{}\ga_T^{}$ dominates over that of the
longitudinal final state $Z_L^{}\ga_T^{}$\,.

\vspace*{1mm}	

The contributions of the dimension-8 operator include
${\cal O}(\Lambda^{-4})$ and ${\cal O}(\Lambda^{-8})$ terms.
The terms of ${\cal O}(\Lambda^{-4})$ arises from the interference between
the dimension-8 operator contribution and the SM contribution, and takes the forms
\beqs
\label{eq:Tsm-T8}
\begin{eqnarray}
\label{eq:Tsm-T8-ZL}
2\Re\texttt{e}\!
\(\overline{\mathcal{T}_{\text{sm}}^{}\mathcal{T}_{(8)}^*}\)[Z_L^{}\ga_T^{}]
&\,=\,&
\pm \frac{\,e^2\!\(1\!-\!4 s_W^2\)\,}{\,2 s_W^{}c_W^{}\,}
\frac{\,M_Z^2\, s\,}{\Lambda^4}\,,
\label{1L}
\\[1mm]
\label{eq:Tsm-T8-ZT}
2 \Re\texttt{e}\!\(\overline{\mathcal{T}_{\text{sm}}^{}\mathcal{T}_{(8)}^*}\)
[Z_T^{}\ga_T^{}]
&=&
\pm \frac{\,e^2\!\(1\!-\!4 s_W^2\)\,}{\,2s_W^{}c_W^{}}
\frac{\,M_Z^4\,}{\Lambda^4}\,,
\end{eqnarray}
\eeqs
which are consistent with results in the literature \cite{Degrande:2013kka}.
[Here the $\pm$ signs of the ${\cal O}(\cut^{\!-4})$ term correspond to
the two possible signs of a given dimension-8 operator,
$\text{sign}(c_j^{})=\pm$,\, as shown in Eq.\eqref{cj}.]
We see that the contribution to the
$Z_L^{}\ga_T^{}$ production channel is enhanced relative to that of the
$Z_T^{}\ga_T^{}$ production channel by a factor of $\,s/M_Z^2\,$
at ${\cal O}(\Lambda^{-4})$.\,

\vspace*{1mm}

The ${\cal O}(\Lambda^{-8})$ terms originate from the pure dimension-8 contributions,
{\small
\beqs
\label{eq:T(8)^2}
\begin{eqnarray}
\overline{|\mathcal{T}_{(8)}^{}|^2}[Z_L^{}\ga_T^{}]
&\,=\,&
\frac{\,(8 s_W^4\!-\!4 s_W^2\!+\!1)(\cos\!2\theta\!+\!3)\,}{\,32\,}
\frac{\,M_Z^2(s\!-\!M^2_Z)^2s\,}{\cut^8}\,,
\\[2mm]
\overline{|\mathcal{T}_{(8)}^{}|^2}[Z_T^{}\ga_T^{}]
&\,=\,&
\frac{\,(8 s_W^4\!-\!4 s_W^2\!+\!1)\sin^2\!\theta\,}
{\,8\,}
\frac{\,M_Z^4(s\!-\!M^2_Z)^2\,}{\cut^8}\,.
\end{eqnarray}
\eeqs}
The energy dependence in the above formulas can be directly understood by power counting,
\vspace*{-4mm}
{\small
\beqs
\label{eq:T8-count}
\beqa
\label{eq:T8-ZL-count}
\mathcal{T}_{(8)}^{}[Z_L^{}\ga_T^{}] &\simeq&
\mathcal{T}_{(8)}^{}[\pi^0\ga_T^{}]
\sim\frac{\,M_Z^{}\,s^{\frac{3}{2}}\,}{\cut^4}\,,
\\
\label{eq:T8-ZT-count}
\mathcal{T}_{(8)}^{}[Z_T^{}\ga_T^{}]&\sim&\frac{\,M_Z^2\,s\,}{\cut^4}\,,
\,
\eeqa
\eeqs
}
which explains the asymptotic high-energy behaviors in Eq.\eqref{eq:T(8)^2}
when $s\!\gg\! M_Z^2$\,.\,
We see that at ${\cal O}(\Lambda^{-8})$ the $Z_L^{}\ga_T^{}$ production channel dominates
over the $Z_T^{}\ga_T^{}$ production channel at high energies $s\gg M_Z^2$\,.\,

\vspace*{1mm}

We can understand further the asymptotic behavior of the interference terms
\eqref{eq:Tsm-T8} for $s\gg M_Z^2$.\,
In the case of the final state $Z_L^{}\ga_T^{}$, since we have
$\,\mathcal{T}_{\text{sm}}^{}[Z_L^{}\ga_T^{}]\sim \frac{M_Z^{}}{\sqrt{s\,}\,}$\,
[Eq.\eqref{eq:SM-ZL-ET}] and
$\,\mathcal{T}_{(8)}^{}[Z_L^{}\ga_T^{}]
 \sim\frac{\,M_Z^{}\,s^{\frac{3}{2}}\,}{\cut^4}\,$
[Eq.\,\eqref{eq:T8-ZL-count}], we find that their interference term
behaves as
$\,\overline{\mathcal{T}_{\text{sm}}^{}\mathcal{T}_{(8)}^*}[Z_L^{}\ga_T^{}]
   \sim\frac{\,M_Z^2\,s\,}{\cut^4}\,$.\, This explains nicely the asymptotic behavior
of Eq.\eqref{eq:Tsm-T8-ZL}.
However, for the final state $Z_T^{}\ga_T^{}$,
using the naive power counting from Eqs.\eqref{eq:SM-A-ZT} and \eqref{eq:T8-ZT-count}
we infer the asymptotic behaviors,
$\,\mathcal{T}_{\text{sm}}^{}[Z_T^{}\ga_T^{}]\sim s^0$\, and
$\,\mathcal{T}_{(8)}^{}[Z_T^{}\ga_T^{}]
 \sim\frac{\,M_Z^2\,s\,}{\cut^4}\,$.\,
Combining these would lead to the following behavior for their interference:
$\,\overline{\mathcal{T}_{\text{sm}}^{}\mathcal{T}_{(8)}^*}[Z_T^{}\ga_T^{}]
   \sim\frac{\,M_Z^2\,s\,}{\cut^4}\,$.\,
However, this naive power counting contradicts Eq.\eqref{eq:Tsm-T8-ZT}, where we see that
$\,\overline{\mathcal{T}_{\text{sm}}^{}\mathcal{T}_{(8)}^*}[Z_T^{}\ga_T^{}]
 \sim\frac{\,M_Z^4\,s^0\,}{\cut^4}\,$.\,
Naive power counting fails in this case for a nontrivial reason,
which is connected to the special structure of the helicity amplitude
$\mathcal{T}_{(8)}^{}[Z_T^{}\ga_T^{}]$.\, We see from Eqs.\eqref{eq:T8-T} and \eqref{eq:T8T-off=0}
of Appendix~\ref{sec:A1} that the off-diagonal helicity amplitides
$\mathcal{T}_{(8)}^{}[Z_T^{}\ga_T^{}]$ with $\lambda\lambda'=+-,-+$ vanish
because of the antisymmetric tensor
$\,\ep^{\mu\nu\al\be}\,$ contained in the $Z\ga Z^*$ vertex
[Eq.\eqref{eq:ZAZ*-vertex}]. Hence, the energy dependence of
$\,\overline{\mathcal{T}_{\text{sm}}^{}\mathcal{T}_{(8)}^*}[Z_T^{}\ga_T^{}]$\,
is determined by the diagonal helicity amplitudes with $\lambda\lambda'=++,--$.
The SM amplitude
$\mathcal{T}_{\text{sm}}^{}[Z_T^{}\ga_T^{}]$ has a negative power of energy
$\propto s^{-1}$\,
in its diagonal helicity amplitudes as shown in Eq.\eqref{eq:Tsm-T}.
This explains neatly the high-energy behavior $\,\overline{\mathcal{T}_{\text{sm}}^{}\mathcal{T}_{(8)}^*}[Z_T^{}\ga_T^{}]
 \sim\frac{\,M_Z^4\,s^0\,}{\cut^4}\,$,\,
 in agreement with Eq.\eqref{eq:Tsm-T8-ZT}.

\vspace*{2mm}	
\section{\hspace{-6mm}.~Probing New Physics in the \boldmath{$ZV\gamma$} Coupling at
\boldmath{$e^+e^-$} Colliders}
\label{sec:3}

In this section, we first analyze the kinematical structure of the reaction
$\,e^+e^-\!\!\to Z\ga\,$ followed by $Z$ decays into pairs of charged leptons $\ell^\pm$.
We then propose suitable kinematical cuts to suppress effectively the
SM backgrounds, and derive the optimal sensitivity reach for the scale of
the new physics in the $ZV\gamma$ coupling.
In Section~\ref{sec:3.1}, we analyze the angular observables
for $Z\ga\,$ production with $Z\!\to\! \ell^+\ell^-$,
and then study probes of the new physics contributions
at ${\cal O}(\cut^{-4})$ in Section~\ref{sec:3.2.1}
and at ${\cal O}(\cut^{-8})$ in Section~\ref{sec:3.2.2},
making use of angular observables
to suppress the SM backgrounds
for the specific $e^+e^-$ collision energy $\sqrt{s}=3$\,TeV.
Then, we extend the analysis to other collider energies
$\sqrt{s}=(250,\,500,\,1000,\,5000)$\,GeV in
Section~\ref{sec:3.2.3}, showing the increase in sensitivity obtainable from
increasing the collider energy. Finally, in Section~\ref{sec:3.3}, we present
a more complete background analysis including
additional non-resonant SM backgrounds with the same final state $\ell^-\ell^+\ga$
(but $\ell^-\ell^+$ not coming from $Z$ decay).

\vspace*{1mm}	
\subsection{\hspace{-5.1mm}.\,Analysis of Angular Observables}
\label{sec:3.1}	
\vspace*{1mm}

\begin{figure}[t]
\begin{center}
\includegraphics[width=13cm,height=7.7cm]{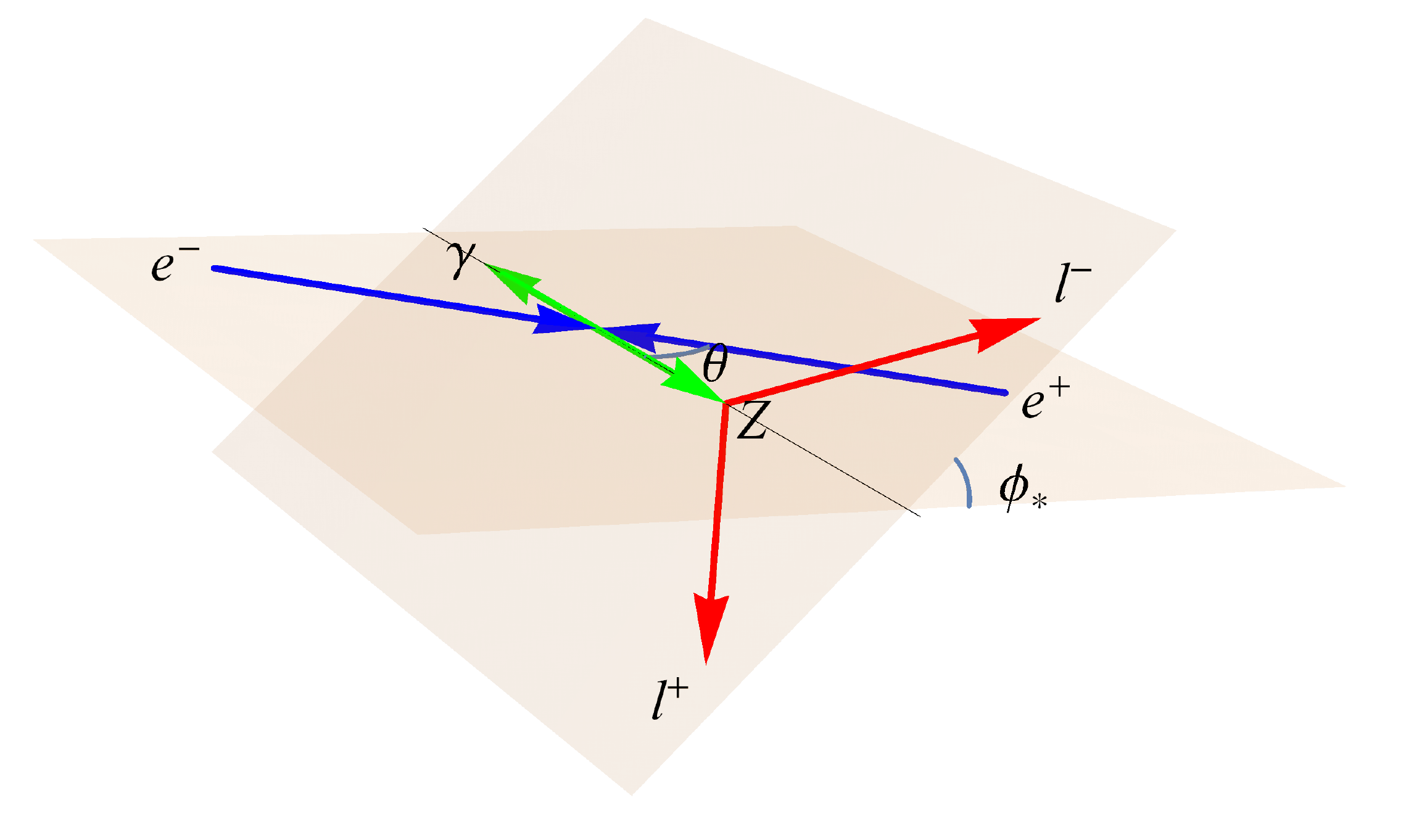}
\vspace*{-6mm}
\caption{\it Illustration of the kinematical structure of the reaction
$\,e^+e^-\!\!\to Z\ga\,$ followed by the leptonic decay $\,Z\!\to \ell^+\ell^-$\,
in the laboratory frame ($\,e^-e^+$\! collision frame).}
\label{figtheta}
\label{fig:1}
\vspace*{-2mm}
\end{center}
\end{figure}

In this subsection, we analyze the kinematical observables for the reaction
$\,e^+e^-\!\!\to Z\gamma\,$ followed by the leptonic decays $\,Z\!\to\! \ell^+\ell^-$.\,
We illustrate the kinematics in Fig.\,\ref{fig:1}, where
the scattering plane is determined by the incident $e^-e^+$
and the outgoing $\,Z\gamma$\, in the collision frame
(with scattering angle $\theta$\,),
and the directions of the final-state leptons $\ell^-\ell^+$ determine the
decay plane. We denote the angle between the two planes as $\phi$ in the laboratory frame
(which is equal to $\phi_*^{}$ in the $Z$ rest frame).

\vspace*{1mm}

In order to study the leptonic final states
$\,Z(q_1^{})\!\to \ell^-(k_1^{})\ell^+(k_2^{})$,\,
we denote the lepton momenta as follows in the $Z$ rest frame:
\beqs
\begin{eqnarray}
k_1^{} &=&  \frac{\,M_Z^{}\,}{2}
\(1,\,\sin\!\theta_*^{}\!\cos\!\phi_*^{},\,
      \sin\!\theta_*^{}\!\sin\!\phi_*^{},\,\cos\!\theta_*^{}\),
\\[1mm]
k_2^{} &=&  \frac{\,M_Z^{}\,}{2}
\(1,\,-\sin\!\theta_*^{}\!\cos\!\phi_*^{},\,-\sin\!\theta_*^{}
    \!\sin\!\phi_*^{},\,-\cos\!\theta_*^{}\).
\end{eqnarray}
\eeqs
Here the positive $z_*^{}$ direction in the $Z$ rest frame is chosen to be
opposite to the final-state photon direction in the laboratory frame,
and $\theta_*^{}$ denotes the angle between the positive $z_*^{}$ direction
and the $\ell^-$ direction in the $Z$ rest frame.
When boosted back to the $e^-e^+$ collision frame (laboratory frame),
the angle $\theta_*^{}$ changes but the azimuthal angle $\phi_*^{}$ is invariant.
This is why the angle $\phi_*^{}$ is equal to the angle $\phi$ between the scattering plane
(defined by the incoming $e^-e^+$ directions and the outgoing $Z\gamma$ directions)
and $Z$ decay plane (defined by the outgoing $\ell^-$ and $\ell^+$ directions)
in the $e^-e^+$ collision frame.

\vspace*{1mm}	

Imposing a lower cut on the scattering angle in the laboratory frame,
$\,\sin\theta>\sin\delta\,$ (where $\delta\ll 1$),
will correspond to a lower cut on the transverse momentum
of the final-state photon $\,P_T^\gamma > q\sin\delta$\,.\,
With this lower cut,
we find the following total cross section for $Z\gamma$ production:
%
\beqs
\begin{eqnarray}
\sigma(Z\gamma)
&\,=\,&
\frac{\,e^4(c_L^2\!+\!c_R^2)\!\!\left[-(s\!-\!M^2_Z)^2\!-\!2(s^2\!+\!M_Z^4)
	\ln\!\(\sin\!\frac{\delta }{2}\)\right]\,}
{\,8\pi s_W^2c_W^2(s\!-\!M^2_Z)s^2\,}
\hspace*{15mm}
\nn\\[1mm]
\label{Zgamma0}
&&
\pm\frac{\,e^2 (c_L^2\!-\!c_R^2)M_Z^2\!\(s\!-\!M_Z^2\)\!\!\(s\!+\!M^2_Z\)\,}
{\,8\pi s_W^{}c_W^{}\cut^4 s^2\,}
\\[1mm]
&& +\frac{\,(c_L^2\!+\!c_R^2)M_Z^2\!\(s\!+\!M^2_Z\)\!\!\(s\!-\!M^2_Z\)^{\!3}\,}
{\,48\pi\cut^8 s^2\,} + O(\delta)\,,
\nn
\\[1.5mm]
&\,=\,&
\frac{\,e^4(1\!-\!4 s_W^2\!+\!8s_W^4)\!\!\left[-(s\!-\!M^2_Z)^2\!-\!2(s^2\!+\!M_Z^4)
\ln\!\(\sin\!\frac{\delta }{2}\)\right]\,}
{\,32\pi s_W^2c_W^2(s\!-\!M^2_Z)s^2\,}
\hspace*{15mm}
\nn\\[1mm]
&&
\pm\frac{\,e^2 (1\!-\!4s_W^2)M_Z^2\!\(s\!-\!M_Z^2\)\!\!\(s\!+\!M^2_Z\)\,}
{\,32\pi s_W^{}c_W^{}\cut^4 s^2\,}
\\[1mm]
&& +\frac{\,(1\!-\!4s_W^2\!+\!8s_W^4)M_Z^2\!\(s\!+\!M^2_Z\)\!\!\(s\!-\!M^2_Z\)^{\!3}\,}
{\,192\pi\cut^8 s^2\,} + O(\delta)\,,
\nn
\label{Zgamma}
\end{eqnarray}
\eeqs
where the $\pm$ signs of the $O(\cut^{-4})$ interference term come from
$\,\text{sign}(c_j^{})\!=\!\pm$\, for each given dimension-8 operator,
as shown in Eq.\eqref{cj}.
In Eq.\eqref{Zgamma0},
\,$(c_L^{},c_R^{})=(s_W^2-\fr{1}{2},\, s_W^2)$\, denote the
$Z$ gauge couplings to the (left,\,right)-handed electron.

\vspace*{1mm}

We compute numerically the exact cross sections
for $e^+e^-\!\to Z\gamma$,\footnote{Since the
leptonic vector coupling of $Z$ boson is proportional to
($1-4s_W^2$), it is sensitive to the value of $s_W^2$.\,
Here we use the $\overline{\text{MS}}$ value
$\,s_W^2= 0.23122\pm 0.00003$ ($\mu =M_Z^{}$) \cite{PDG}.}
as a function of the new physics scale $\cut$ and for different collider energies.
imposing a photon transverse momentum cut
$\,P_T^{\ga}=q\sin\delta$\, with $\delta>0.2$\,:
{\small
\beqs
\label{eq:sigma-all-d0.2}
\begin{eqnarray}
\sqrt{s\,}\!=250\text{GeV}, &~~~&
\sigma(Z\gamma) = \left[7749
\pm 8.90\!\left(\!\frac{\,0.5\text{TeV}\,}{\Lambda}\!\right)^{\!\!4}\!
+ 1.98\!\left(\!\frac{\,0.5\text{TeV}\,}{\Lambda}\!\right)^{\!\!8}
\right]\!\text{fb}\,,~~~~~~~~
\\[1mm]
\sqrt{s\,}\!=500\text{GeV}, &~~~&
\sigma(Z\gamma)
= \left[1624 \pm 1.38\!\(\!\frac{\,0.8\text{TeV}\,}{\Lambda}\!\)^{\!\!4}\!
+0.929\!\(\!\frac{\,0.8\text{TeV}\,}{\Lambda}\!\)^{\!\!8}\right]\!\text{fb}\,,
\\[1mm]
\sqrt{s\,}\!=1\text{TeV}, &~~~&
\sigma(Z\gamma)  = \left[390 \pm 0.566\!\(\!\frac{\,\text{TeV}\,}{\Lambda}\!\)^{\!\!4}
+2.62\!\(\!\frac{\,\text{TeV}\,}{\Lambda}\!\)^{\!\!8}\right]\!\text{fb}\,,
\\[1mm]
\sqrt{s\,}\!=3\text{TeV}, &~~~&
\sigma(Z\gamma)
= \left[42.9 \pm 0.0354\!\(\!\frac{\,2\text{TeV}\,}{\Lambda}\!\)^{\!\!4}
+0.843\!\(\!\frac{\,2\text{TeV}\,}{\Lambda}\!\)^{\!\!8}\right]
\!\text{fb}\,,
\label{eq:sigma-3TeV-d0.2}
\\[1mm]
\sqrt{s\,}\!=5\text{TeV}, &~~~&
\sigma(Z\gamma) = \left[15.4 \pm 0.0145\!\(\!\frac{\,2.5\text{TeV}\,}{\Lambda}\!\)^{\!\!4}
+1.09\!\(\!\frac{\,2.5\text{TeV}\,}{\Lambda}\!\)^{\!\!8}\right]\!\text{fb}\,.
~~~~~~~~~~
\end{eqnarray}
\eeqs
}
\hspace*{-2.5mm}
As we show in Section\,\ref{sec:3.2}-\ref{sec:3.3}
(cf.\ Table\,\ref{tab:1}),
the sensitivity reach of $\,\cut$\, in each case is such that
on the right-hand-side of the corresponding formula above,
the ratio inside each $\[{\cdots}\]$\, is ${\cal O}(1)$.
Thus, we see from Eq.\eqref{eq:sigma-all-d0.2}
that, for the relevant sensitivity reaches of $\cut$,
the contributions of the dimension-8 operator
are always much smaller than the SM contributions,
so the perturbation expansion is valid.
Also, Eq.\eqref{eq:sigma-all-d0.2} shows that for $\sqrt{s\,}<1$~TeV
the ${\cal O}(\Lambda^{\!-4})$ contribution is dominant, whereas for
$\sqrt{s\,}\gtrsim 1$~TeV, the ${\cal O}(\Lambda^{\!-8})$ contribution
becomes dominant. This is because the ${\cal O}(\Lambda^{\!-8})$ contributions
have higher energy dependence than the ${\cal O}(\Lambda^{\!-4})$
contributions, as shown in Eqs.\eqref{eq:Tsm-T8}-\eqref{eq:T(8)^2}.

\vspace*{1mm}	

The total cross section for
\,$e^+e^-\!\to Z\gamma\to\ell^+\ell^-\gamma$\, is given by the product
\begin{eqnarray}
\sigma(\ell^+\ell^-\gamma)\,=\, \sigma(Z\gamma)\times \text{Br}(\ell^+\ell^-) \,.
\label{llgamma}
\end{eqnarray}
The differential cross section is a function of the three kinematical angles
$(\theta,\,\theta_*^{},\,\phi_*^{})$, and is computed from the helicity
amplitudes \eqref{eq:T-llgamma}-\eqref{eq:T-llgamma-sum} in Appendix-\ref{sec:A2}.
We define the normalized angular distribution function as
\begin{eqnarray}
f_\xi^j \,=\, \frac{\di\sigma_{\!j}^{}}{\,\sigma_{\!j}^{}\di\xi\,}\,,
\end{eqnarray}
where $\,\xi =\theta,\,\theta_*^{},\,\phi_*^{}$,\,
and $\,\sigma_j^{}$ (with $j=0,1,2$) represents the SM contribution ($\sigma_0^{}$),
the ${\cal O}(\cut^{-4})$ contribution ($\sigma_1^{}$), and the
${\cal O}(\cut^{-8})$ contribution ($\sigma_2^{}$), respectively.

\begin{figure}[t]
\includegraphics[width=7.8cm,height=5.5cm]{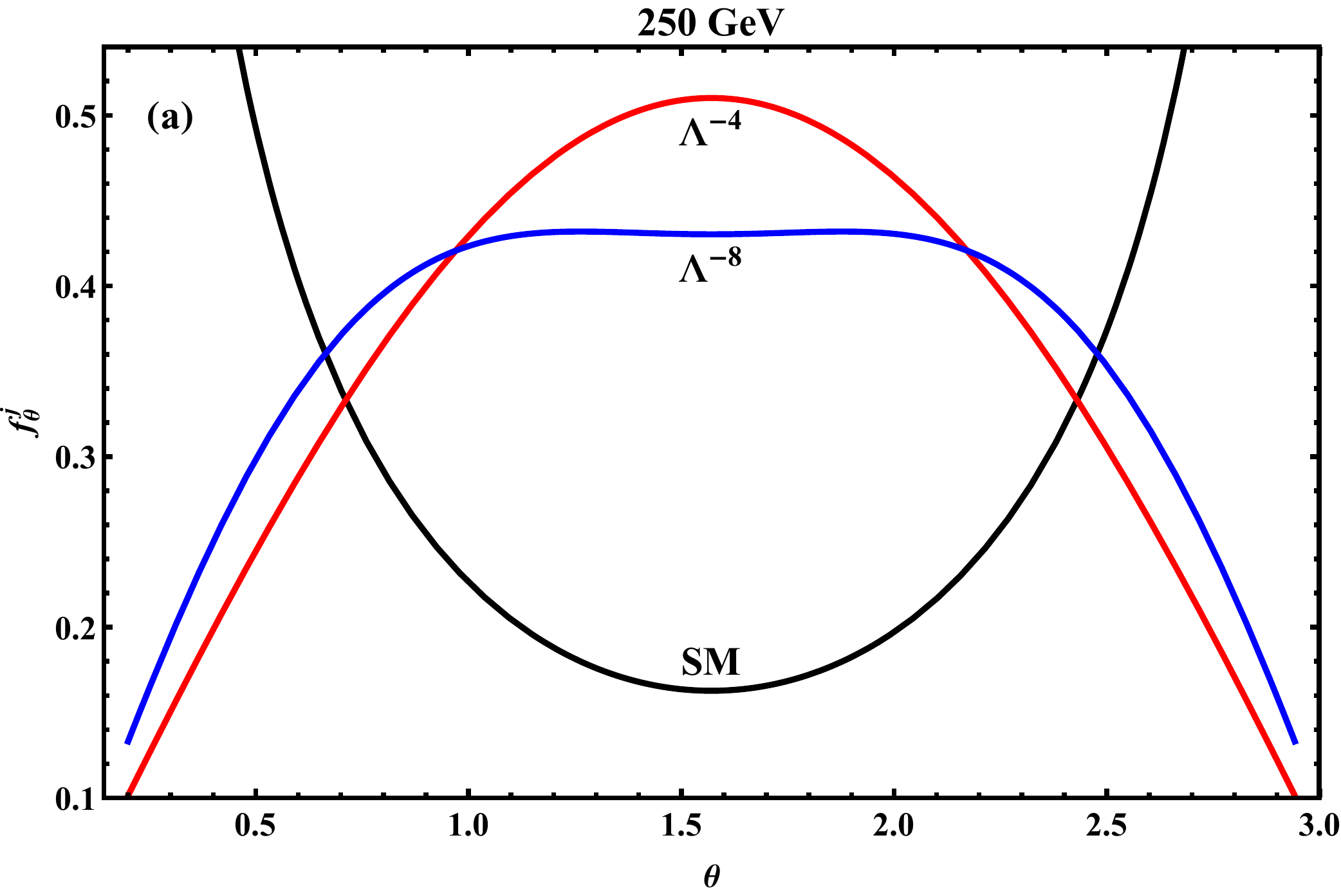}
\includegraphics[width=7.8cm,height=5.5cm]{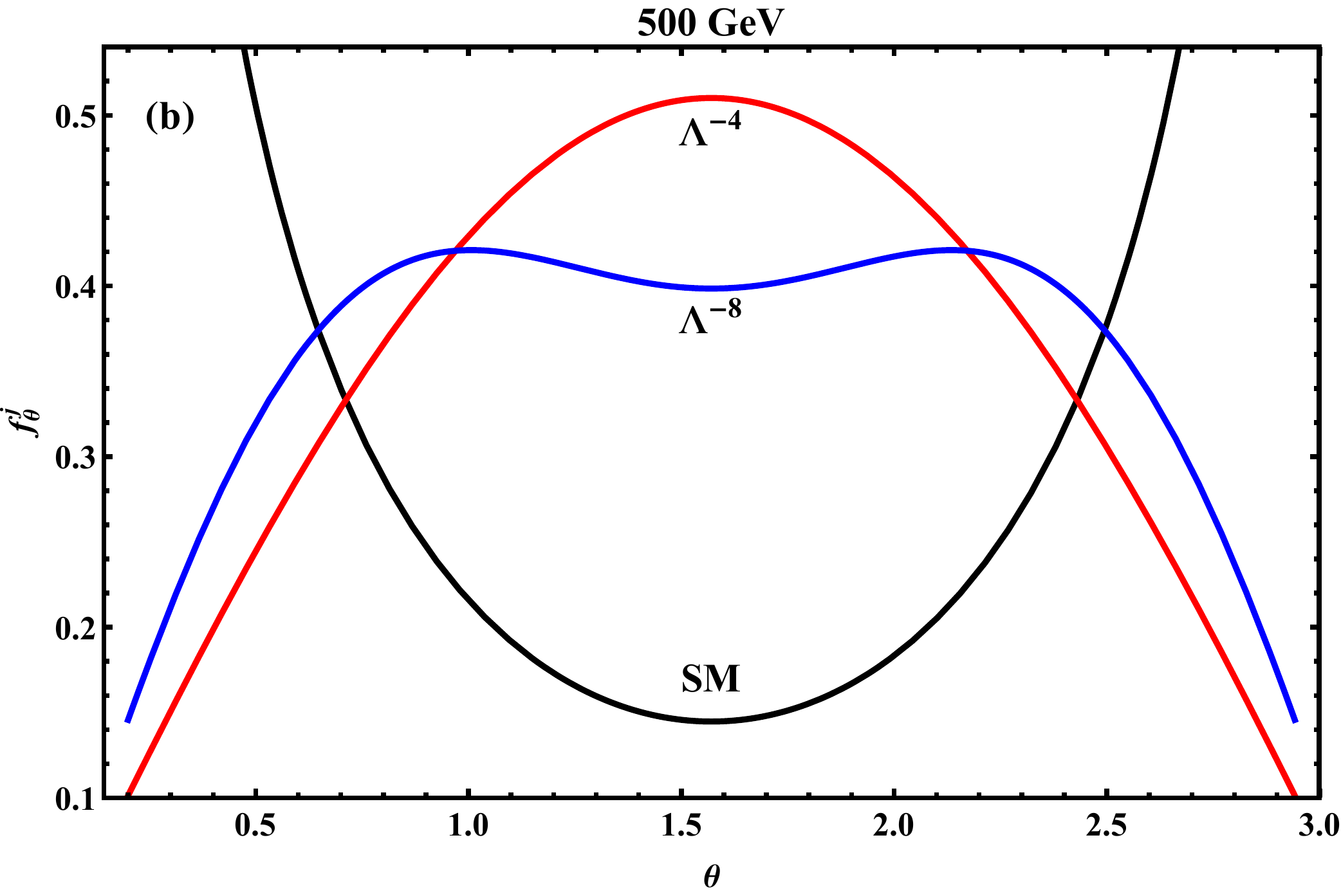}
		\\[3mm]
\includegraphics[width=7.8cm,height=5.5cm]{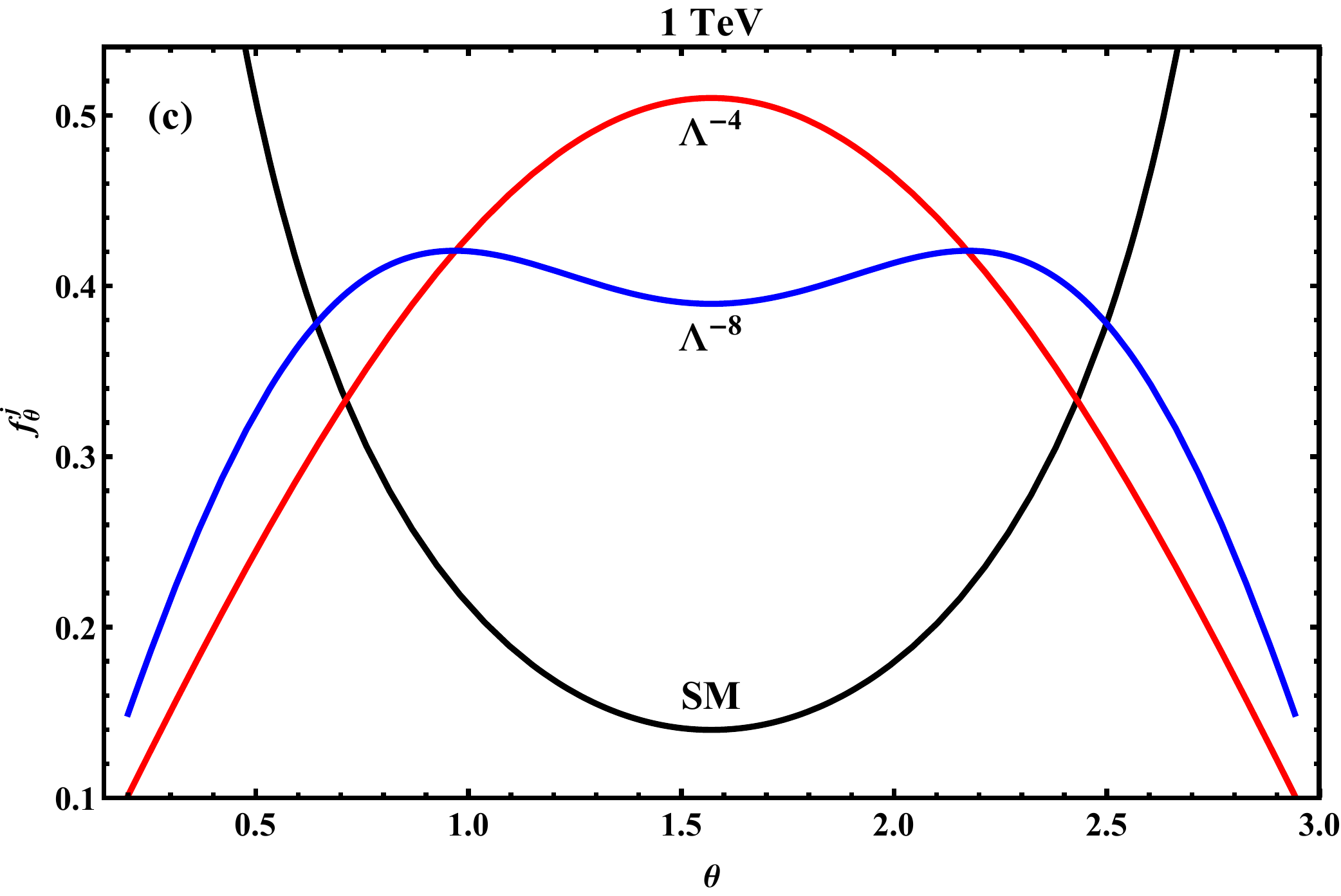}
\includegraphics[width=7.8cm,height=5.5cm]{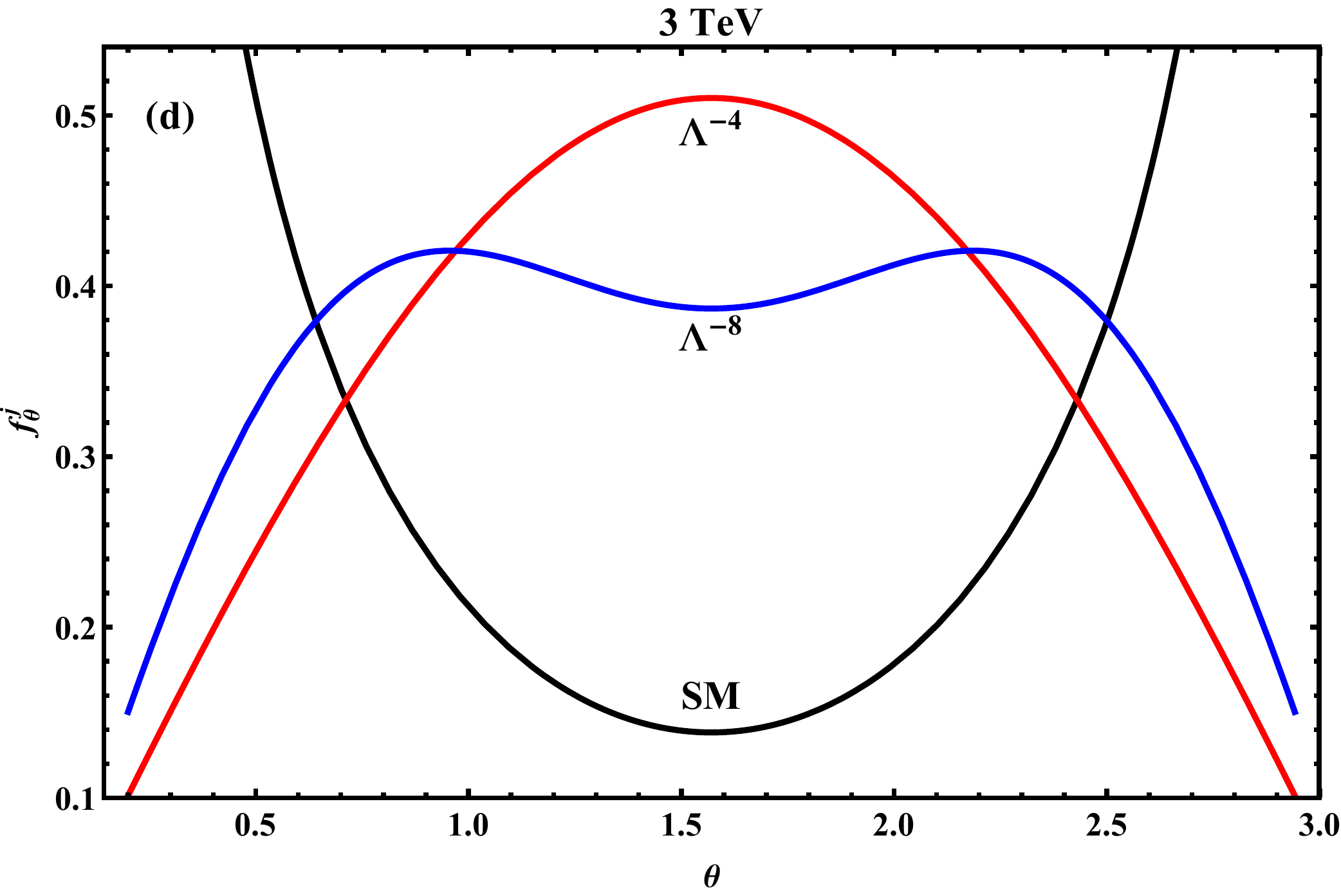}
\vspace*{-7mm}
\caption{{\it Normalized angular distributions in the polar scattering angle $\theta$
in the laboratory frame for different collision energies,}
$\sqrt{s}=(250\,\text{GeV}, 500\,\text{GeV}, 1\,\text{TeV}, 3\,\text{TeV})$.
{\it In each plot, the black, red and blue curves denote the contributions from the
SM, the ${\cal O}(\cut^{-4})$ and ${\cal O}(\cut^{-8})$ terms, respectively.
We use a polar angle cut $\,\delta=0.2\,$ for illustration.}}
\label{figtheta}
\label{fig:22}
\vspace*{2mm}
\end{figure}

\vspace*{1mm}

We find the following normalized polar angular distribution functions
$\,f_\theta^j\,$ and $\,f_{\theta_*}^j$,
\vspace*{-3mm}
{\small
\beqs
\begin{eqnarray}
f_\theta^{0} &=&
-\frac{\,\csc\!\theta\!\left[3 s^2\! +\cos\!2\theta (s\!-\!M_Z^2)^2\!
       +2 M_Z^2s +3 M_Z^4\right]\,}
      {\,4\!\left[(s\!-\!M_Z^2)^2\!+2(s^2\!+\!M_Z^4)\!\ln\!\(\sin\frac{\delta}{2}\)\right]\,}\,,
\\
f_\theta^{1} &=& \frac{1}{2}\sin\theta\,,
\\[1.5mm]
f_\theta^{2} &=&
\frac{\,3\sin\!\theta\!\left[3s+\cos\!2\theta(s\!-\!2 M_Z^2)+ 2M_Z^2\right]\,}
      {\,16 (s+\!M_Z^2)\,} \,;
\end{eqnarray}
\eeqs
}
and
{\small
\beqs
\begin{eqnarray}
f_{\theta_*}^{0} &=&
\frac{\,3\sin\!\theta_*^{}(3\!+\!\cos\!2\theta_*^{})\,}{16}
+\frac{3\sin\!\theta_*^{}(1+3\cos\!2\theta_*^{})M_Z^2\,s\,}
      {\,8\!\left[(s\!-\!M_Z^2)^2 \!+\! 2(s^2\!+\!M_Z^4)
      \ln\!\(\!\sin\!\frac{\delta}{2}\)\right]}
      +O(\delta) \,,
\hspace*{10mm}
\\[1mm]
f_{\theta_*}^{1} &=&
\frac{\,3\sin\!\theta_*^{}\!\left[2s \!-\! \cos\!2\theta_*^{}(2s\!-\!M_Z^2)\!+\!3 M_Z^2\right]}
     {16(s+\!M_Z^2)} + O(\delta) \,,
\\[1mm]
f_{\theta^*}^{2} &=&
\frac{\,3 \sin\!\theta_*^{}\!\left[2s \!-\!\cos\!2\theta_*^{}\!
     (2s\!-\!M_Z^2)\!+\!3 M_Z^2\right]}{16(s+M_Z^2)}
+ O(\delta)\,.
\end{eqnarray}
\eeqs
}
Then, we compute the normalized azimuthal angular distribution functions
$\,f_{\phi_*}^j\,$ as follows,
{\small
\beqs
\label{eq:f-phi*}
\begin{eqnarray}
\hspace*{-8mm}
f_{\phi_*}^{0} &=&
\frac{1}{2\pi} +\frac{3\pi^2(c_L^2\!-\!c_R^2)^2M_Z^{}\sqrt{s}\,(s\!+\!M_Z^2)\cos\!\phi_*^{}\!
-8(c_L^2\!+\!c_R^2)^2M_Z^2\,s \cos\!2\phi_*^{}\,}
{\,16 \pi(c_L^2\!+\!c_R^2)^2\!\left[(s\!-\!M_Z^2)^2\!+2(s^2\!+\!M_Z^4)\!\ln\!\(\!\sin\frac{\delta}{2}\)
 \right]\,}+O(\delta),
\hspace*{4mm}
\label{f0}
\\[2mm]
\hspace*{-8mm}
f_{\phi_*}^{1} &=& \frac{1}{2\pi} -
\frac{\,9\pi^2\!\sqrt{s}\,(s\!+\!M_Z^2)\cos\!\phi_*^{}\!-\!
      32M_Z^{}\,s \cos\!2\phi_*^{}\,}
     {\,128\pi M_Z^{}(s\!+\!M_Z^2)\,} + O(\delta) ,
\label{f1}
\\[2mm]
\hspace*{-8mm}
f_{\phi_*}^{2} &=&
\frac{1}{2\pi} - \frac{\,9\pi (c_L^2\!-\!c_R^2)^2M_Z^{}\sqrt{s}\cos\!\phi_*^{}\,}
                      {\,128(c_L^2\!+\!c_R^2)^2(s\!+\!M_Z^2)\,} + O(\delta) \,,
\label{f2}
\end{eqnarray}
\eeqs
}
\hspace*{-2.5mm}
where the coefficients \,$(c_L^{},\,c_R^{})=(s_W^2\!-\!\fr{1}{2},\, s_W^2)$\, are the
gauge couplings of the $Z$ boson to the (left, right)-handed leptons.
Here we have again chosen a lower cutoff $\,\delta\ll 1$\, on the polar scattering angle $\theta$,\,
which corresponds to a lower cut
on the transverse momentum of the final state photon, $\,P_T^\gamma> q\sin\!\delta$.\,

\begin{figure}[t]
\includegraphics[width=7.7cm,height=5.5cm]{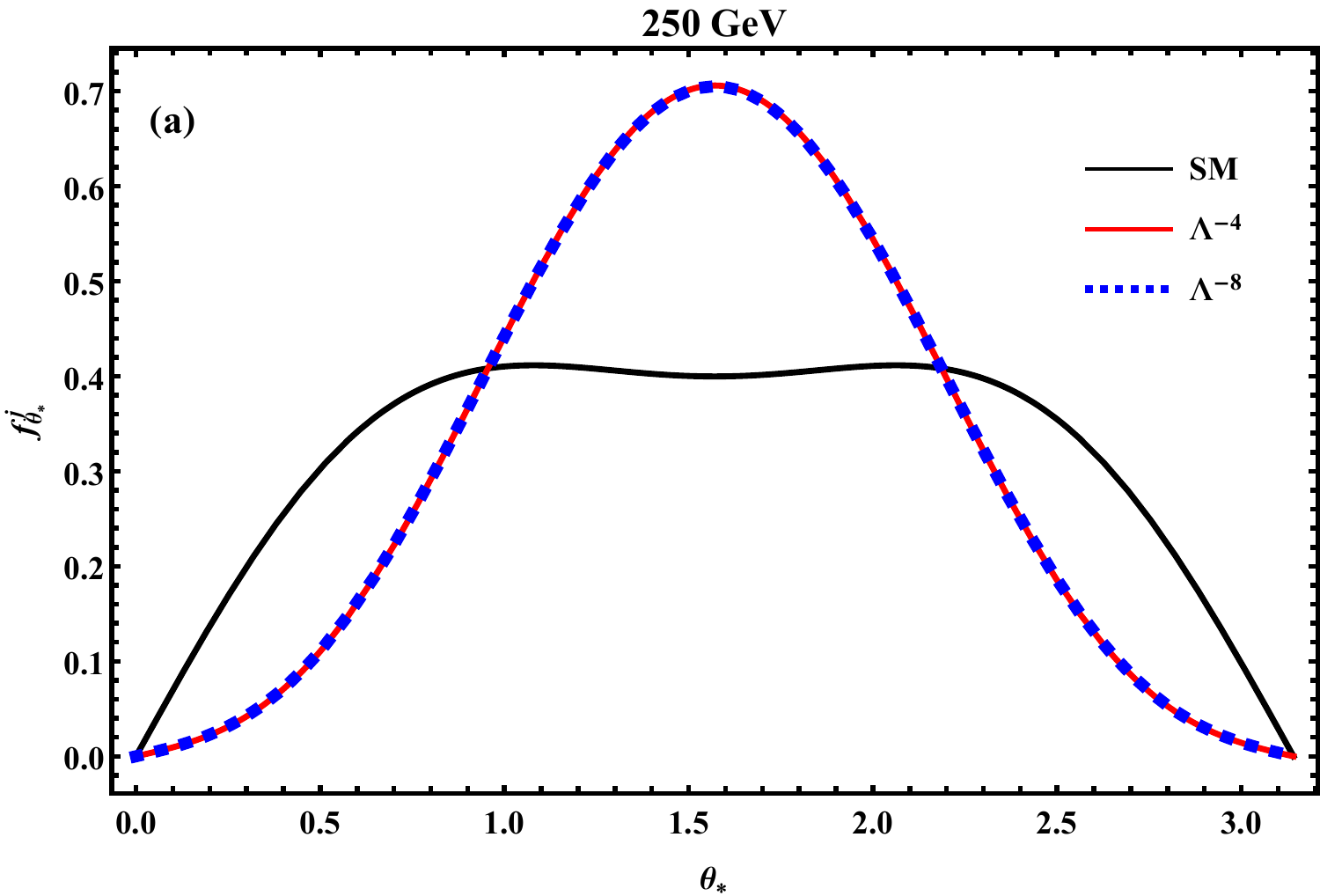}
\includegraphics[width=7.7cm,height=5.5cm]{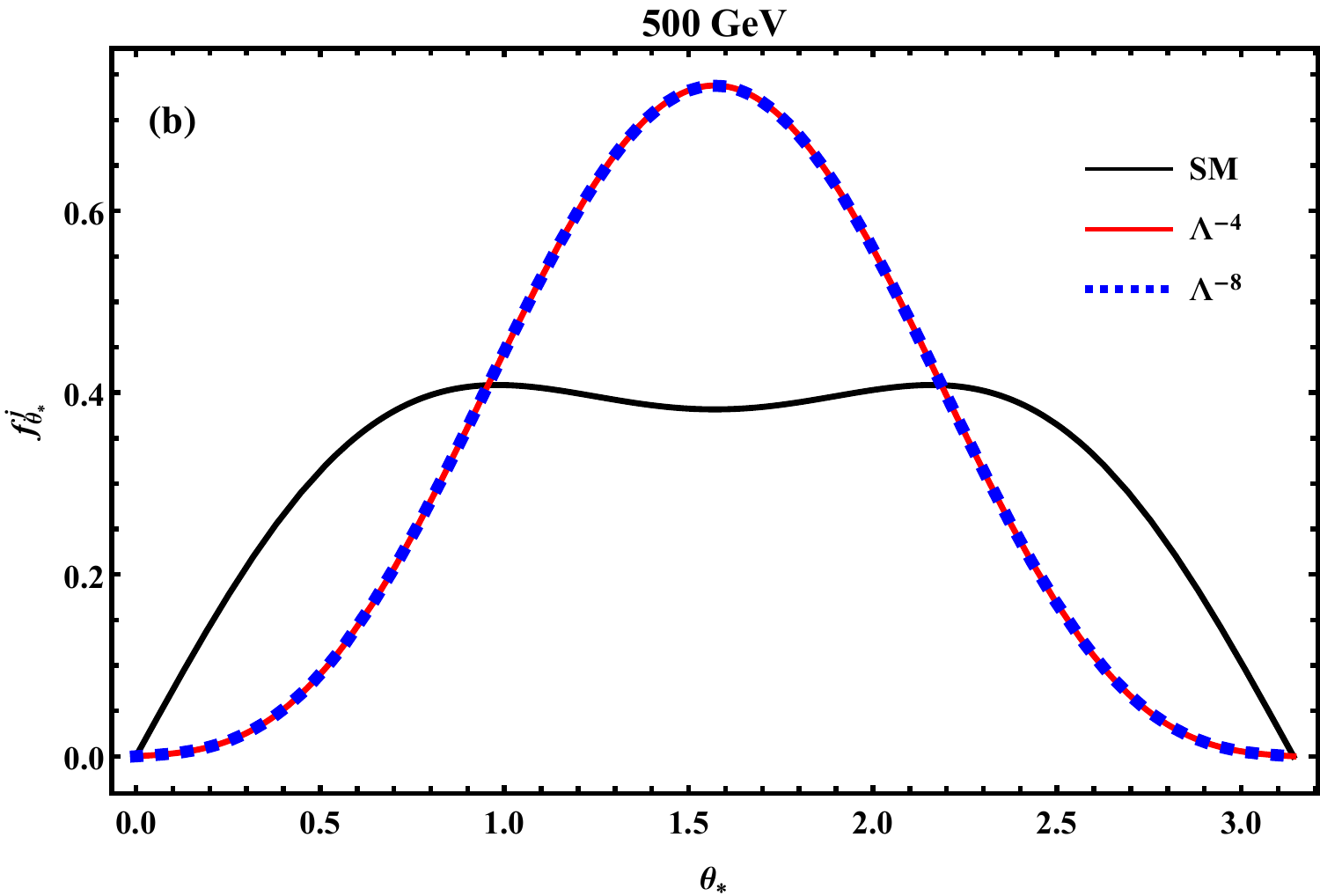}
\\[3mm]
\includegraphics[width=7.7cm,height=5.5cm]{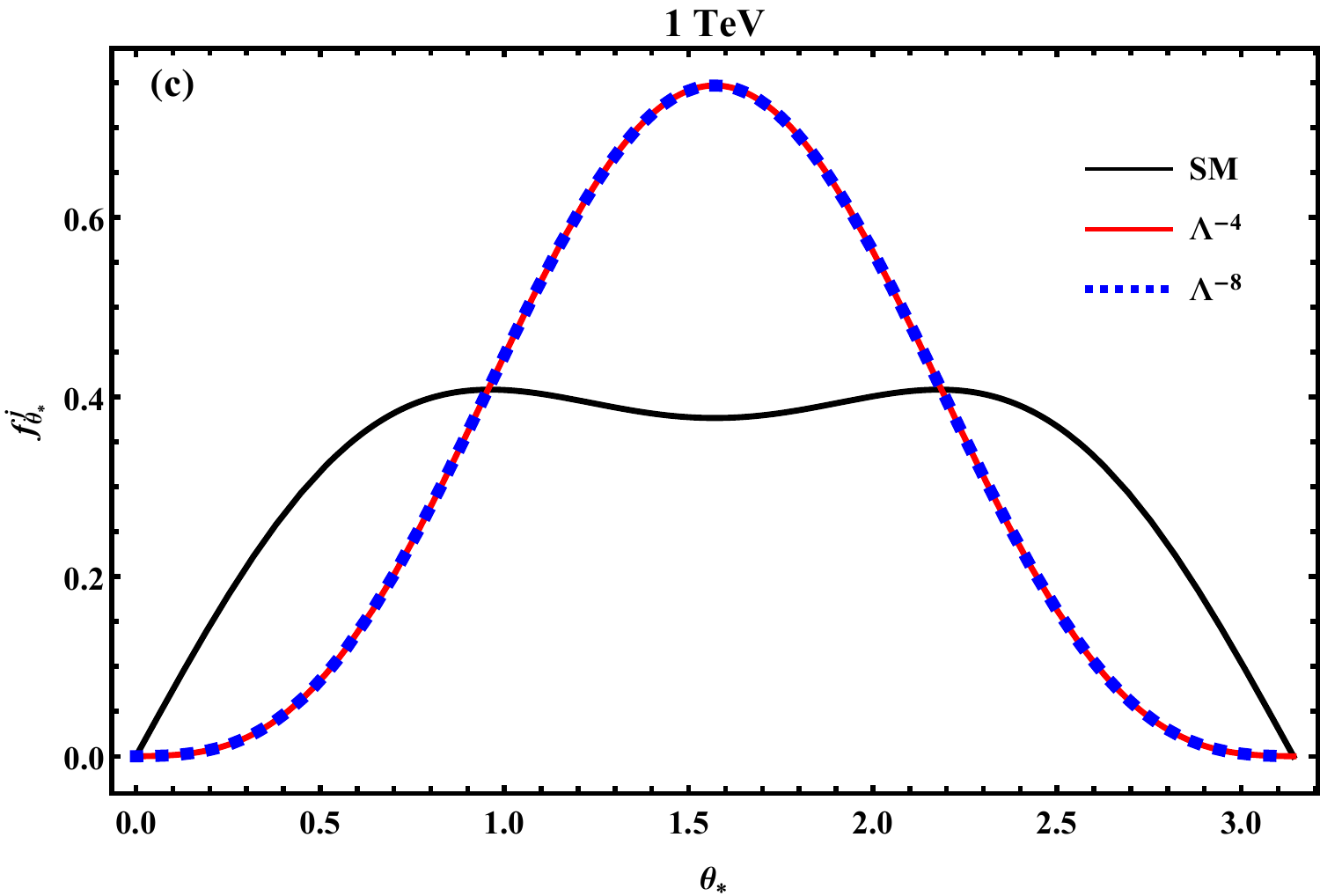}
\includegraphics[width=7.7cm,height=5.5cm]{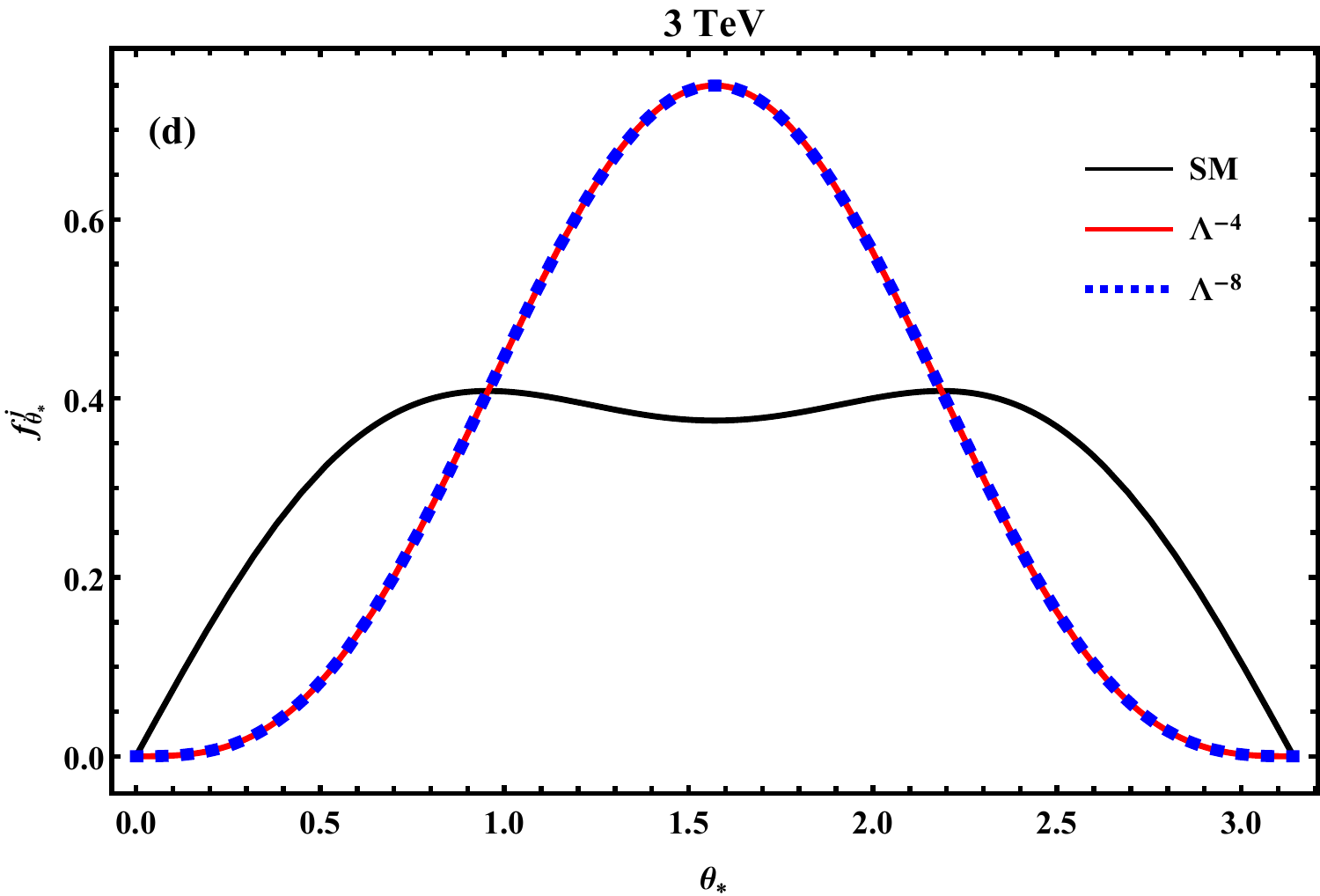}
\vspace*{-7mm}
\caption{{\it Normalized angular distribution in the polar angle $\theta_*^{}$
in the $Z$ decay frame for different collision energies,}
$\sqrt{s}=(250\,\text{GeV}, 500\,\text{GeV}, 1\,\text{TeV}, 3\,\text{TeV})$.
{\it In each plot, the black, red and blue curves denote the contributions from the
SM, the ${\cal O}(\cut^{-4})$ and ${\cal O}(\cut^{-8})$ terms, respectively, where the
red and blue curves exactly overlap.
We use a laboratory polar angle cut $\delta=0.2$ for illustration.}
}	
\label{figtz}
\label{fig:2}
\label{fig:33}
\end{figure}

\vspace*{1mm}

As a side remark, we note that if the $Z$ boson were a stable particle,
one could in principle measure its polarization
directly to extract the new physics signal of the dimension-8 operator.
However, since the $Z$ decays rapidly into fermion pairs,
the contributions of the out-going longitudinal $Z_L^{}$ and transverse $Z_T^{}$
intermediate states interfere in the angular distributions of the fermions produced in
$Z_L^{}$ and $Z_T^{}$ decays. Such interference effects appear as the $\phi_*^{}$
angular dependence in Eq.\eqref{eq:f-phi*}.

\vspace*{1mm}

\begin{figure}[t]
\includegraphics[width=7.7cm,height=5.5cm]{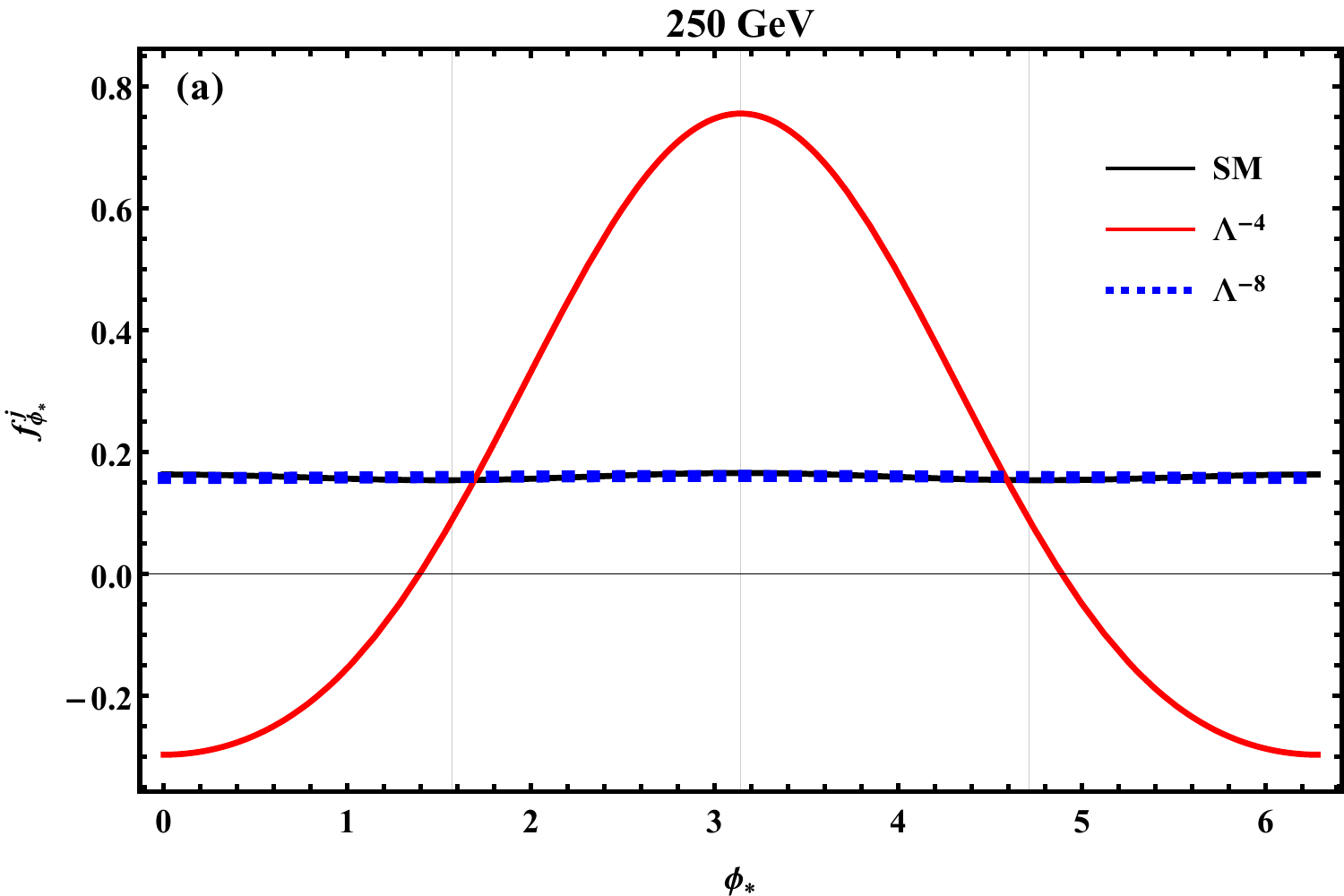}
\includegraphics[width=7.7cm,height=5.5cm]{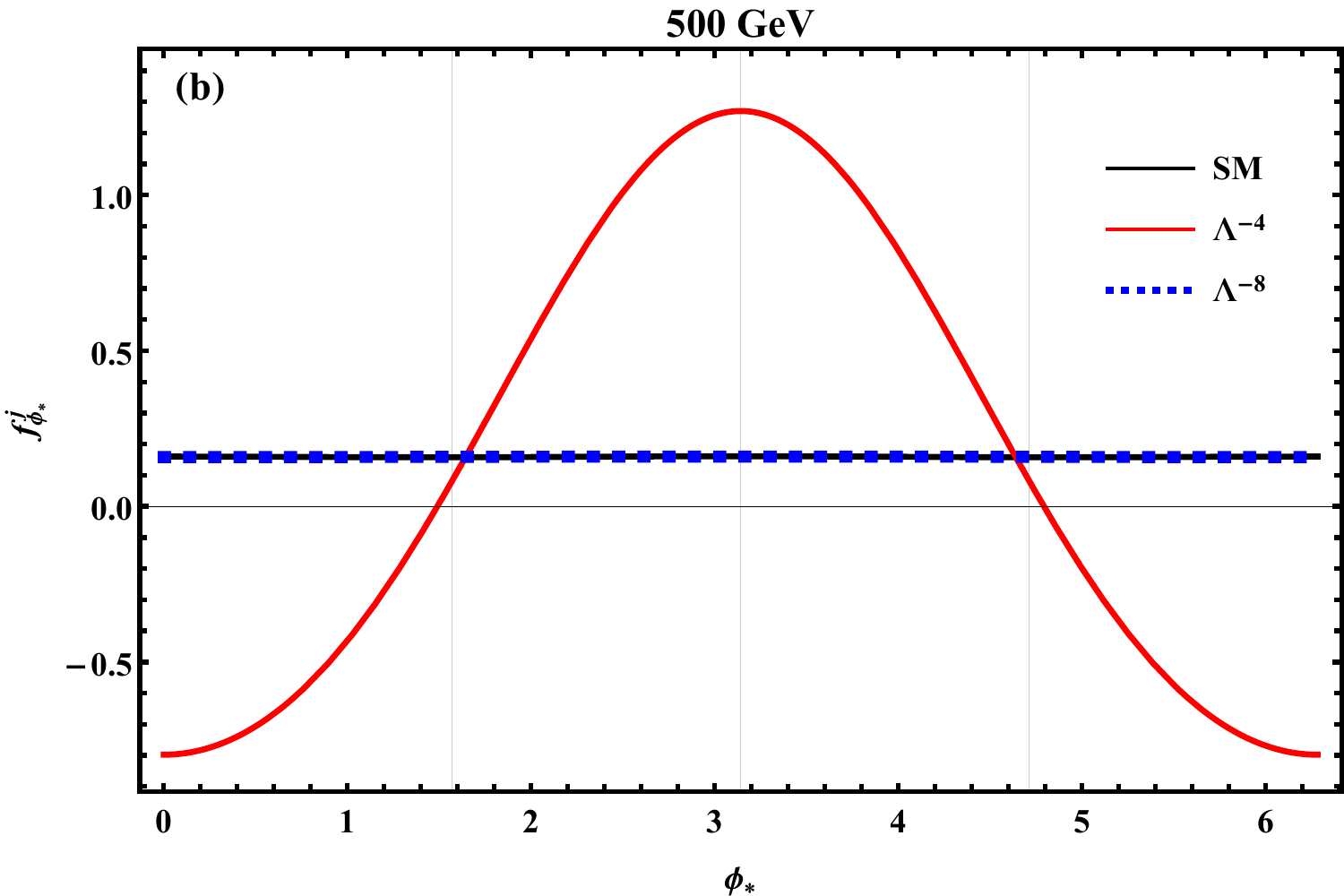}
\\[3mm]
\includegraphics[width=7.7cm,height=5.5cm]{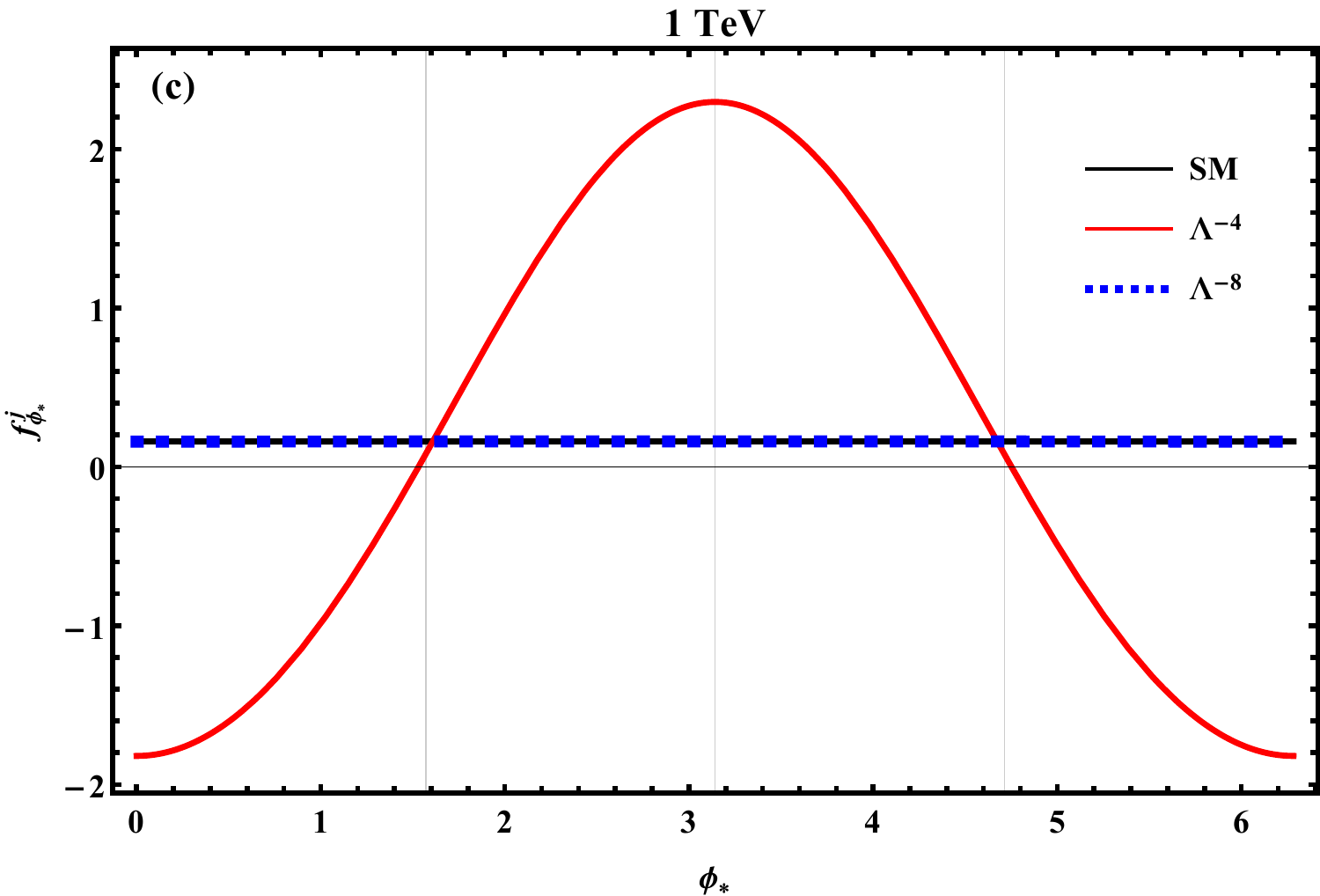}
\includegraphics[width=7.7cm,height=5.5cm]{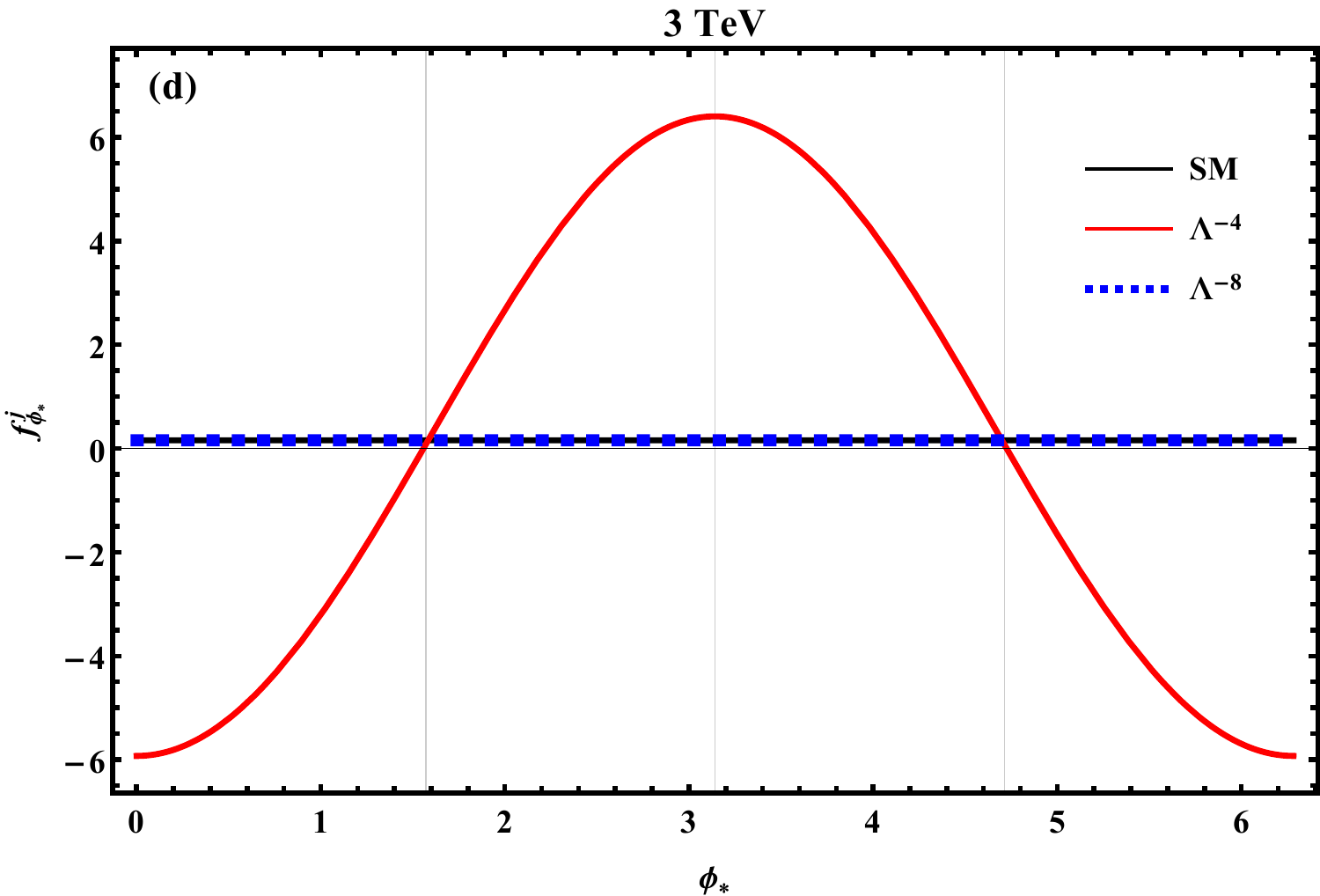}
\vspace*{-8mm}
\caption{{\it Normalized angular distribution in the azimuthal angle $\phi_*^{}$
for different collision energies,}
$\sqrt{s}=(250\,\text{GeV}, 500\,\text{GeV}, 1\,\text{TeV}, 3\,\text{TeV})$.
{\it In each plot, the black, red and blue curves denote the contributions from the
SM, the ${\cal O}(\cut^{-4})$ and ${\cal O}(\cut^{-8})$ terms, respectively,
where the blue and black curves nearly overlap.
We use a laboratory polar angle cut $\delta=0.2$ for illustration.}}
\label{figphi}
\label{fig:3}
\label{fig:44}
\end{figure}

Using the above results, we present numerical results
for the normalized angular distribution functions
of $\,\theta$,\, $\theta_*^{}$, and $\phi_*^{}$,\,
in Fig.\,\ref{figtheta}, Fig.\,\ref{figtz}, and Fig.\,\ref{figphi}, respectively.
In each figure, the four plots have input different collision energies
$\sqrt{s}=(250\,\text{GeV}, 500\,\text{GeV}, 1\,\text{TeV},$ $3\,\text{TeV})$,
corresponding to the expected collision energies of the ILC, CEPC and FCC-ee, of possible ILC energy upgrades,
and the design energy of CLIC.
In each plot, we use black, blue, and red curves to denote the contributions from the
SM, the ${\cal O}(\cut^{-4})$ term and the ${\cal O}(\cut^{-8})$ term, respectively.
In this analysis, we set a lower cut $\,\sin\theta > \sin\delta$
(with $\delta=0.2$)\, for illustration,
which corresponds to a lower cut on the photon transverse momentum
$\,P_T^\gamma > q\sin\delta\simeq 0.2q\,$.

\vspace*{1mm}

It is of interest to examine the behaviours of the angular distribution functions
$\,f_{\xi}^j$\, in the high-energy limit $\,s\gg M_Z^2$\,.\,
For all the functions $f_{\theta}^j$ and $f_{\theta_*}^j$,
the coefficients of all trigonometric functions approach constants.
This is why Figs.\,\ref{fig:22} and \ref{fig:33} show that
the distributions in $\theta$ and $\theta_*^{}$
are not sensitive to the collision energy $\sqrt{s\,}$,
as we vary the collision energy
$\sqrt{s}=(250\,\text{GeV}, 500\,\text{GeV}, 1\,\text{TeV}, 3\,\text{TeV})$
in the four plots.
For the angular functions $f_{\phi_*}^0$ and $f_{\phi_*}^2$,
the coefficients of $\,\cos\phi_*^{}$\, are suppressed by $\,M_Z^{}/\!\sqrt{s}$\,,\,
so they approach the constant term $\frac{1}{\,2\pi\,}$
for $\,s\gg M_Z^2$.\,
This is why in Fig.\,\ref{figphi} the angular functions
$f_{\phi_*}^0$ and $f_{\phi_*}^2$ (shown as the black and blue curves)
appear fairly flat and largely overlap each other.
In contrast, for the angular function $\,f_{\phi_*}^1$,\,
the coefficient of $\cos \phi_*^{}$ is enhanced by an energy factor $\sqrt{s}/M_Z^{}$,
and can be much larger than the constant term.
For $s\gg M_Z^2$,\, we can approximate Eq.\eqref{f1} in the following form:
\beqa
\label{eq:f1-phi-asymp}
f_{\phi_*}^1 &=&
\frac{1}{2\pi}\!\(1\!+\frac{1}{2}\!\cos\!2\phi_*^{}\)
-\frac{9\pi\sqrt{s\,}}{\,128M_Z^{}\,}\cos\!\phi_*^{}
+ {\cal O}\!\!\(\!\frac{M_Z^2}{s},\delta\!\)
\nn\\[2mm]
&\simeq& 0.159\!\(1\!+\frac{1}{2}\!\cos\!2\phi_*^{}\)
- 0.606\!\(\!\frac{\sqrt{s\,}}{250\,\text{GeV}}\!\)\!\cos\!\phi_*^{}\!
+ {\cal O}\!\!\(\!\frac{M_Z^2}{s},\delta\!\)\!.
\hspace*{15mm}
\eeqa
We see that the $\cos\!\phi_*^{}$ term dominates $\,f_{\phi_*}^1$
for $\sqrt{s\,}\!>\! 250\,$GeV.\,
This also explains why in Fig.\,\ref{figphi} the magnitude of the angular function
$f_{\phi_*}^1$ (red curve) grows almost linearly with
the collision energy $\sqrt{s\,}$,\, and has its maximum at
$\,\phi_*^{}\!= \pi\,$ and minima at
$\,\phi_*^{}\!= 0,2\pi\,$.\,

\vspace*{1mm}

In the laboratory frame, the opening angle $\Delta\theta_{\ell\ell}^{}$
between the two outgoing leptons from
$Z$ decay is a function of $\,\theta_*^{}$.\,
For $\sqrt{s}\!\sim\! M_Z^{}$, we expect $\Delta\theta_{\ell\ell}^{} \!\sim\!\pi$,\,
while for $s\gg M_Z^2$,\, we have $\Delta\theta_{\ell\ell}^{}\!\to 0$\,.

\vspace*{2mm}	
\subsection{\hspace{-5mm}.\,Probing the New Physics Scale in the \boldmath{$ZZ\gamma$} Coupling}
\label{sec:3.2}
\vspace*{1mm}

In this subsection, we analyze how to probe the new physics scale
with ${\cal O}(\cut^{-4})$ contributions (Section\,\ref{sec:3.2.1})
and including up to ${\cal O}(\cut^{-8})$ contributions (Section\,\ref{sec:3.2.2})
for the $e^+e^-$ collision energy $\sqrt{s}=3$\,TeV.
We demonstrate that making use of the angular observables
can suppress the SM backgrounds efficiently.
Extension of this analysis to other collision energies will be presented
in Section\,\ref{sec:3.2.3}.

\vspace*{1mm}
\subsubsection{\hspace{-5mm}.\,Analysis of the \boldmath{${\cal O}(\Lambda^{\!-4})$} Contribution}
\label{sec:3.2.1}
\vspace*{1mm}

In order to analyze the sensitivity to the new physics scale $\cut$,\,
we are strongly motivated by the $\phi_*^{}$ distributions in Fig.\,\ref{figphi}.
Inspecting Fig.\,\ref{figphi}, we divide the range of $\phi_*^{}$
into two regions (bins) ---
the regions\,$(a)$ and\,$(b)$. Region\,$(a)$ includes the ranges
$\left[0,\,\fr{\pi}{2}\right]\bigcup\left[\frac{3\pi}{2},\, 2\pi\right)$
and region\,$(b)$ is the range
$\(\frac{\pi}{2},\, \frac{3\pi}{2}\)$.\,
The sum of the areas of regions $(a)$ and $(b)$ ($S^a_{j}$ and $S^b_{j}$) is
$\,S^a_{j}\!+S^b_{j}=1$,\, because the angular function is normalized to unity.
However, the difference $\,S^b_{1}\!-S^a_{1}\gg 1$\,
for the angular function $f_{\phi_*}^1$,\,
while  $S^b_{0}\!-S^a_{0}$ is subject to a strong cancellation in the SM contribution
$f_{\phi_*}^0$ since $f_{\phi_*}^0$ is rather flat.
We can make use of this feature to suppress the SM background
and enhance significantly the ${\cal O}(\Lambda^{\!-4})$
signal at the same time. To this end, we define the functions
\begin{eqnarray}
\label{eq:Oj}
\mathbb{O}_j^{} \, \equiv \,
\sigma_j^{}
\(\int_{\frac{\pi}{2}}^{\frac{3\pi}{2}}-\int_0^{\frac\pi 2}-\int_{\frac{3\pi}{2}}^{2\pi}
  \) \!f^j_{\phi_*^{}}\di\phi_*^{}
=\,\sigma_j^{}(S^b_j-S^a_j) \,,
\end{eqnarray}
for $j=0,1,2$.\,
Furthermore, Fig.\,\ref{fig:44} shows that
the ${\cal O}(\Lambda^{\!-8})$ distribution $f_{\phi_*}^2$ is fairly flat and largely overlaps
with $f_{\phi_*}^0$ of the SM.
Thus, the ${\cal O}(\Lambda^{\!-8})$ contributions to $\,S^b_{2}\!-S^a_{2}$
also cancel strongly and become negligible.
Hence, for the signal analysis, here we only need to consider the ${\cal O}(\Lambda^{\!-4})$
contributions.

\vspace*{1mm}

We define the signal and background event numbers as follows:
\beqs
\label{eq:SI-BI}
\vspace*{-1.5mm}
\begin{eqnarray}
\label{eq:SI}
S_I^{} &\,\simeq\,& \mathbb{O}^1_{} \!\times \mathcal{L}\times\epsilon
= N_1^b\!-\!N_1^a
\,,
\\
\label{eq:BI}
B_I^{} &\,=\,& \mathbb{O}^0_{} \!\times \mathcal{L}\times\epsilon
= N_0^b-N_0^a
\,,
\end{eqnarray}
\eeqs
where $\,\mathcal{L}\,$ denotes the luminosity and $\,\epsilon\,$ is the detection efficiency.
In the above, $(N_1^a,\, N_1^b)$ denote the signal event numbers in regions (a) and (b),
respectively, which are given by the $\mathcal{O}(\cut^{-4})$ contributions
since the $\mathcal{O}(\cut^{-8})$ contributions strongly cancel as to be negligible.
Besides, $N_0^a$ and $N_0^b$ denote the SM event numbers in regions $(a)$ and $(b)$,
respectively.
We note that for the observable \eqref{eq:Oj} the nonzero signals actually come from the
$\,\cos\!\phi_*^{}\,$ term in the $f_{\phi_*^{}}^{1}$ distribution \eqref{f1},
\beqa
\label{eq:SI-cos(phi)}
\,S_I^{} \simeq N_1^b\!-\!N_1^a
\propto\int_0^{2\pi}\!\!\di\phi_*^{}\,|\!\cos\!\phi_*^{}|\,,
\eeqa
whereas the contributions to $\,S_I^{}$\, from all the other terms of $\,f_{\phi_*^{}}^{1}$
vanish after integration over $\phi_*^{}$.

\vspace*{1mm}

Next, we note that the SM background distribution $f_{\phi_*^{}}^{0}$ in Eq.\eqref{f0}
and Fig.\,\ref{fig:44}  is dominated by the constant term $\,\fr{1}{\,2\pi\,}$.\,
Thus, the SM background $B_I^{}$ is small
due to the large cancellation between regions $(a)$ and $(b)$,
while its combined statistical error $\Delta_B^{}$ from the two regions
receives no cancellation:
\beqs
\label{eq:BI-DeltaBI}
\vspace*{-2mm}
\begin{eqnarray}
B_I^{} &\,=\,& N_0^b-N_0^a\ll N_0^{a,b}\,,
\label{eq:BI}
\\[1mm]
\Delta_{B_I}^{} \!&=&\!
\sqrt{\Delta_a^2\!+\!\Delta_b^2\,}
=\sqrt{N_0^{a}\!+\!N_0^{b}\,}
=\sqrt{\sigma_0^{}\!\times\!\mathcal{L}\!\times\epsilon\,}\,.
\label{eq:Delta-BI}
\end{eqnarray}
\eeqs
Thus, despite that the SM background $B_I^{}$ receives large cancellation,
its statistical error $\Delta_{B_I}^{}$ does not
since it arises from the $\phi_*^{}$-independent constant term of
$\,f_{\phi_*^{}}^0$ via
$\Delta_{B_I}^{2}\!=\!N_0^{a}\!+\!N_0^{b}\propto\!
 \int_0^{2\pi}\!\!f_{\phi_*^{}}^0\di\phi_*^{}$.\,
With these,  we estimate the signal significance by
\begin{eqnarray}
\label{eq:Z4}
\mathcal{Z}_{4}^{}
=\frac{|S_I^{}|}{\,\Delta_{B_I}^{}\,}
=\frac{|\mathbb{O}^1(Z\gamma)|}{\sqrt{\sigma^0_{}(Z\gamma)\,}\,}
\!\times\!
\sqrt{\text{Br}(Z\!\!\to\!\ell\ell)\!\times\!\mathcal{L}\!\times\!\epsilon\,} \, .
\end{eqnarray}

\vspace*{1mm}

We note that $\mathbb{O}^1_{}$ and $\sigma^0_{}$
are functions of the angular cut $\delta$ (corresponding to the photon transverse momentum
cut $P_T^\gamma >q\sin\delta$\,).
Fig.\,\ref{figphi} shows that the magnitude $|f_{\phi_*}^1|$ is very small around
$\phi_*^{}=\frac{\pi}{2},\frac{3\pi}{2}$\, since it is dominated
by the $\cos\!\phi_*^{}$ term as in Eq.\eqref{eq:f1-phi-asymp}.
So we may cut off the nearby area to reduce the SM backgrounds.
For this, we introduce a cut parameter $\,0<\phicc<\fr{\pi}{2}\,$,\,
using which region $(a)$ reduces to
$\left[0,\, \phic^{}\right]\bigcup\left[2\pi\!-\!\phi_c^{},\, 2\pi\right)$
and region $(b)$ becomes $\,\left[\pi\!-\!\phic^{},\, \pi\!+\!\phic^{}\right]\,$.\,
We then compute the corresponding signal observable
$\,\mathbb{O}^1_c\,$ and the background fluctuation $\sqrt{\sigma^0_c\,}\,$.\,
In Fig.\,\ref{fig:3}, the angular function $f_{\phi_*}^0$ appears rather flat,
so we obtain a simple expression for $\sigma^0_c$\,, as follows:
\beqs
\label{eq:O1c-sigma0c}
\begin{eqnarray}	
|\mathbb{O}^1_c| &=&
|\sigma_1^{}|\!\(
\int_{\pi-\phi_c^{}}^{\pi+\phi_c^{}}
- \int_0^{\phi_c^{}} - \int_{2\pi-\phi_c^{}}^{2\pi}
\) \!f^1_{\phi_*}\di\phi_*^{}
\nn\\[1mm]
&\simeq&\frac{\,3\alpha (1\!-\!4 s_W^2)M_Z^{}(s\!-\!M_Z^2)\!
\left[3 (\pi \!-\!2 \delta )(s\!+\!M_Z^2)
- \!(s\!-\!3 M_Z^2)\sin\!2\delta\right]\sin\!\phi_c^{}\,}
{\,256\, s_W^{} c_W^{}\, \cut^4\, s^{\frac 3 2}_{}} \,,
\hspace*{15mm}
\label{eq:O1c}
\\[1.5mm]
\sigma^0_c
&\simeq& \frac{\,2\phi_c^{}\,}{\pi}\sigma^0
\nn\\[1mm]
&=&
\frac{\,\alpha^2(1\!-\!4s_W^2\!+\!8s_W^4)\!
\left[-\cos\!\delta(s\!-\!M_Z^2)^2\!+\!2(s^2\!+\!M_Z^4)\!
\ln\!\(\!\cot\!\frac{\delta}{2}\)\right]\!\phi_c^{}\,}
{c_W^2s_W^2(s\!-\!M_Z^2)s^2}\,,
\label{eq:sigma0c}
\end{eqnarray}
\eeqs
where $\,\alpha \!=e^2/4\pi$\, is the fine structure constant.

\vspace*{1mm}

For our analysis, we choose the values of
$\,\phi_c^{}\!=\phicm$\, and $\,\delta = \delta_{\text{m}}^{}\,$
such that the signal significance
$\,\mathcal{Z}_4^{}=(|\mathbb{O}^1_c|/\!\sqrt{\sigma^0_c\,})\!
   \sqrt{\text{Br}(Z\!\to\!\ell\ell)\!\times\!\LL\!\times\!\ep\,}$\,
is maximized.
Thus, $\,\phi_c^{}=\phicm\,$ corresponds to the maximum of the function
$\,\sin\phicc/\!\sqrt{\phicc}$\,,\,
and we derive $\,\phicm\simeq 1.17$,\,
which is independent of the collision energy $\!\sqrt{s\,}$.\,
The value of $\,\delta_{\text{m}}^{}\,$ required to obtain the maximal significance of $\,\mathcal{Z}_4^{}\propto (|\mathbb{O}^1_c|/\!\sqrt{\sigma^0_c\,})$\,
depends on the collision energy $\sqrt{s\,}$\,.\,
For high energies $s\gg M_Z^2$\,,\,  we find
$\,\delta_{\text{m}}^{}\simeq 0.329$\,.\,

\begin{figure}[t]
\includegraphics[width=7.4cm,height=5.5cm]{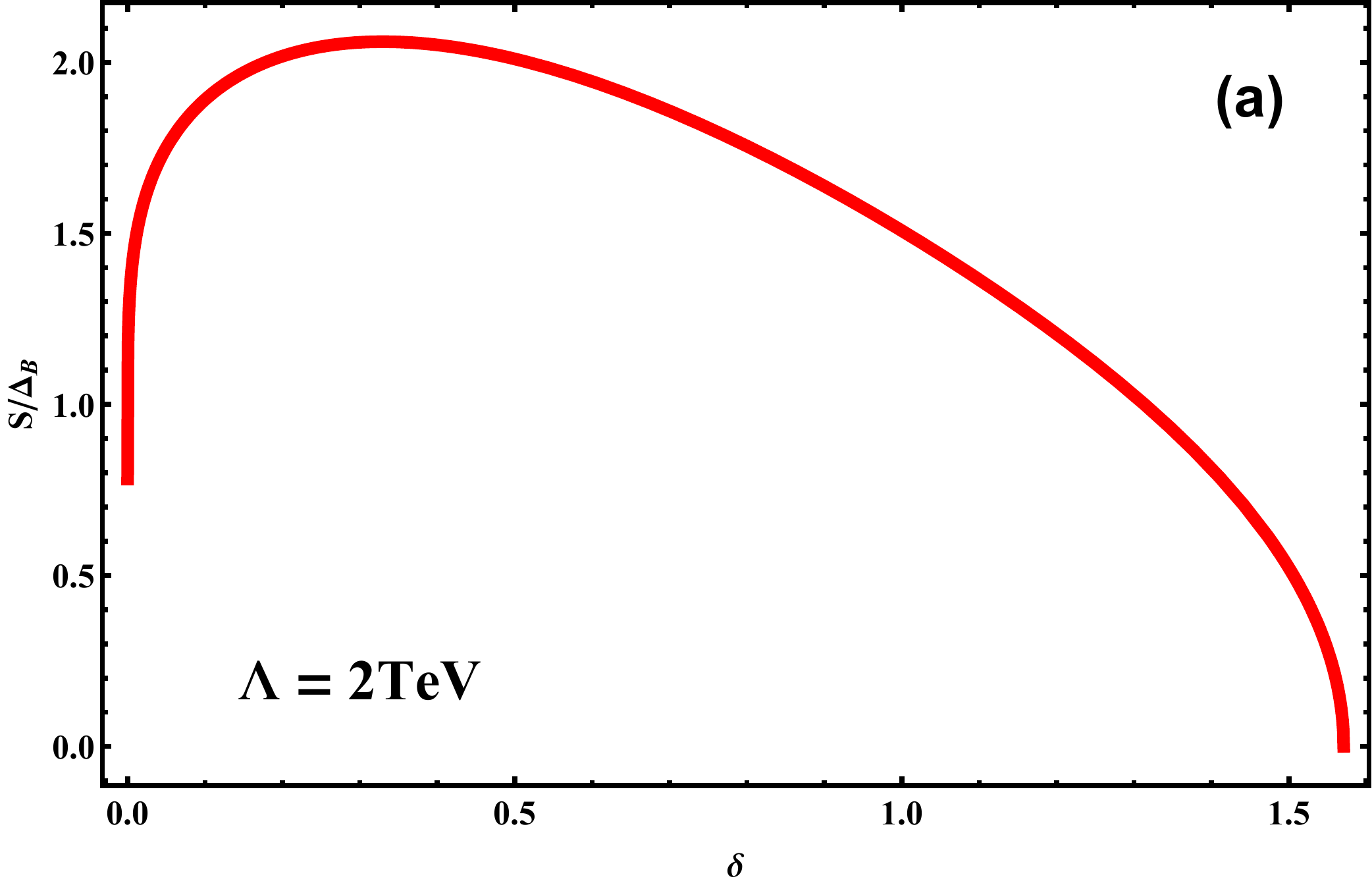}
\hspace*{2mm}
\includegraphics[width=7.4cm,height=5.5cm]{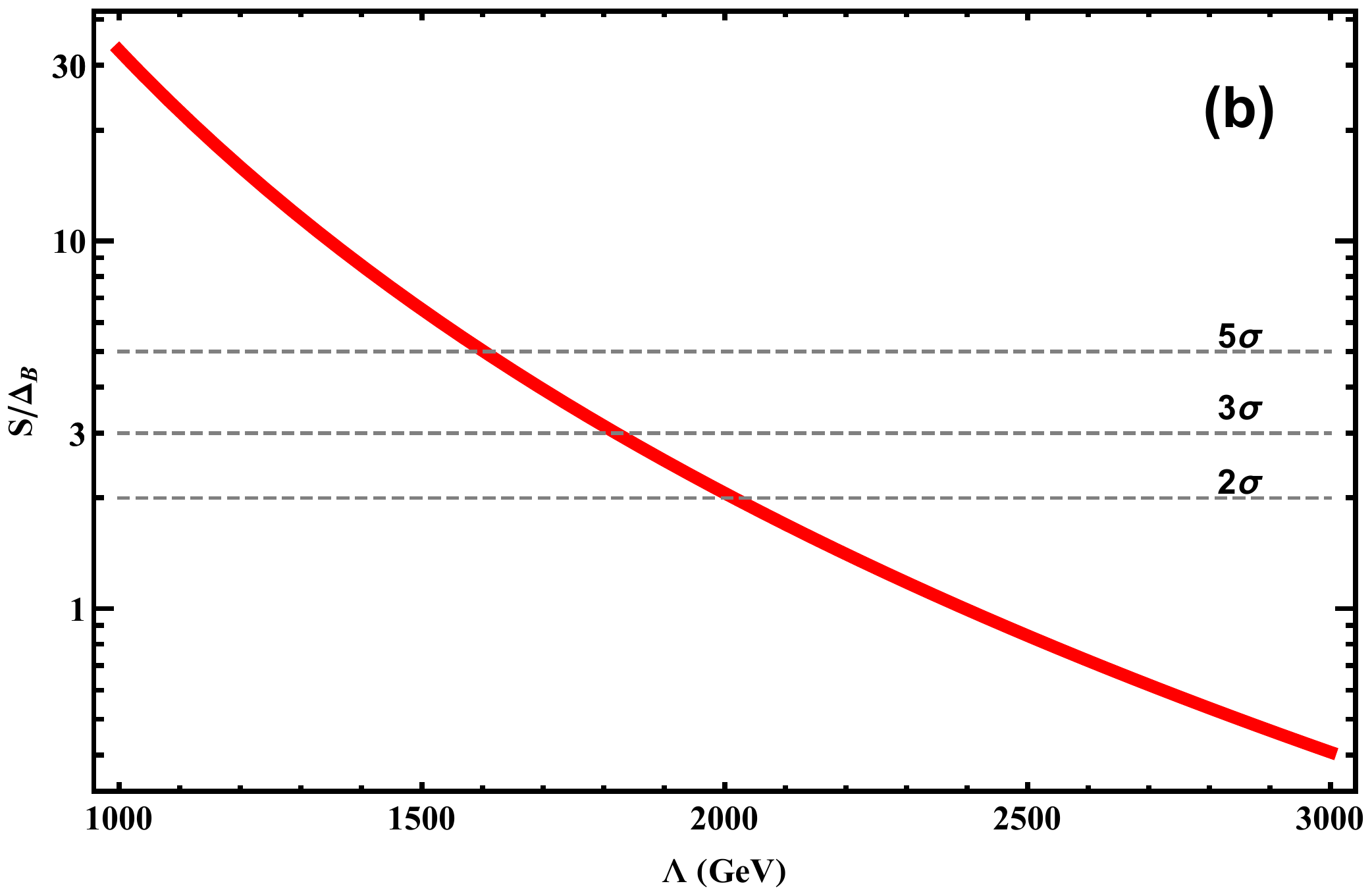}
\caption{{\it Analysis of the significance $\mathcal{Z}_{4}^{}\!=S/\Delta_B^{}$.
Plot\,(a) depicts $\mathcal{Z}_{4}^{}$ as a function of $\,\delta$\,.
Plot\,(b) presents $\,\mathcal{Z}_{4}^{}$\, for $\,\delta=\delta_{\text{m}}^{}$\,
as a function of the new physics scale $\Lambda$\,.\, For illustration, we choose
the collision energy} $\sqrt{s\,}=3\,\text{TeV}$ {\it and the integrated luminosity}
$\,\mathcal{L}=2\,\text{ab}^{-1}$.
}
\label{Z}
\label{fig:55}
\vspace*{3mm}
\end{figure}

\vspace*{1mm}

We present in Fig.\,\ref{Z} the signal significance obtained in this way
for the collision energy $\sqrt{s}=3$\,TeV.
We input the total leptonic branching fraction $\text{Br}(Z\!\to\!\ell^-\ell^+)\simeq 0.10$,\,
and assume an integrated luminosity $\mathcal{L}=2\,\text{ab}^{-1}$
and an ideal detection efficiency $\,\epsilon =1\,$ for simplicity.
In Fig.\,\ref{Z}(a), we depict the significance $\,\mathcal{Z}_4^{}$\,
as a function of the angular cut $\delta$, which exhibits the maximum at
$\,\delta_{\text{m}}^{}\simeq 0.33$\, for $\cut=1$\,TeV,\, as expected.
Thus, under the angular cuts
$\,(\phi_c^{},\,\delta)\!=(\phicm,\,\delta_{\text{m}}^{})$,\,
we derive
\begin{eqnarray}
(\sigma^0_c,\,\mathbb{O}^1_c) &\simeq&
\left(23.1,\, 11.1\!\left(\!\frac{\text{TeV}}{\Lambda}\right)^{\!\!4}\right)\!
\text{fb}\,.
\label{s0o1}
\eeqa
For illustration, we may then use Eq.\eqref{eq:Z4} to estimate
the signal significance as follows:
\beqa
\label{eq:Z4-E3TeV}
\mathcal{Z}_{4}^{} \,\simeq\,
32.7\!\left(\!\frac{\,\text{TeV}\,}{\Lambda}\!\right)^{\!\!4}
\simeq\,
5.0\!\left(\!\frac{\,1.60\,\text{TeV}\,}{\Lambda}\!\right)^{\!\!4},
\end{eqnarray}
which is plotted in Fig.\,\ref{fig:55}(b).
From this, we find that the probe of the new physics scale can reach
$\,\cut= (2.0,\,1.8,\,1.6)$~TeV at $(2\sigma,\, 3\sigma,\, 5\sigma)$ level,
respectively.

\vspace*{1mm}

We note that the practical detection efficiency would be smaller than 100\%,
so the actual sensitivity may be somewhat weaker.
But, as we show later in Eqs.\eqref{lam4} and \eqref{lam8} of Section\,\ref{sec:3.3},
the sensitivity reach for $\cut$ has rather weak dependences
on the integrated luminosity and detection efficiency,
namely, $\,\cut\propto (\mathcal{L}\,\ep)^{\frac{1}{8}}\,$
at ${\cal O}(\cut^{-4})$ and $\,\cut\propto (\mathcal{L}\,\ep)^{\frac{1}{16}}\,$ at
${\cal O}(\cut^{-8})$.
Hence, increasing $\,\mathcal{L}\,$ or $\,\ep\,$
only has minor effect on the sensitivity reach
of the new physics scale $\cut$.
In contrast, raising the collision energy $\sqrt{s\,}$ can do more to improve the
sensitivity reach of $\cut$ because
$\,\cut\propto (\!\sqrt{s\,})^{\frac{1}{2}}\,$
at ${\cal O}(\cut^{-4})$ and
$\,\cut\propto (\!\sqrt{s\,})^{\frac{5}{8}}\,$
at ${\cal O}(\cut^{-8})$.\,

\vspace*{1mm}
\subsubsection{\hspace{-5mm}.\,Analysis Including the \boldmath{${\cal O}(\Lambda^{\!-8})$} Contribution}
\label{sec:3.2.2}
\vspace*{1mm}

In this subsection, we present an analysis to include the contribution of
${\cal O}(\Lambda^{\!-8})$ without invoking strong cancellation on it.
Since the ${\cal O}(\Lambda^{\!-8})$ term has a higher power of energy dependence,
it may have better sensitivity when
the collision energy is higher, e.g., $\sqrt{s}=3$\,TeV, even though it is suppressed by $\Lambda^{\!-8}$.\, 
In effective theory analyses considering one operator at time, it is the customary to
compute the full \,$|\text{amplitude}|^2$ including both the interference term with the SM
and the squared term of the effective operator when calculating the cross section.
In this way the full information
concerning a single effective operator is retained.

\vspace*{1mm}

We see from Fig.\,\ref{fig:3} that both the distributions
$f_{\phi^*}^0$ and $f_{\phi^*}^2$ are rather flat. Thus it is hard to discriminate the
${\cal O}(\Lambda^{\!-8})$ contribution from the $\phi_*^{}$ distributions.
Hence, in order to enhance the signal sensitivity to the ${\cal O}(\Lambda^{\!-8})$
contribution, we study instead the distributions in $\theta$ and $\theta_*^{}$.
For this, we choose the region $\,\theta\in [\delta,\pi\!-\delta]$\, and
$\,\theta_*^{}\!\in [\delta_*^{},\pi\!-\delta_*^{}]$.\,
With the angular cuts $(\delta,\,\delta_*^{})$, we compute the SM contribution
$\,\sigma_c^0(Z\ga)$,\,
the ${\cal O}(\cut^{-4})$ contribution $\,\sigma_c^1(Z\ga)$,\,
and the ${\cal O}(\cut^{-8})$ contribution $\,\sigma_c^2(Z\ga)\,$
as follows:
{\small
\beqs
\label{eq:sigma-c-012}
\begin{eqnarray}
\sigma_c^0(Z\gamma) &=&
\frac{\,e^4\!\(8 s_W^4\!-\!4 s_W^2\!+\!1\)\,}
{\,32\pi s_W^2 c_W^2\!\(s\!-\!M_Z^2\)\!s^2\,}\!\times\!\frac{1}{16}\!
\left[4\cos\!\delta\(9\cos\!\delta_*\!-\cos\!3\delta_*\)\!M_Z^2\,s\,  \right.
\\
&& \hspace*{2mm}
\left.
-\(15 \cos\!\delta_*\!+\cos\!3\delta_*\)\!(s^2\!+\!M_Z^4)\!\!
 \(\!\!\cos\!\delta \!+\! 2\ln\tan\!\frac{\delta}{2}\) \!\right]\!,
\hspace*{10mm}
\nn
\\[2mm]
\sigma^1_c(Z\gamma) &=&
\pm\frac{\,e^2(1\!-\!4s_W^2) M_Z^2\!
\(s\!-\!M_Z^2\)\,}{32\pi s_W^{}c_W^{}\cut^{\!4}s^2}
\frac{\,\left[2(5\!-\!\cos\!2\delta_*^{})s\!+\!
\(\cos\!2\delta_*^{}\!\!+\!7\)\!M_Z^2 \right]\!\cos\!\delta \!\cos\!\delta_*^{}\,}{8},
\hspace*{15mm}
\\[2mm]
\sigma_c^2(Z\gamma)&=&
\frac{\,\(8 s_W^4\!-\!4 s_W^2\!+\!1\)\!M_Z^2\!\(s\!-\!M_Z^2\)^{\!3}\,}
{192\pi\cut^{\!8} s^2}
\\
&&
\times
\frac{\,\cos\!\delta\,}{64}\!
\left[(7\!+\!\cos\!2\delta)\!\(9\cos\!\delta_*\!-\!\cos\!3\delta_*\!\)\!s
\!+\! (5\!-\!\cos\!2\delta)\!\(15\cos\!\delta_*\!+\!\cos\!3\delta_*\)\!M_Z^2
\right]\!,
\nn
\end{eqnarray}
\eeqs
}
\hspace*{-2.5mm}
where the $\pm$ signs of the ${\cal O}(\cut^{\!-4})$ term correspond to
the two possible signs of a given dimension-8 operator,
$\text{sign}(c_j^{})=\pm$,\,
as shown in Eq.\eqref{cj}.
If we take the limit $(\delta,\,\delta_*^{})\!\to\! 0$\, in the above formulas,
we find that they reduce consistently to Eq.\eqref{Zgamma}, as expected.

\begin{figure}
\centering
\includegraphics[width=10cm,height=7cm]{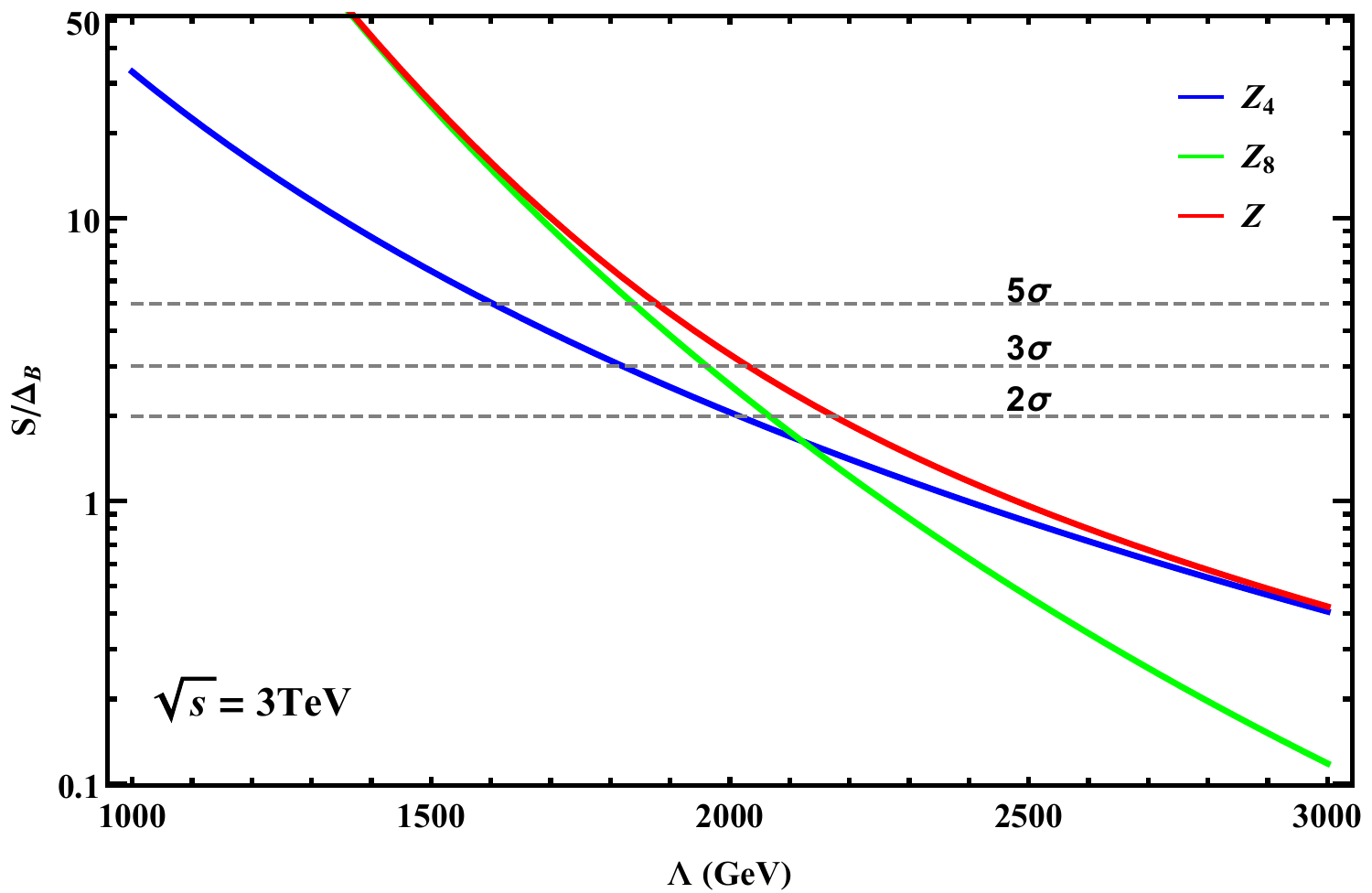}
\vspace*{-3mm}
\caption{{\it The significance $\,\SZZ_{8}^{}\!=S/\!\Delta_B^{}$\,
is presented as a function of $\,\Lambda\,$ as a green curve
for the cut parameters $(\delta,\,\delta_{*})=(\delta_m^{},\,\delta_{*m})$.\,
The significance $\,\SZZ_{4}^{}$ [from Fig.\,\ref{fig:55}(a)] is shown
as the blue curve for comparison. The combined significance
$\,\SZZ\!=\!\sqrt{\SZZ_4^2\!+\!\SZZ_8^2\,}$\, is depicted by the red curve.
For illustration, we choose}
$\sqrt{s\,}=3\,\text{TeV}$ {\it and} $\,\mathcal{L}=2\,\text{ab}^{-1}$.
}
\label{Z8}
\label{fig:66}
\end{figure}

\vspace*{1mm}

We can then estimate the corresponding signal significance to be
\begin{eqnarray}
\label{eq:Z48}
\mathcal{Z}_{8}^{} =
\frac{|S_{II}^{}|}{\,\Delta_{B_{II}}^{}\,}
= \frac{\,|\sigma_c^1(Z\gamma)\!+\!\sigma_c^2(Z\gamma)|\,}{\!\sqrt{\sigma_c^0(Z\gamma)\,}\,}
  \!\times\! \sqrt{\text{Br}(Z\!\!\to\!\ell\ell)\!\times\!\mathcal{L}\!\times\!\epsilon\,} \, .
\end{eqnarray}
For comparison with our two-bin analysis of $\mathcal{Z}_{4}^{}$ in Sec.\,\ref{sec:3.2.1},
we could use the same definition of the two bins (regions), (a) and (b), for the
$\phi_*^{}$ distribution (Fig.\,\ref{fig:44}), as given above Eq.\eqref{eq:Oj}.
But, the signal and background cross sections in Eq.\eqref{eq:Z48}
correspond to integrating the distributions $f_{\phi_*^{}}^{j}$ ($j=0,1,2$)
over the full range of $\,\phi_*^{}\in [0,\,2\pi)$,\,
for which all the $\cos\!\phi_*^{}$ and $\cos\!2\phi_*^{}$ terms vanish
and only the constant terms $\,\frac{1}{2\pi}\,$ in Eq.\eqref{eq:f-phi*} survive.
This means that we {\it sum up the two bins (a) and (b),} rather than
{\it taking the difference of the two bins} (as we did for $\mathcal{Z}_{4}^{}$
in Sec.\,\ref{sec:3.2.1}).
Namely, the current signal $S_{II}^{}$, the SM background $B_{II}^{}$ and its
statistical error $\Delta_{B_{II}}^{}$ take the following forms,
\beqs
\label{eq:SII-BII-DBII}
\beqa
\label{eq:SII}
S_{II}^{} &=&
\left[\sigma_c^1(Z\gamma)\!+\!\sigma_c^2(Z\gamma)\right]\!\times\!
\left[\text{Br}(Z\!\!\to\!\ell\ell)\!\times\!\mathcal{L}\!\times\!\epsilon\right]
= \over{N}_{\!12}^{a}+\over{N}_{\!12}^{b}\,,
\\[1mm]
\label{eq:BII}
B_{II}^{} &=&
\sigma_c^0(Z\gamma)\!\times\!
\left[\text{Br}(Z\!\!\to\!\ell\ell)\!\times\!\mathcal{L}\!\times\!\epsilon\right]
= \over{N}_0^a + \over{N}_0^b \,,
\\[1mm]
\label{eq:DBII}
\Delta_{B_{II}}^{} \!&=&\! \sqrt{B_{II}^{}}
= (\over{N}_0^a + \over{N}_0^b)^{\!{1}/{2}}_{}\,\,.
\eeqa
\eeqs
Here $(\over{N}_{\!12}^{a},\,\over{N}_{\!12}^{b})$ denote the signal event numbers in the
regions (a) and (b) of the distributions $f_{\phi_*^{}}^{1,2}$,\,
in which only the constant terms $1/(2\pi)$ contribute to the sum
$\,\over{N}_{\!12}^{a}\!+\!\over{N}_{\!12}^{b}$
under the uniform integration over full range of $\phi_*^{}$.\,
Also, $\over{N}_0^a\!+\!\over{N}_0^b$ denotes
the sum of the background event numbers in the
regions (a) and (b) of the distribution $f_{\phi_*^{}}^{0}$.\,
(Here our notations of the background event numbers $(\over{N}_0^a,\,\over{N}_0^b)$
differ from $({N}_0^a,\,{N}_0^b)$ in the other independent analysis of
Sec.\,\ref{sec:3.2.1} because we have added different $\theta$ cuts separately
for the two analyses in Sec.\,\ref{sec:3.2.1} and Sec.\,\ref{sec:3.2.2}.)

\vspace*{1mm}

To obtain the maximal signal significance, we derive the corresponding values
of the angular cuts
$(\delta,\,\delta_*^{})=(\delta_{\text{m}}^{},\,\delta_{*\text{m}}^{})$,
which are $(\delta_{\text{m}}^{},\,\delta_{*\text{m}}^{})\simeq (0.623,\,0.820)$.
Inputting $\,\text{Br}(Z\!\!\to\!\ell\ell)\!\simeq 0.10$\, and choosing
$\sqrt{s}=3\,$TeV, $\mathcal{L}=2\,\text{ab}^{-1}$ and $\epsilon=1$,\,
we compute the cross section for $Z\ga$ production:
%
{\small
\begin{eqnarray}
\label{eq:sigma-Zga-3TeV}
\sigma(Z\gamma)
= \left[10.1 \pm 0.0251\!\(\!\frac{\,2\text{TeV}\,}{\Lambda}\!\)^{\!\!4}
	+0.554\!\(\!\frac{\,2\text{TeV}\,}{\Lambda}\!\)^{\!\!8}\right]
\!\text{fb}\,.
\end{eqnarray}
}
Thus, from Eq.\eqref{eq:Z48} we estimate the signal significance
{\small
\begin{eqnarray}
\label{eq:Z8-E3TeV}
\mathcal{Z}_{8}^{}
\simeq\,
\left|\pm 1.79\!\(\!\frac{\,\text{TeV}\,}{\Lambda}\!\)^{\!\!4}+
631\!\(\!\frac{\,\text{TeV}\,}{\Lambda}\!\)^{\!\!8}\right|
\,\simeq\,
\left|
\pm 0.112\!\(\!\frac{\,2\,\text{TeV}\,}{\Lambda}\!\)^{\!\!4}+
2.46\!\left(\!\frac{\,2\,\text{TeV}\,}{\Lambda}\!\right)^{\!\!8}\right|\!.
\hspace*{12mm}
\end{eqnarray}
}
\hspace*{-3.5mm}
We see that for the collision energy $\sqrt{s}=3\,$TeV,
the ${\cal O}(\cut^{\!-8})$ term dominates the sensitivity.

\vspace*{1mm}

In Appendix\,\ref{sec:B}, we have systematically proven that the possible
correlation between the two significances $\mathcal{Z}_{4}^{}$ and $\mathcal{Z}_{8}^{}$
are fully negligible. Hence, it is well justified to combine
$\mathcal{Z}_{4}^{}$\, and $\,\mathcal{Z}_{8}^{}$\,
for achieving an improved sensitivity reach to the new physics scale $\cut$\,,
\vspace*{-2mm}
\beqa
\mathcal{Z} = \!\sqrt{\mathcal{Z}_{4}^2\!+\!\mathcal{Z}_{8}^2\,}\,.\,
\eeqa
This is depicted by the red curve in Fig.\,\ref{fig:66}.
In this way, we find that the new physics scale can be probed up to
$\,\cut \simeq (2.2,\,2.0,\,1.9)$~TeV at the $(2\sigma,\, 3\sigma,\, 5\sigma)$ levels,
respectively. These numbers apply for both $\pm$ signs in Eq.\eqref{eq:Z8-E3TeV},
since we find that the case of minus sign in Eq.\eqref{eq:Z8-E3TeV}
only causes a tiny difference in the $\cut$ bound by less than 1\%.
Hence, the ${\cal O}(\cut^{\!-4})$ term in Eq.\eqref{eq:Z8-E3TeV}
has negligible effect for the collider energy $\sqrt{s\,}=3$\,TeV.\,
As we will show in Section\,\ref{sec:3.2.3}, this feature applies to
all cases with $\sqrt{s\,}\gtrsim 1$\,TeV.\,

{
\tabcolsep 1pt
\begin{center}
\begin{table}[t]
\vspace*{-2mm}
\begin{tabular}{c|ccccc}
\hline\hline
&&&&&
\\[-4.1mm]
$\sqrt{s\,}$~(GeV) & 250 & 500 & 1000~~ & 3000~ & 5000
\\[0.3mm]
\hline
\\[-4.4mm]
$\SZZ_{4}^{}\!:~(\delta_\mm^{(4)},\,\phi_{c}^{\mm})$
& (0.368,\,1.17) & (0.340,\,1.17) & (0.332,\,1.17) & (0.329,\,1.17)) & (0.329,\,1.17)
\\[-4.5mm]
\\
\hline
\\[-4.4mm]
~$\SZZ_{8}\!:~(\delta_\mm^{(8)},\, \delta_{*\mm}^{})$~ &
\,(0.608,\,0.692) & (0.616,\,0.790) & (0.621,\,0.814) &
(0.623,\,0.820) & (0.623,\,0.821)\,
\\[0.3mm]
\hline\hline
\end{tabular}
\vspace*{2mm}
\caption{{\it Summary of the optimal angular cuts for realizing the maximal signal
significance. For the signal significance $\SZZ_4^{}$,
we impose the cuts} $(\delta_\mm^{(4)},\,\phi_{c}^{\mm})$
{\it on the angular distributions of} $(\theta,\,\phi_*^{})$;
{\it whereas for the signal significance $\SZZ_8^{}$,
we set the cuts} $(\delta_\mm^{(8)},\, \delta_{*\mm}^{})$
{\it on the angular distributions of $(\theta,\,\theta_*^{})$.\,
These cuts correspond to the allowed angular ranges,
$\delta_\mm^{}\!<\theta <\pi-\delta_\mm^{}$,\,
$\delta_{*\mm}^{}\!<\!\theta_*^{}\!<\pi\!-\delta_{*\mm}^{}$,\, and
$\,\phi\in \left[0,\,\phi_c^{\mm}\right] \bigcup
\left[2\pi\!-\!\phi_c^{\mm},\, 2\pi\right) \bigcup
\left[\pi\!-\!\phic^{},\, \pi\!+\!\phic^{}\right]
$,\, where $\delta_{\mm}^{}\!=\!\delta_{\mm}^{(4)}\!$ or $\delta_{\mm}^{(8)}$.
}}
\label{tab:cut}
\label{tab:11}
\end{table}
\end{center}
}

\vspace*{-10mm}
\subsubsection{\hspace{-5mm}.\,Analysis of Different Collision Energies}
\vspace*{1mm}
\label{sec:3.2.3}

In this subsection, we further extend our analysis of $\sqrt{s\,}\!=\!3$\,TeV case to
different collision energies $\sqrt{s\,}\!=(250,\,500,\,1000,\,5000)$\,GeV,
in each case with a sample integrated luminosity $\mathcal{L}=2\,\text{ab}^{-1}$.

\vspace*{1mm}

Using the same method as we presented in Section\,\ref{sec:3.2.1}-\ref{sec:3.2.2},
we analyze the SM backgrounds and signal contributions for different collider energies.
For each given collider energy $\sqrt{s\,}$,\, we derive the optimal angular cuts
for realizing the maximal signal significance $\SZZ_4^{}$ and $\SZZ_8^{}$.\,
Namely, for the analysis of $\SZZ_4^{}$, we use the angular cuts
$(\delta_\mm^{},\,\phi_c^\mm)$ for angles $(\theta,\,\phi_*^{})$;\,
while for the analysis of $\SZZ_8^{}$, we use the angular cuts
$(\delta_\mm^{},\,\delta_{*\mm}^{})$ for angles $(\theta,\,\theta_*^{})$.\,
We summarize the optimal angular cuts for different collider energies in Table\,\ref{tab:cut}.
As we noted below Eq.\eqref{eq:O1c-sigma0c}, the dominant contribution to
the signal significance $\SZZ_4^{}$ depends on the cut $\phi_c^{}$ only through
a simple function $\,\sin\phicc/\!\sqrt{\phicc}$\,,\,
which does not depend on energy $\!\sqrt{s\,}$ and reaches its maximum
at $\phi_c^{\mm}\simeq 1.17$.\, This is why the optimal cut
$\phi_c^{\mm}$ is nearly independent of the collider energy $\sqrt{s\,}$,
as shown in Table\,\ref{tab:cut}.
Here, for the $\delta_\mm^{}$ cut, we have added the superscript to
$\delta_\mm^{(4)}$ for $\SZZ_4^{}$ and the superscript to
$\delta_\mm^{(8)}$ for $\SZZ_8^{}$, to indicate that the two cuts
$\delta_\mm^{(4)}$ and $\delta_\mm^{(8)}$ are imposed (optimized) for the
observables of $\SZZ_4^{}$ and $\SZZ_8^{}$ separately and independently.
We note that this $\theta$ cut is a type of basic selection cuts.
Table\,\ref{tab:cut} shows that the optimized $\theta$ cut
has little change when the collider energy $\!\sqrt{s\,}$ varies from 250\,GeV
all the way up to 5\,TeV for the analysis of either $\mathcal{Z}_4^{}$ or $\mathcal{Z}_8^{}$.\,
In fact, the angular $\theta$ cuts in Table\,\ref{tab:cut} cannot cause any
correlation between $\SZZ_4^{}$ and $\SZZ_8^{}$.\, This is fully clarified
in Appendix\,\ref{sec:B}.

\vspace*{1mm}

With the optimal kinematical cuts in Table\,\ref{tab:cut},
we compute the SM contributions and the ${\cal O}(\Lambda^{-4})$
contributions for different collider energies,

%
{\small
\beqs
\label{eq:Z4-all}
\begin{eqnarray}
\sqrt{s}=250\,\text{GeV}, &~~~&
(\sigma^0_c,\, \mathbb{O}^1_c)
=\left(\!3936,\, 0.913\!\left(\!\frac{\text{TeV}}{\Lambda}\right)^{\!\!4}\right)
\!\text{fb}\,,
\hspace*{12mm}
\\
\sqrt{s}=500\,\text{GeV}, &~~~&
(\sigma^0_c,\, \mathbb{O}^1_c)
=\left(\!860,\, 1.85\!\left(\!\frac{\text{TeV}}{\Lambda}\right)^{\!\!4}\right)
\!\text{fb}\,,
\\
\sqrt{s}=1\,\text{TeV}, &~~~&
(\sigma^0_c,\, \mathbb{O}^1_c)
= \left(\!209,\, 3.71\!\left(\!\frac{\text{TeV}}{\Lambda}\right)^{\!\!4}\right)
\!\text{fb}\,,
\\
\sqrt{s}=3\,\text{TeV}, &~~~&
(\sigma^0_c,\, \mathbb{O}^1_c)
= \left(\!23.1,\, 11.1\!\left(\!\frac{\text{TeV}}{\Lambda}\right)^{\!\!4}\right)\!
\text{fb}\,,
\\
\sqrt{s}=5\,\text{TeV}, &~~~&
(\sigma^0_c,\, \mathbb{O}^1_c)
= \left(\!8.30,\, 18.5\!\left(\!\frac{\text{TeV}}{\Lambda}\right)^{\!\!4}\right)
\!\text{fb}\,,
\end{eqnarray}
\eeqs
}
%
where we include the case of $\sqrt{s}=3$\,TeV from Eq.\eqref{s0o1} for comparison.

\vspace*{1mm}

With these, we derive the following signal significances at each collision energy,
for the leptonic branching fraction $\,\text{Br}(Z\!\!\to\!\ell\ell)\!\simeq 0.10$\, and
an integrated luminosity $\mathcal{L}=2\,\text{ab}^{-1}$:\,
{\small
\beqs
\vspace*{-4mm}
\label{eq:Z4-other}
\label{eq:Z4-all}
\begin{eqnarray}
\sqrt{s}=250\,\text{GeV}, &~~~&
\mathcal{Z}_{4}^{} =\,
3.29\!\(\!\frac{\,0.5\text{TeV}\,}{\Lambda}\!\)^{\!\!4}\!\times
\!\sqrt{\epsilon\,}\,,
\hspace*{12mm}
\\
\sqrt{s}=500\,\text{GeV}, &~~~&
\mathcal{Z}_{4}^{} =\,
2.18\!\left(\!\frac{\,0.8\text{TeV}\,}{\Lambda}\!\)^{\!\!4}\! \times
\!\sqrt{\epsilon\,}\,,
\\
\sqrt{s}=1\,\text{TeV}, &~~~&
\mathcal{Z}_{4}^{} =\,
 3.62\!\(\!\frac{\,\text{TeV}\,}{\Lambda}\!\)^{\!\!4}\!\times
\!\sqrt{\epsilon\,}\,,
\\
\sqrt{s}=3\,\text{TeV}, &~~~&
\mathcal{Z}_{4}^{} \,\simeq\,
2.05\!\left(\!\frac{\,2\text{TeV}\,}{\Lambda}\!\right)^{\!\!4}\!\times
\!\sqrt{\epsilon\,}\,,
\\
\sqrt{s}=5\,\text{TeV}, &~~~&
\mathcal{Z}_{4}^{} =\,
2.33\!\(\!\frac{\,2.5\text{TeV}\,}{\Lambda}\!\)^{\!\!4}\! \times
\!\sqrt{\epsilon\,}\,.
\end{eqnarray}
\vspace*{-3mm}
\eeqs
}

We note that the signal significance is nearly proportional to
the squared centre-of-mass collision energy $\,(\!\sqrt{s\,})^2\,$.\,
At high energies $s\!\gg\! M_Z^2$, we have
\begin{eqnarray}
\mathcal{Z}_{4}^{}\propto
\frac{M_Z^{}\, s}{\,\Lambda^4\,}\sqrt{\LL\!\times\!\epsilon\,} \,.
\end{eqnarray}
Thus, for a given significance $\SZZ_{4}^{}$, the corresponding reach of
the new physics scale $\Lambda$ is
\begin{eqnarray}
\cut \propto
\(\!\frac{\,M_Z^{}\sqrt{\mathcal{L}\!\times\!\epsilon\,}\,}
         {\,\mathcal{Z}_{4}^{}}\)^{\!\!\!\frac{1}{4}}
\!\!\times\!\(\!\sqrt{s\,}\)^{\!\frac{1}{2}}_{} .
\label{lam4}
\end{eqnarray}
We see that the collision energy $\sqrt{s\,}$\, has the most sensitive effect
on the reach of the new physics scale $\Lambda$\,.\,
For instance, raising the collision energy from
$\,\sqrt{s\,}=250$~GeV to $\sqrt{s\,}=3$~TeV,
the reach of the new physics scale is improved by a significant factor
$\,\cut(3~\text{TeV})/\cut(250~\text{GeV})\simeq 3.46\,$.\,
On the other hand,
$\cut$ has a rather weak dependence on the significance,
$\,\cut\propto\SZZ_4^{-\frac{1}{4}}\,$,\, so the $5\sigma$ reach is only
slightly weaker than the $2\sigma$ reach:
$\cut(5\sigma)/\cut(2\sigma) \simeq 1/1.26$\,.
Furthermore, we note that $\cut$ depends much more weakly on the integrated luminosity
and the detection efficiency,
$\,\cut\propto (\mathcal{L}\!\times\!\epsilon)^{\frac{1}{8}}$.\,
For instance, increasing the integrated luminosity from $\mathcal{L}=2\,\text{ab}^{-1}$
to $\mathcal{L}=6\,\text{ab}^{-1}$, would enhance the reach of the new physics scale by
only a factor of
$\,\cut(6\text{ab}^{-1})/\cut(2\text{ab}^{-1}) \simeq 1.15$\,.\,
Also, if the detection efficiency is increased from $\,\ep=40\%$\, to \,$\ep=90\%$,\,
the reach of the new physics scale would be only slightly extended by a factor of
$\,\cut(90\%)/\cut(40\%) \simeq 1.11$\,.\,

\vspace*{1.5mm}

Next, extending Section\,\ref{sec:3.2.2} to different collision energies, we include
contributions up to ${\cal O}(\cut^{\!-8})$ in a similar manner.
We apply the optimal kinematical cuts as in Table\,\ref{tab:cut} and compute the
cross sections of $\,e^+e^-\!\!\to\! Z\gamma$\, as follows:
{\small
\beqs
\begin{eqnarray}
\sqrt{s\,}\!=250~\text{GeV}, &~~~&
\sigma(Z\gamma) = \left[2427
\pm 6.62\!\left(\!\frac{\,0.5\text{TeV}\,}{\Lambda}\!\right)^{\!\!4}\!
+ 1.39\!\left(\!\frac{\,0.5\text{TeV}\,}{\Lambda}\!\right)^{\!\!8}
\right]\!\text{fb}\,,~~~~~~~~
\\[1mm]
\sqrt{s\,}\!=500~\text{GeV}, &~~~&
\sigma(Z\gamma)
= \left[417 \pm 0.996\!\(\!\frac{\,0.8\text{TeV}\,}{\Lambda}\!\)^{\!\!4}\!
           +0.624\!\(\!\frac{\,0.8\text{TeV}\,}{\Lambda}\!\)^{\!\!8}\right]\!\text{fb}\,,
\\[1mm]
\sqrt{s\,}\!=1~\text{TeV}, &~~~&
\sigma(Z\gamma)  = \left[94.0 \pm 0.404\!\(\!\frac{\,\text{TeV}\,}{\Lambda}\!\)^{\!\!4}
+1.73\!\(\!\frac{\,\text{TeV}\,}{\Lambda}\!\)^{\!\!8}\right]\!\text{fb}\,,
\\[1mm]
\sqrt{s\,}\!=3~\text{TeV}, &~~~&
\sigma(Z\gamma)
= \left[10.1 \pm 0.0252\!\(\!\frac{\,2\text{TeV}\,}{\Lambda}\!\)^{\!\!4}
	+0.554\!\(\!\frac{\,2\text{TeV}\,}{\Lambda}\!\)^{\!\!8}\right]
\!\text{fb}\,,
\\[1mm]
\sqrt{s\,}\!=5~\text{TeV}, &~~~&
\sigma(Z\gamma) = \left[3.63 \pm 0.0103\!\(\!\frac{\,2.5\text{TeV}\,}{\Lambda}\!\)^{\!\!4}
+0.718\!\(\!\frac{\,2.5\text{TeV}\,}{\Lambda}\!\)^{\!\!8}\right]\!\text{fb}\,,
~~~~~~~~~~
\end{eqnarray}
\eeqs
}
\hspace*{-3.5mm}
where for comparison we have also included the result from Eq.\eqref{eq:sigma-Zga-3TeV}
for the case of $\sqrt{s\,}\!=3\,\text{TeV}$.\,
The above can be compared to the cross sections
\eqref{eq:sigma-all-d0.2} with only a preliminary angular cut $\,\delta>0.2$\,
(corresponding to a lower cut on the photon transverse momentum $\,P_T^{\ga}=q\sin\delta$\,).
We see that under the final angular cuts on $(\theta,\,\theta_*^{},\,\phi_*^{})$
the SM contribution is substantially reduced in each case, whereas the signal contributions
at ${\cal O}(\cut^{\!-4})$ and ${\cal O}(\cut^{\!-8})$ are little changed.

\vspace*{1mm}

Uisng the above, we analyze the signal significance
up to ${\cal O}(\Lambda^{\!-8})$,\, for different collider energies.
With inputs of the leptonic branching fraction
$\,\text{Br}(Z\!\!\to\!\ell\ell)\!\simeq 0.10$\, and
an integrated luminosity $\mathcal{L}\!=\!2\,\text{ab}^{-1}$,\, we arrive at
{\small
\beqs
\label{eq:Z8-other}
\label{eq:Z8-all}
\begin{eqnarray}
\sqrt{s}=250\,\text{GeV}, &~~~&
\SZZ_{8}^{}
= \left|\pm
1.90\!\(\!\frac{\,0.5\text{TeV}\,}{\Lambda}\!\)^{\!\!4} +
0.400\!\(\!\frac{\,0.5\text{TeV}\,}{\Lambda}\!\)^{\!\!8}\right|
\!\times\! \sqrt{\ep\,}\,, \hspace*{25mm}
\\[0mm]
\sqrt{s}=500\,\text{GeV}, &~~~&
\SZZ_{8}^{} =
\left|\pm
0.689\(\!\frac{\,0.8\text{TeV}\,}{\Lambda}\!\)^{\!\!4} +
0.432\!\(\!\frac{\,0.8\text{TeV}\,}{\Lambda}\!\)^{\!\!8}\right|
\!\times\!\sqrt{\ep\,}\,,
\\[0mm]
\sqrt{s}=1\,\text{TeV}, &~~~&
\SZZ_{8}^{} =
\left|\pm
0.589\!\(\!\frac{\,\text{TeV}\,}{\Lambda}\!\)^{\!\!4} +
2.53\!\(\!\frac{\,\text{TeV}\,}{\Lambda}\!\)^{\!\!8}\right|
\!\times\!\sqrt{\ep\,}\,,
\\[0mm]
\sqrt{s}=3\,\text{TeV}, &~~~&
\SZZ_{8}^{} =
\left|\pm
0.112\!\(\!\frac{\,2\,\text{TeV}\,}{\Lambda}\!\)^{\!\!4}+
2.46\!\left(\!\frac{\,2\,\text{TeV}\,}{\Lambda}\!\right)^{\!\!8} \right|
\!\times\!\sqrt{\ep\,}\,,
\\[0mm]
\sqrt{s}=5\,\text{TeV}, &~~~&
\SZZ_{8}^{} =
\left|\pm
0.0764\!\(\!\frac{\,2.5\text{TeV}\,}{\Lambda}\!\)^{\!\!4} +
5.32\!\(\!\frac{\,2.5\text{TeV}\,}{\Lambda}\!\)^{\!\!8}\right|
\!\times\! \sqrt{\ep\,} \,.
\end{eqnarray}
\eeqs
}
\hspace*{-3mm}
From the above, we note that for the relevant reach of $\,\cut$\,,\,
the ${\cal O}(\Lambda^{\!-4})$
terms give the dominant contributions for collision energies
$\sqrt{s}<\! 1$\,TeV,\, while the ${\cal O}(\Lambda^{\!-8})$ terms become dominant
for  $\sqrt{s}\gtrsim\! 1$\,TeV.\,
When ${\cal O}(\Lambda^{\!-8})$ becomes dominant at high energies, we have
$\,\cut\propto\SZZ_8^{-\frac{1}{8}}$.\, In such cases, the reach in $\cut$
becomes rather insensitive to the significance $\SZZ_8^{}$.\, For instance,
at high energies $\sqrt{s}\gtrsim\! 1$~TeV,\, we have
$\cut(5\sigma)/\cut(2\sigma) \simeq 1/1.12$\, for $\SZZ_8^{}$,\,
whereas we previously found $\cut(5\sigma)/\cut(2\sigma) \simeq 1/1.26$\,
for $\SZZ_4^{}$.\,

\vspace*{1mm}

At high energies $\,s\gtrsim (1\text{TeV})^2\gg\!M_Z^2$,\,
the ${\cal O}(\cut^{\!-8})$ terms become dominant,
so we have the approximate relation
\begin{eqnarray}
\SZZ_{8}^{}\propto
\frac{\,M_Z^2 (\!\sqrt{s\,})^{5}\,}
{\,\Lambda^8\,}\sqrt{\LL\!\times\!\ep\,} \,,
\end{eqnarray}
and hence
\begin{eqnarray}
\Lambda \,\propto\,
\(\!\!\frac{\,M_Z^2\sqrt{\LL\,\ep\,}\,}{\,\SZZ_8^{}\,}\!\)^{\!\!\!\frac{1}{8}}\!
\(\!\sqrt{s\,}\)^{\!\frac{5}{8}} .
\label{lam8}
\end{eqnarray}
We see from Eqs.\eqref{lam4} and \eqref{lam8}
that the sensitivity to $\Lambda$ increases with the collision energy
with the power $\!(\!\sqrt{s\,})^{\frac{1}{2}}_{}$ or $\!(\!\sqrt{s\,})^{\frac{5}{8}}_{}$,\,
a relatively slow rate of increase.
We note also that the sensitivity to the new physics scale $\cut$
is rather insensitive to the integrated luminosity $\LL$ and the detection efficiency
$\,\ep\,$, owing to their small power-law dependence $\(\LL\,\ep\)^{\!\frac{1}{16}}_{}$.\,

\vspace*{1mm}

\begin{table}[t]
\begin{center}
\begin{tabular}{c|ccccc}
\hline\hline
$\sqrt s $ (GeV) & 250 & 500 & 1000~~ & 3000~ & 5000 \\
\hline
\\[-4.4mm]
$\Lambda_{2\sigma}^{}$\,(TeV)& 0.59(0.58) & 0.84(0.82) & 1.2~~ & 2.2~ & 2.9
\\[-4.4mm]
\\
\hline
\\[-4.4mm]
$\Lambda_{5\sigma}^{}$\,(TeV) & 0.48(0.46) & 0.68(0.65) & 1.0~~ & 1.9~ & 2.6 \\
\hline\hline
\end{tabular}
\end{center}
\vspace*{-3.7mm}
\caption{{\it Combined sensitivity reaches to the new physics scale $\cut$
at the $2\sigma$ and $5\sigma$ levels, for different collider energies.
Here the two numbers in the parentheses correspond to the case of the
dimension-8 operator whose coefficient has a minus sign, while in all other
entries the effects due to the coefficient having a minus sign are negligible.
For illustration, we have input a fixed representative integrated luminosity}
$\,\LL=2\,\text{ab}^{-1}\!$ {\it and an ideal detection efficiency $\,\ep=100\%$.}
}
\label{tab:1}
\label{tab:22}
\end{table}

Finally, we compute from Eqs.\eqref{eq:Z4-all} and \eqref{eq:Z8-all},
the combined significance, $\SZZ\!=\sqrt{\SZZ_4^2\!+\!\SZZ_8^2\,}$,\,
for each collider energy.
With these, in Table~\ref{tab:1}
we present the corresponding combined sensitivity reaches to the new physics scale $\cut$
at different $e^+ e^-$ collider energies.
In the last row of this Table,
the two numbers in the parentheses correspond to the case of the
dimension-8 operator whose coefficient has a minus sign, whereas in all other
entries the effects due to the coefficient having a minus sign are negligible.
Note also that in current analysis we have chosen a universal integrated luminosity
$\,\LL=2\,\text{ab}^{-1}\!$ for illustration. But it is straightforward
to obtain the significance $\SZZ$ for any other given integrated luminosity
$\,\LL\,$ by a simple rescaling due to $\,\SZZ\!\propto\! \sqrt{\!\LL\,}$.\,

\vspace*{1.5mm}

Before concluding this Section, we mention that we have performed
a numerical Monte Carlo simulation based on the analytical formula
\eqref{eq:O1c-sigma0c}.
We used for this purpose {\tt CUDAlink} in {\tt Mathematica}, so as to exploit
the {\tt CUDA} parallel computing architecture on Graphical Processing Units
(GPUs), which can generate millions of events in seconds.
We have computed the probability density function
of $\theta$, $\theta_*^{}$ and $\phi_*^{}$ for the case of
$\sqrt{s\,} =3\,$TeV and $\Lambda=2$\,TeV.
Eq.\eqref{eq:sigma-3TeV-d0.2}
shows that the SM contribution dominates
the total cross section.
According to Eq.\eqref{s0o1}, we have
$\,\mathbb{O}_1^c/\sigma_0^c\simeq 0.03$\,.\,
For comparison, our Monte Carlo simulation yielded the following
event counts: $\,|N_a^{}\!-\!N_b^{}|=3054$ and
$\,N_a^{}\!+\!N_b^{}=104752$,
corresponding to $\,|N_a^{}\!-\!N_b^{}|/(N_a^{}\!+\!N_b^{}) \simeq 0.029$.\,
This agrees well with the ratio $\mathbb{O}_1^c/\sigma_0^c\simeq 0.03$
inferred from our analytical formula \eqref{eq:O1c-sigma0c}, serving
as a consistency check on our Eq.\eqref{s0o1}.
Our Monte Carlo simulation package may be used to generate other
distributions and quantities that may be of interest for
experiments.\footnote{Further details may be obtained from RQX.}

\vspace*{1mm}

In passing, we note that subsequent to Refs.\,\cite{nTGC1,Degrande:2013kka}
some other follow-up studies discussed testing nTGCs at
ILC(800GeV)\,\cite{India1} and ILC(500GeV)\,\cite{India2}
using the form factors for the nTGC\,\cite{nTGC1}.
Unlike 
Refs.\,\cite{India1}\cite{India2},
we study the fully gauge-invariant dimension-8 operators \eqref{eq:dim8H} for the nTGCs
which are much more restrictive and result in just one independent operator
$\mathcal{O}_{\widetilde{B}W}$.  Only the gauge-invariant formulation of nTGCs
via dimension-8 operators can identify the power-dependence on the associated
UV cutoff $\cut$, making possible the probe of the new physics scale $\cut$\,.
We perform a systematic study and comparison,
probing the sensitivity to $\cut$ of nTGC measurements at different collider energies
and covering all future $e^+e^-$ colliders being planned,
including CEPC, FCCee, ILC and CLIC, rather than studying only a specific collider energy
(800GeV\,\cite{India1} or 500GeV\,\cite{India2}).
Our independent study presents the complete angular distributions
both analytically and numerically
for all future $e^+e^-$ colliders, which allow us to construct the
angular observables in Sec.\,\ref{sec:3.1}.
Refs.\,\cite{India1}\cite{India2} considered some angular observables, but did
not provide any complete angular distributions.
In particular, they did not show any
$\phi_*^{}$ distributions, and they also did not give any analytical formulas
for angular differential cross sections. Hence, it is difficult to compare with
our independent work even for the special collider energy 500\,GeV.
%
We study systematically the impact of $e^\mp$ beam polarizations on probing
the new physics scale of nTGCs in Sec.\,\ref{sec:55},
unlike Ref.\,\cite{India2}, which did not consider the beam polarizations.
We study both leptonic $Z$-decays (Sec.\,\ref{sec:3} and \ref{sec:55})
and invisible $Z$-decays (Sec.\,\ref{sec:44} and \ref{sec:55}).
Ref.\,\cite{India1} even did not consider the final state $Z$-decays
for a realistic analysis of signals and backgrounds,
while \cite{India2} did not study the invisible $Z$-decay channels.

\newpage
\subsection{\hspace{-5mm}.\,Non-Resonant Backgrounds}
\label{sec:3.3}	
\vspace*{1mm}

\begin{figure}[t]
\centering
\vspace*{-5mm}
\hspace*{-5mm}
	\includegraphics[width=3.8cm]{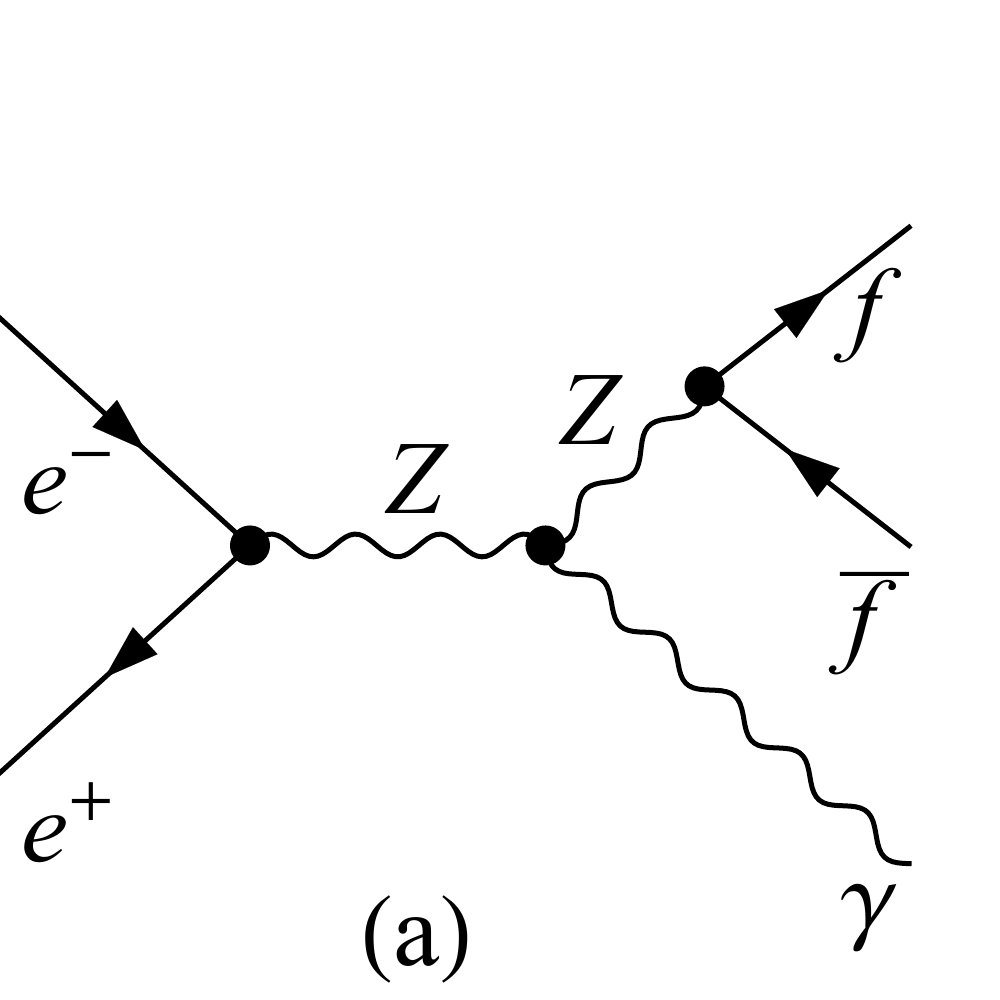}
	\includegraphics[width=3.8cm]{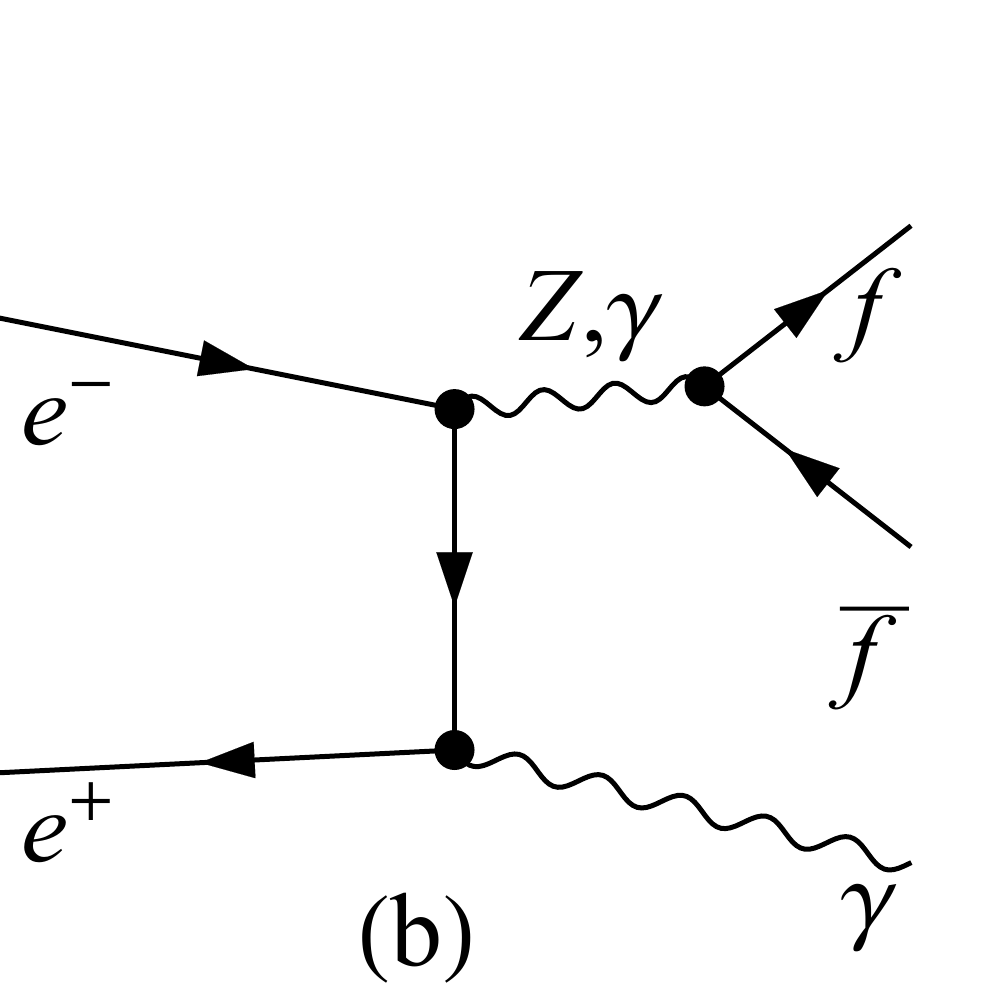}
	\\[-6mm]
	\includegraphics[width=3.8cm]{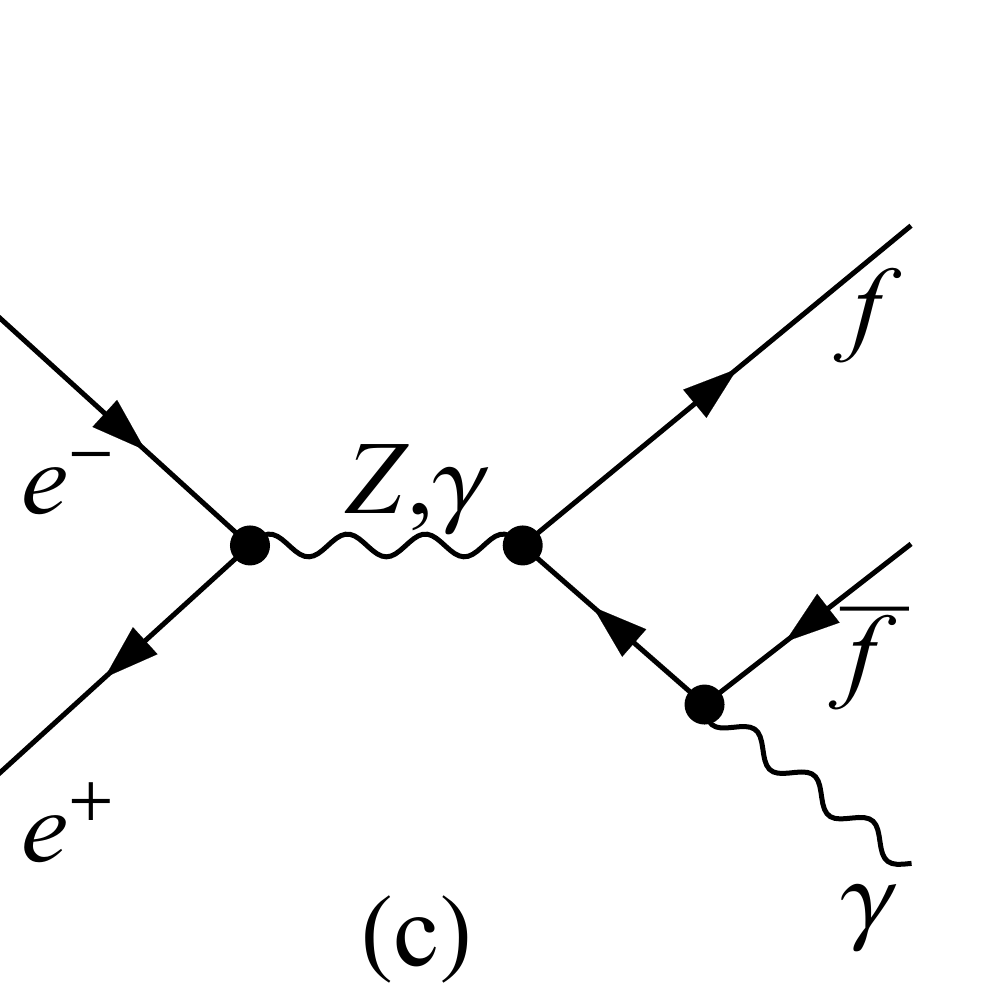}
	\includegraphics[width=3.8cm]{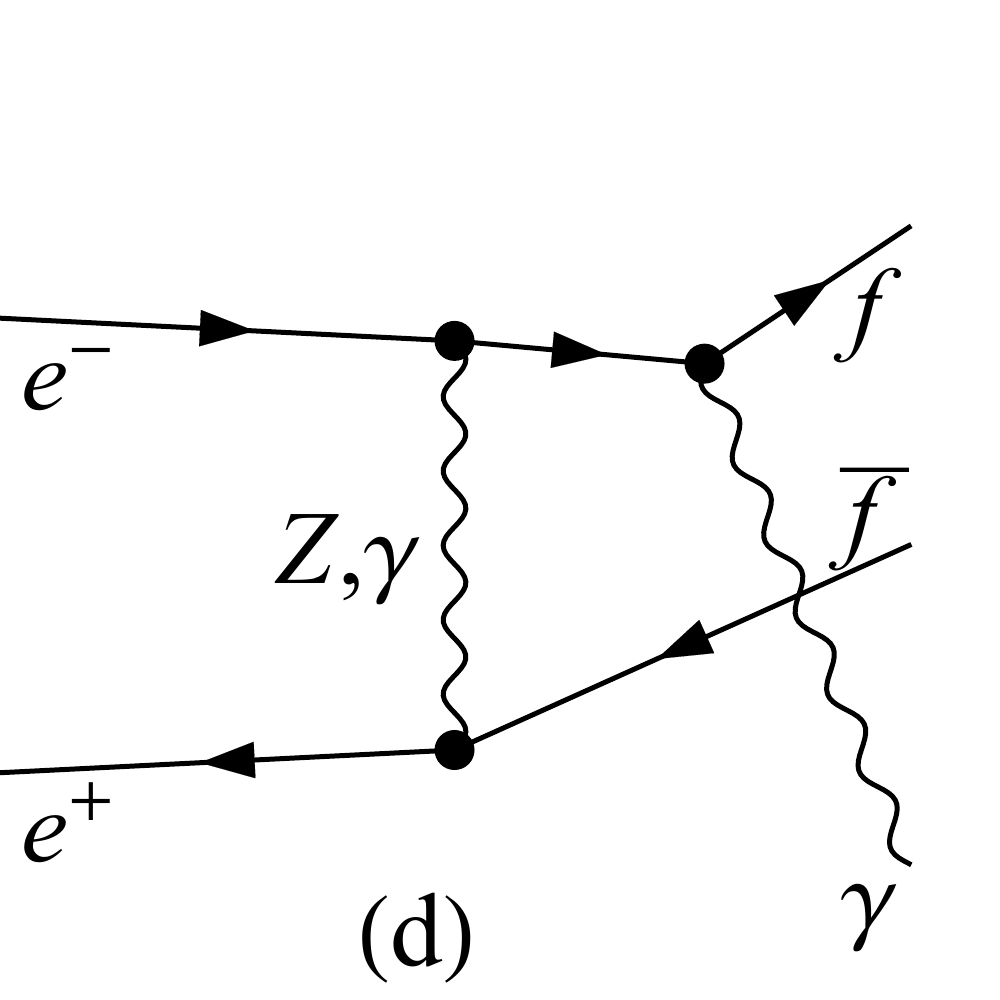}
	\includegraphics[width=3.8cm]{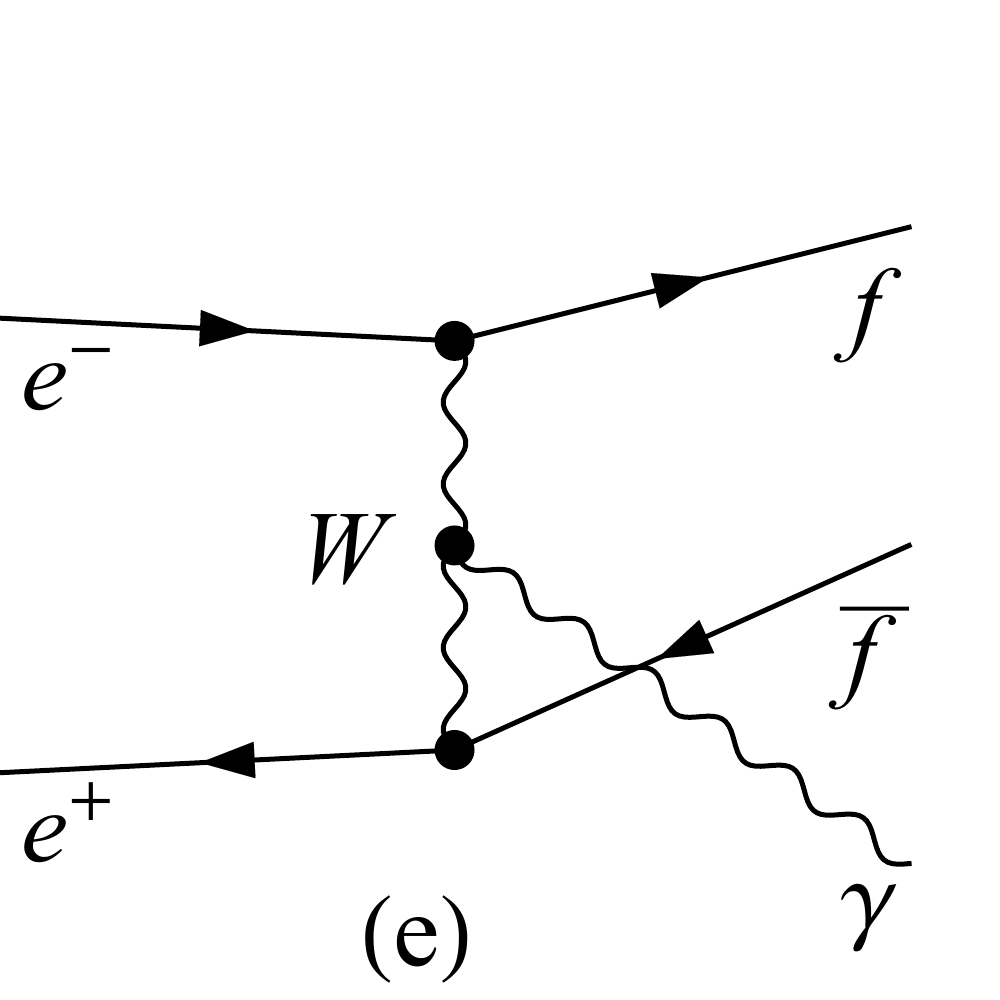}
	\vspace*{-2mm}
\caption{\it Five types of Feynman diagrams which contribute to the process
$e^-e^+\!\!\to\!\! f\bar{f}\gamma$\, (with $f\!=\!\ell,\nu$).
Type\,(a) is our signal with a $Z^*Z\gamma$ vertex solely from the dimension-8
operator \eqref{eq:obtw}, and types\,(b),(c),(d),(e) are the SM backgrounds.
The type\,(b) process $\,e^-e^+\!\!\to\! V\gamma\!\to\!\gamma f \bar f$\, (with $V\!\!=Z,\ga^*$)
gives an irreducible background (and there is a similar $u$-channel diagram).
Type\,(c) is $s$-channel gauge boson exchange ($V\!=Z^*,\ga^*$) with final-state
$\gamma$ radiation.
Type\,(d) is $t$-channel $V$ exchange ($V\!\!=\!Z^*,\ga^*$)
with the final-state $\gamma$ radiated from $e^\pm$ (in the final or initial state).
Type\,(e) is for the $\nu\bar{\nu}\ga$ final state with
$t$-channel $W^*\!$-exchange and the $\ga$ radiated from either the $W^*$ or
the initial-state $e^\pm$.\,
}
\label{feyndiag}
\label{SB}
\label{fig:Fdiag}
\label{fig:8}
\label{fig:77}
\end{figure}

In the analyses so far, we have considered
$e^-e^+\!\!\to\! Z\ga$\, production with on-shell $Z$ decays
($Z\!\!\to\!\ell^-\ell^+$). This means that we have considered only
the signal contribution Fig.\,\ref{SB}(a)
and the irreducible background Fig.\,\ref{SB}(b).
There are other (non-resonant) SM backgrounds with the same final state of $f\bar{f}\ga$
($f\!\!=\!\ell, \nu$), but having very different topology
where $\,\ga$\, is radiated either from the final-state fermions [Fig.\,\ref{SB}(c)-(d)],
or from a $t$-channel $W$ boson [Fig.\,\ref{SB}(e)] in the case of
$\,Z\!\to\!\nu\bar{\nu}$\, decay (which will be studied in Section\,\ref{sec:44}).
These backgrounds may give visible but small contributions
after proper kinematic cuts.

\vspace*{1mm}

For the backgrounds with an $e^-e^+\ga$ final state, we have type\,(b) (with 4 diagrams),
type\,(c) (with 4 diagrams) and type\,(d) (with 8 diagrams), for a total of 16 diagrams.
For the other $f\bar{f}\ga$ final states with $f=\mu,\tau$,
we have type\,(b) (with 4 diagrams) and type\,(c) (with 8 diagrams),
which amount to 12 diagrams.
For the final state $\nu\bar{\nu}\ga$,\, we have the SM backgrounds from
type\,(b)(with 2 diagrams) and type\,(e) (with 3 diagrams).
For each of these diagrams, there are $\,2^3\!=8$\, helicity combinations.
Rather than writing explicitly the cross sections for all these combinations
and calculating them analytically,
we have computed the cross section numerically using a
Monte Carlo method. The relative accuracy of numerical Monte Carlo integration
is $\,{O}(1\!/\!\sqrt{\!N})$,\, where $N$ is the number of samples, and
one needs a large sample to obtain precise results.
We use {\tt FeynArts}\,\cite{Hahn:2000kx} to generate all the background diagrams
and then compute them by {\tt FeynCalc}\,\cite{Shtabovenko:2016sxi,Mertig:1990an}.
Finally, we convert the expressions to $C$ form and use {\tt CUDAlink}
for numerical integrations to compute the cross sections and other observables.
All steps are done in {\tt Mathematica} and the numerical computation speed is
${O}(10^7)$\,diagrams/$s$.\footnote{%
We perform the numerical integrations using GPU parallel computing, which is faster
than standard CPU computing by a factor of ${O}(10^3)$.
Further details may be obtained from RQX.}

\vspace*{1mm}

The diagrams of Fig.\ref{feyndiag}(c)-(d) have additional soft and collinear divergences,
which can be removed by imposing lower cuts on the photon transverse momentum
$\,P_T^{\ga}>0.2P_{\ga}^{}$\, and on the lepton-photon invariant mass
$\,M(\ell\gamma)>0.1\sqrt{s}\,$.\,
We further require $|M(\ell\ell)\!-\!M_Z^{}|\!<\! 10$\,GeV\,
so as to be close to the $Z$ boson mass-shell.
Applying these cuts together with those in Table\,\ref{tab:cut},
we first compute the observables for the reaction channel
$e^-e^+\!\!\to\!\gamma e^-e^+$\,
by including the additional backgrounds in Fig.\,\ref{feyndiag}(c)-(d),
{\small
	\beqs
	\vspace*{-1mm}
	\label{eq:Z4-e}
	\begin{eqnarray}
	\sqrt{s}=250\,\text{GeV}, &~~~&
	(\sigma^{0e}_c,\, \mathbb{O}^{1e}_c)
	=\left(\!141,\, 0.0256\!\left(\!\frac{\text{TeV}}{\Lambda}\right)^{\!\!4}\right)
	\!\text{fb}\,,
	\hspace*{12mm}
	\\
	\sqrt{s}=500\,\text{GeV}, &~~~&
	(\sigma^{0e}_c,\, \mathbb{O}^{1e}_c)
	=\left(\!26.6,\, 0.0524\!\left(\!\frac{\text{TeV}}{\Lambda}\right)^{\!\!4}\right)
	\!\text{fb}\,,
	\\
	\sqrt{s}=1\,\text{TeV}, &~~~&
	(\sigma^{0e}_c,\, \mathbb{O}^{1e}_c)
	= \left(\!6.15,\, 0.109\!\left(\!\frac{\text{TeV}}{\Lambda}\right)^{\!\!4}\right)
	\!\text{fb}\,,
	\\
	\sqrt{s}=3\,\text{TeV}, &~~~&
	(\sigma^{0e}_c,\, \mathbb{O}^{1e}_c)
	= \left(\!0.691,\, 0.340\!\left(\!\frac{\text{TeV}}{\Lambda}\right)^{\!\!4}\right)\!
	\text{fb}\,,
	\\
	\sqrt{s}=5\,\text{TeV}, &~~~&
	(\sigma^{0e}_c,\, \mathbb{O}^{1e}_c)
	= \left(\!0.250,\, 0.567\!\left(\!\frac{\text{TeV}}{\Lambda}\right)^{\!\!4}\right)
	\!\text{fb}\,.
	\end{eqnarray}
	\eeqs
}
\hspace*{-3mm}
We then derive the following signal significance $\SZZ_4^e$ at each collision energy,
assuming an integrated luminosity $\LL =2\,\abinv$:
{\small
\beqs
\vspace*{-1mm}
\label{eq:Z4-E}
\begin{eqnarray}
	\sqrt{s}=250\,\text{GeV}, &~~~&
	\mathcal{Z}_{4}^e =\,
	1.54\!\(\!\frac{\,0.5\text{TeV}\,}{\Lambda}\!\)^{\!\!4}\!\times
	\!\sqrt{\epsilon\,}\,,
	\hspace*{12mm}
	\\
	\sqrt{s}=500\,\text{GeV}, &~~~&
	\mathcal{Z}_{4}^{e} =\,
	1.11\!\left(\!\frac{\,0.8\text{TeV}\,}{\Lambda}\!\)^{\!\!4}\! \times
	\!\sqrt{\epsilon\,}\,,
	\\
	\sqrt{s}=1\,\text{TeV}, &~~~&
	\mathcal{Z}_{4}^{e} =\,
	1.97\!\(\!\frac{\,\text{TeV}\,}{\Lambda}\!\)^{\!\!4}\!\times
	\!\sqrt{\epsilon\,}\,,
	\\
	\sqrt{s}=3\,\text{TeV}, &~~~&
	\mathcal{Z}_{4}^{e} \,=\,
	1.14\!\left(\!\frac{\,2\text{TeV}\,}{\Lambda}\!\right)^{\!\!4}\!\times
	\!\sqrt{\epsilon\,}\,,
	\\
	\sqrt{s}=5\,\text{TeV}, &~~~&
	\mathcal{Z}_{4}^{e} =\,
	1.30\!\(\!\frac{\,2.5\text{TeV}\,}{\Lambda}\!\)^{\!\!4}\! \times
	\!\sqrt{\epsilon\,}\,.
\end{eqnarray}
\eeqs}
Next, we extend the analysis of Section\,\ref{sec:3.2.2} by including the $O(\cut^{\!-8})$
contributions and the additional backgrounds in Fig.\,\ref{feyndiag}(c)-(d).
Thus, we arrive at
{\small
	\beqs
	\label{eq:Z8-e}
	\begin{eqnarray}
	\sqrt{s\,}\!=250\,\text{GeV}, &~~~&
	\sigma(ee\gamma) = \left[85.0
	\pm 0.20\!\left(\!\frac{\,0.5\text{TeV}\,}{\Lambda}\!\right)^{\!\!4}\!
	+ 0.0418\!\left(\!\frac{\,0.5\text{TeV}\,}{\Lambda}\!\right)^{\!\!8}
	\right]\!\text{fb}\,,~~~~~~~~
	\\[1mm]
	\sqrt{s\,}\!=500\,\text{GeV}, &~~~&
	\sigma(ee\gamma)
	= \left[13.6 \pm 0.32\!\(\!\frac{\,0.8\text{TeV}\,}{\Lambda}\!\)^{\!\!4}\!
	+0.0192\!\(\!\frac{\,0.8\text{TeV}\,}{\Lambda}\!\)^{\!\!8}\right]\!\text{fb}\,,
	\\[1mm]
	\sqrt{s\,}\!=1\,\text{TeV}, &~~~&
	\sigma(ee\gamma)  = \left[3.03 \pm 0.13\!\(\!\frac{\,\text{TeV}\,}{\Lambda}\!\)^{\!\!4}
	+0.0536\!\(\!\frac{\,\text{TeV}\,}{\Lambda}\!\)^{\!\!8}\right]\!\text{fb}\,,
	\\[1mm]
	\sqrt{s\,}\!=3\,\text{TeV}, &~~~&
	\sigma(ee\gamma)
	= \left[0.325 \pm 0.0008\!\(\!\frac{\,2\text{TeV}\,}{\Lambda}\!\)^{\!\!4}
	+0.0172\!\(\!\frac{\,2\text{TeV}\,}{\Lambda}\!\)^{\!\!8}\right]
	\!\text{fb}\,,
	\\[1mm]
	\sqrt{s\,}\!=5\,\text{TeV}, &~~~&
	\sigma(ee\gamma) = \left[0.116 \pm 0.0004\!\(\!\frac{\,2.5\text{TeV}\,}{\Lambda}\!\)^{\!\!4}
	+0.0222\!\(\!\frac{\,2.5\text{TeV}\,}{\Lambda}\!\)^{\!\!8}\right]\!\text{fb}\,.
	~~~~~~~~~~
	\end{eqnarray}
	\eeqs	
}
With these, we derive the signal significance $\SZZ_8^e$ for the $ee\ga$ channel,
{\small
	\beqs
	\label{eq:Z8-E}
	\begin{eqnarray}
	\sqrt{s}=250\,\text{GeV}, &~~~&
	\SZZ_{8}^{e}
	= \left|\pm
	0.96\!\(\!\frac{\,0.5\text{TeV}\,}{\Lambda}\!\)^{\!\!4} +
	0.203\!\(\!\frac{\,0.5\text{TeV}\,}{\Lambda}\!\)^{\!\!8}\right|
	\!\times\! \sqrt{\ep\,}\,, \hspace*{25mm}
	\\[0mm]
	\sqrt{s}=500\,\text{GeV}, &~~~&
	\SZZ_{8}^{e} =
	\left|\pm
	0.38\(\!\frac{\,0.8\text{TeV}\,}{\Lambda}\!\)^{\!\!4} +
	0.232\!\(\!\frac{\,0.8\text{TeV}\,}{\Lambda}\!\)^{\!\!8}\right|
	\!\times\!\sqrt{\ep\,}\,,
	\\[0mm]
	\sqrt{s}=1\,\text{TeV}, &~~~&
	\SZZ_{8}^{e} =
	\left|\pm
	0.31\!\(\!\frac{\,\text{TeV}\,}{\Lambda}\!\)^{\!\!4} +
	1.38\!\(\!\frac{\,\text{TeV}\,}{\Lambda}\!\)^{\!\!8}\right|
	\!\times\!\sqrt{\ep\,}\,,
	\\[0mm]
	\sqrt{s}=3\,\text{TeV}, &~~~&
	\SZZ_{8}^{e} =
	\left|\pm
	0.06\!\(\!\frac{\,2\,\text{TeV}\,}{\Lambda}\!\)^{\!\!4}+
	1.35\!\left(\!\frac{\,2\,\text{TeV}\,}{\Lambda}\!\right)^{\!\!8} \right|
	\!\times\!\sqrt{\ep\,}\,,
	\\[0mm]
	\sqrt{s}=5\,\text{TeV}, &~~~&
	\SZZ_{8}^{e} =
	\left|\pm
	0.03\!\(\!\frac{\,2.5\text{TeV}\,}{\Lambda}\!\)^{\!\!4} +
	2.90\!\(\!\frac{\,2.5\text{TeV}\,}{\Lambda}\!\)^{\!\!8}\right|
	\!\times\! \sqrt{\ep\,} \,.
	\end{eqnarray}
	\eeqs}
Then, we analyze the reaction channel
$\,e^-e^+\!\!\!\to\!\gamma\,\mu^-\mu^+$\, under the same cuts and including the
additional backgrounds as in Fig.\,\ref{feyndiag}(c).
With these we obtain the following,
{\small
	\beqs
	\vspace*{-4mm}
	\label{eq:Z4-mu}
	\begin{eqnarray}
	\sqrt{s}=250\,\text{GeV}, &~~~&
	(\sigma^{0\mu}_c,\, \mathbb{O}^{1\mu}_c)
	=\left(\!112,\, 0.0256\!\left(\!\frac{\text{TeV}}{\Lambda}\right)^{\!\!4}\right)
	\!\text{fb}\,,
	\hspace*{12mm}
	\\
	\sqrt{s}=500\,\text{GeV}, &~~~&
	(\sigma^{0\mu}_c,\, \mathbb{O}^{1\mu}_c)
	=\left(\!24.1,\, 0.0522\!\left(\!\frac{\text{TeV}}{\Lambda}\right)^{\!\!4}\right)
	\!\text{fb}\,,
	\\
	\sqrt{s}=1\,\text{TeV}, &~~~&
	(\sigma^{0\mu}_c,\, \mathbb{O}^{1\mu}_c)
	= \left(\!6.00,\, 0.109\!\left(\!\frac{\text{TeV}}{\Lambda}\right)^{\!\!4}\right)
	\!\text{fb}\,,
	\\
	\sqrt{s}=3\,\text{TeV}, &~~~&
	(\sigma^{0\mu}_c,\, \mathbb{O}^{1\mu}_c)
	= \left(0.687,\, 0.340\!\left(\!\frac{\text{TeV}}{\Lambda}\right)^{\!\!4}\right)\!
	\text{fb}\,,
	\\
	\sqrt{s}=5\,\text{TeV}, &~~~&
	(\sigma^{0\mu}_c,\, \mathbb{O}^{1\mu}_c)
	= \left(\!0.250,\, 0.567\!\left(\!\frac{\text{TeV}}{\Lambda}\right)^{\!\!4}\right)
	\!\text{fb}\,.
	\end{eqnarray}
	\eeqs
}
\hspace*{-2.5mm}
Thus, we derive the signal significance $\SZZ_4^\mu$\, at each collision energy
and with an integrated luminosity $\LL \!=2\,\abinv$,
{\small
	\beqs
	\vspace*{-1mm}
	\label{eq:Z4-MU}
	\begin{eqnarray}
	\sqrt{s}=250\,\text{GeV}, &~~~&
	\mathcal{Z}_{4}^\mu =\,
	1.74\!\(\!\frac{\,0.5\text{TeV}\,}{\Lambda}\!\)^{\!\!4}\!\times
	\!\sqrt{\epsilon\,}\,,
	\hspace*{12mm}
	\\
	\sqrt{s}=500\,\text{GeV}, &~~~&
	\mathcal{Z}_{4}^{\mu} =\,
	1.16\!\left(\!\frac{\,0.8\text{TeV}\,}{\Lambda}\!\)^{\!\!4}\! \times
	\!\sqrt{\epsilon\,}\,,
	\\
	\sqrt{s}=1\,\text{TeV}, &~~~&
	\mathcal{Z}_{4}^{\mu} =\,
	1.99\!\(\!\frac{\,\text{TeV}\,}{\Lambda}\!\)^{\!\!4}\!\times
	\!\sqrt{\epsilon\,}\,,
	\\
	\sqrt{s}=3\,\text{TeV}, &~~~&
	\mathcal{Z}_{4}^{\mu} \,=\,
	1.14\!\left(\!\frac{\,2\text{TeV}\,}{\Lambda}\!\right)^{\!\!4}\!\times
	\!\sqrt{\epsilon\,}\,,
	\\
	\sqrt{s}=5\,\text{TeV}, &~~~&
	\mathcal{Z}_{4}^{\mu} =\,
	1.30\!\(\!\frac{\,2.5\text{TeV}\,}{\Lambda}\!\)^{\!\!4}\! \times
	\!\sqrt{\epsilon\,}\,.
	\end{eqnarray}
	\eeqs
}

\begin{table}[t]
\begin{center}
\begin{tabular}{c|ccccc}
			\hline\hline
			\\[-4.4mm]
			$\sqrt{s\,}$~(GeV) & 250 & 500 & 1000~~ & 3000~ & 5000 \\[0.4mm]
			\hline\hline
			\\[-4.4mm]
			$\Lambda^{2\sigma}_{\ell\bar\ell}$\,(TeV)& 0.57(0.56) & 0.82(0.80) & 1.2~ & 2.1~ & 2.9
			\\[-4.4mm]
			\\
			\hline
			\\[-4.4mm]
			$\Lambda^{5\sigma}_{\ell\bar\ell}$\,(TeV) & 0.46(0.44) & 0.67(0.64) & 0.98(0.95) & 1.9~ & 2.5
	\\[0.7mm]
	\hline\hline
	\end{tabular}
	\end{center}
	\vspace*{-3.7mm}
	\caption{{\it %
			Sensitivity reaches of the new physics scale $\cut$
			from the $\ell^-\ell^+\gamma$ channel,
            including additional backgrounds [Fig.\ref{feyndiag}(c)-(d)],
			at the $2\sigma$ and $5\sigma$ levels, for different collider energies.
			The numbers in the parentheses correspond to the case of the
			dimension-8 operator whose coefficient is negative, while in the other
			entries the sensitivities for the two signs of the coefficient are indistinguishable.
			These results were obtained assuming a fixed representative integrated luminosity}
		$\,\LL=2\,\text{ab}^{-1}\!$ {\it and an ideal detection efficiency $\,\ep=100\%$.}
	}
\label{tab:3}
\end{table}

\vspace*{-2mm}

Next, similar to Eq.\eqref{eq:Z8-e} for the $ee\ga$ channel,
we compute the cross sections of the $\mu\mu\ga$ channel,
including the $O(\cut^{\!-8})$ contributions and the additional backgrounds
in Fig.\,\ref{feyndiag}(c).
Thus, we arrive at
{\small
	\beqs
	\label{eq:Z8-mu}
	\begin{eqnarray}
	\sqrt{s\,}\!=250~\text{GeV}, &~~~&
	\sigma(\mu\mu\gamma) = \left[75.8
	\pm 0.20\!\left(\!\frac{\,0.5\text{TeV}\,}{\Lambda}\!\right)^{\!\!4}\!
	+ 0.0418\!\left(\!\frac{\,0.5\text{TeV}\,}{\Lambda}\!\right)^{\!\!8}
	\right]\!\text{fb}\,,~~~~~~~~
	\\[1mm]
	\sqrt{s\,}\!=500~\text{GeV}, &~~~&
	\sigma(\mu\mu\gamma)
	= \left[13.2 \pm 0.31\!\(\!\frac{\,0.8\text{TeV}\,}{\Lambda}\!\)^{\!\!4}\!
	+0.0192\!\(\!\frac{\,0.8\text{TeV}\,}{\Lambda}\!\)^{\!\!8}\right]\!\text{fb}\,,
	\\[1mm]
	\sqrt{s\,}\!=1~\text{TeV}, &~~~&
	\sigma(\mu\mu\gamma)  = \left[3.02 \pm 0.013\!\(\!\frac{\,\text{TeV}\,}{\Lambda}\!\)^{\!\!4}
	+0.0536\!\(\!\frac{\,\text{TeV}\,}{\Lambda}\!\)^{\!\!8}\right]\!\text{fb}\,,
	\\[1mm]
	\sqrt{s\,}\!=3~\text{TeV}, &~~~&
	\sigma(\mu\mu\gamma)
	= \left[0.325 \pm 0.0008\!\(\!\frac{\,2\text{TeV}\,}{\Lambda}\!\)^{\!\!4}
	+0.0172\!\(\!\frac{\,2\text{TeV}\,}{\Lambda}\!\)^{\!\!8}\right]
	\!\text{fb}\,,
	\\[1mm]
	\sqrt{s\,}\!=5~\text{TeV}, &~~~&
	\sigma(\mu\mu\gamma) = \left[0.116 \pm 0.0004\!\(\!\frac{\,2.5\text{TeV}\,}{\Lambda}\!\)^{\!\!4}
	+0.0222\!\(\!\frac{\,2.5\text{TeV}\,}{\Lambda}\!\)^{\!\!8}\right]\!\text{fb}\,.
	~~~~~~~~~~
	\end{eqnarray}	
	\eeqs
}
With these, we obtain the signal significance $\SZZ_8^\mu$ for the $\mu\mu\ga$ channel,
{\small
\beqs
\label{eq:Z8-MU}
\begin{eqnarray}
	\sqrt{s}=250\,\text{GeV}, &~~~&
	\SZZ_{8}^{\mu}
	= \left|\pm
	1.0\!\(\!\frac{\,0.5\text{TeV}\,}{\Lambda}\!\)^{\!\!4} +
	0.215\!\(\!\frac{\,0.5\text{TeV}\,}{\Lambda}\!\)^{\!\!8}\right|
	\!\times\! \sqrt{\ep\,}\,, \hspace*{25mm}
	\\[0mm]
	\sqrt{s}=500\,\text{GeV}, &~~~&
	\SZZ_{8}^{\mu} =
	\left|\pm
	0.39\(\!\frac{\,0.8\text{TeV}\,}{\Lambda}\!\)^{\!\!4} +
	0.236\!\(\!\frac{\,0.8\text{TeV}\,}{\Lambda}\!\)^{\!\!8}\right|
	\!\times\!\sqrt{\ep\,}\,,
	\\[0mm]
	\sqrt{s}=1\,\text{TeV}, &~~~&
	\SZZ_{8}^{\mu} =
	\left|\pm
	0.32\!\(\!\frac{\,\text{TeV}\,}{\Lambda}\!\)^{\!\!4} +
	1.38\!\(\!\frac{\,\text{TeV}\,}{\Lambda}\!\)^{\!\!8}\right|
	\!\times\!\sqrt{\ep\,}\,,
	\\[0mm]
	\sqrt{s}=3\,\text{TeV}, &~~~&
	\SZZ_{8}^{\mu} =
	\left|\pm
	0.06\!\(\!\frac{\,2\,\text{TeV}\,}{\Lambda}\!\)^{\!\!4}+
	1.35\!\left(\!\frac{\,2\,\text{TeV}\,}{\Lambda}\!\right)^{\!\!8} \right|
	\!\times\!\sqrt{\ep\,}\,,
	\\[0mm]
	\sqrt{s}=5\,\text{TeV}, &~~~&
	\SZZ_{8}^{\mu} =
	\left|\pm
	0.03\!\(\!\frac{\,2.5\text{TeV}\,}{\Lambda}\!\)^{\!\!4} +
	2.90\!\(\!\frac{\,2.5\text{TeV}\,}{\Lambda}\!\)^{\!\!8}\right|
	\!\times\! \sqrt{\ep\,} \,.
\vspace*{-2.5mm}
\end{eqnarray}
\eeqs
}

The analysis of the $\tau\tau\ga$ channel is the same as that of the $\mu\mu\ga$ channel,
since the $\mu$ and $\tau$ masses are negligible compared to
the collision energy $\sqrt{s\,}$ and the $Z$ boson mass $M_Z^{}$.
Finally, we obtain the combined signal significance:
\beqa
\SZZ_{\ell\ell}^{} =
\sqrt{(\SZZ_4^{e})^2\!+\!(\SZZ_4^{\mu})^2\!+\!(\SZZ_4^{\tau})^2
\!+\!(\SZZ_8^e)^2\!+\!(\SZZ_8^\mu)^2\!+\!(\SZZ_8^\tau)^2\,} \,,
\eeqa
where $\,\SZZ_4^{\tau}\!\simeq\! \SZZ_4^{\mu}\,$ and
$\,\SZZ_8^{\tau}\!\simeq\!\SZZ_8^{\mu}\,$.\,
By requiring the signal significance $\,\SZZ_{\ell\ell}^{}\!=\!2,5$,\,
we derive the $2\sigma$ and $5\sigma$ bounds on
the corresponding new physics scale
$\,\cut=\!\cut_{\ell\ell}^{2\sigma}\!,\cut_{\ell\ell}^{5\sigma}$.\,
We present these bounds in Table\,\ref{tab:3}.
In comparison with Table\,\ref{tab:22}, we see that the refinements on
the $\cut$ bounds are rather minor,
so the results of Section\,\ref{sec:3.2} are little affected.
This is because the additional non-resonant backgrounds
in Fig.\,\ref{fig:8} can be sufficiently suppressed by kinematic cuts
on the photon transverse momentum and the invariant mass of lepton pair.
Furthermore, Eqs.\eqref{lam4} and \eqref{lam8} indicate the relation
$\cut\!\propto\! \SZZ_4^{-\frac{1}{4}}\!\!
 \propto\! (\sigma_c^0)^{\frac{1}{8}}$
 when the $\cut^{-4}$ contribution dominates the signal,
and the relation
$\cut\!\propto\!\SZZ_8^{-\frac{1}{8}}
 \!\!\propto\! (\sigma_c^0)^{\frac{1}{16}}$ when the $\cut^{-8}$
contribution dominates the signal,
which are insensitive to the change of the background cross section
$\sigma_c^0$\,.\,

\begin{figure}[t]
\centering
\hspace*{-5mm}
\includegraphics[width=7.7cm,height=6.8cm]{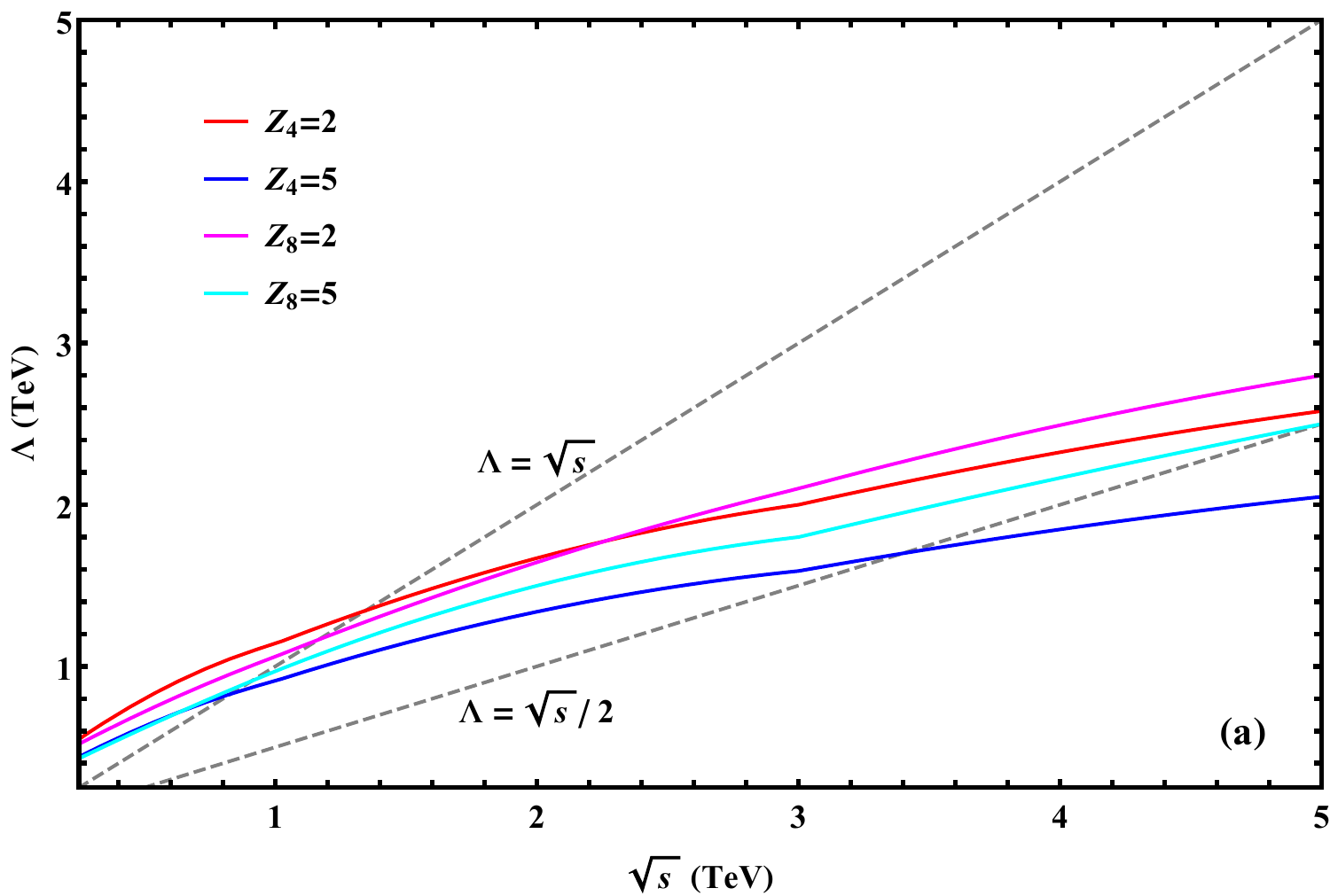}
\includegraphics[width=7.7cm,height=6.8cm]{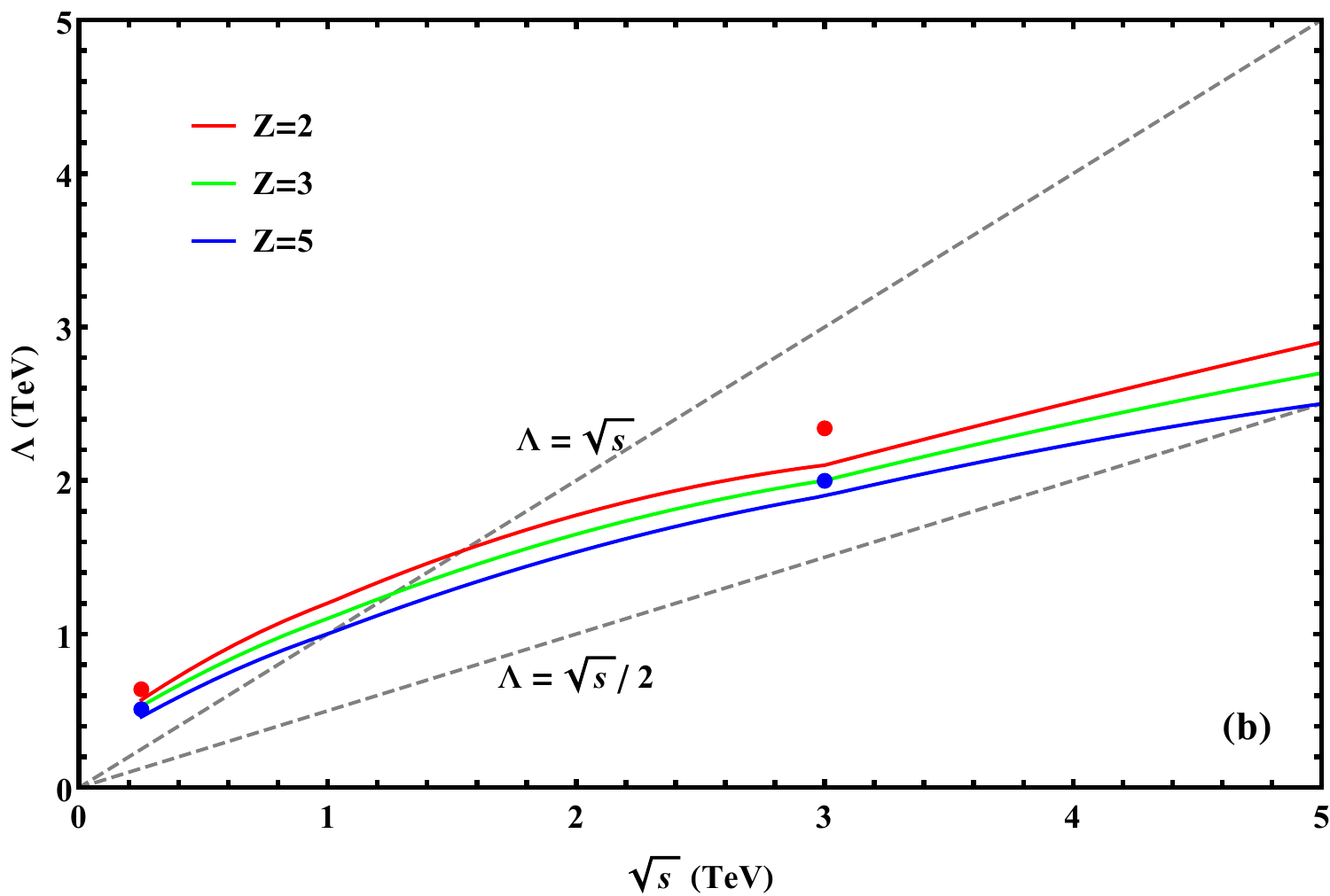}
\vspace*{-3mm}
\caption{\it Reaches for the new physics scale $\cut$ as functions
of the $e^+ e^-$ collision energy $\sqrt{s\,}$
with an integrated luminosity $\,\mL =2\text{ab}^{-1}$.\,
Plot\,(a) shows the $\cut$ reaches for the signal significances $(\SZZ_{4}^{},\,\SZZ_{8}^{})$
at $2\sigma$ and $5\sigma$ levels, respectively.
Plot\,(b) depicts the combined sensitivity
$\,\SZZ =\sqrt{\SZZ_{4}^{2}\!+\!\SZZ_{8}^{2}\,}=(2,\,3,\,5)\sigma$,\,
as shown by the (red, green, blue) curves.
In this plot, the (red, blue) dots show the $(2\si ,\,5\si )$ reaches
with a projected integrated luminosity $\,\mL =5\text{ab}^{-1}$
at $\sqrt{s\,}=\rm{250GeV}$ (CEPC and FCC-ee) and
$\sqrt{s\,}=\rm{3TeV}$ (CLIC).
For reference, we also show the lines
$\,\cut = \!\sqrt{s\,}$\, and $\,\cut = \!\sqrt{s\,}/2$\, in each plot.}
\label{Z4L}
\label{fig:7}
\label{fig:88}
\vspace*{4mm}
\end{figure}

\vspace*{1mm}

Finally, we present in Fig.\,\ref{fig:88}
reaches for the new physics scale $\cut$ as functions
of the $e^+ e^-$ collision energy $\sqrt{s\,}$,\,
with a universal integrated luminosity $\,\mathcal{L} =2\,\text{ab}^{-1}$.\,
In Fig.\,\ref{fig:7}(a), we show the $\cut$ reaches
for the signal significances $(\SZZ_{4}^{},\,\SZZ_{8}^{})$
at $2\sigma$ and $5\sigma$ levels, respectively.
Then, in Fig.\,\ref{fig:7}(b), we depict the combined sensitivity
$\,\SZZ \!=\!\sqrt{\SZZ_{4}^{}\!+\!\SZZ_{8}^{}\,}\!=(2,\,3,\,5)\sigma$\,,\,
shown as the (red,\,green,\,blue) curves.
Furthermore, we present the sensitivity reaches with
a projected integrated luminosity $\,\mathcal{L} =5\,\text{ab}^{-1}$
at $\sqrt{s\,}=\rm{250\,GeV}$ (CEPC and FCC-ee) and $\sqrt{s\,}=\rm{3\,TeV}$ (CLIC),
shown as (red, blue) dots at $(2\sigma ,\,5\sigma )$ level.
For reference, we also show two lines
$\,\cut \!=\! \sqrt{s\,}\,$ and $\,\cut \!=\! \sqrt{s\,}/2\,$ in each plot,
since the effective field theory description may be expected to hold when
$\,\cut \!>\! \sqrt{s\,}\,$ or $\,\cut \!>\! \sqrt{s\,}/2\,$.%
\footnote{%
In the effective theory approach, the exact relation between the cutoff scale
$\cut$ and the mass $M$ of the lowest underlying new state $X$ is unknown, and
one expects $\,M/\cut =O(1)$.\, If the new state $X$ could only be produced in pairs
in the $e^+e^-$ collisions (e.g., the production of dark matter particles),
$\sqrt{s}/2$ would be the appropriate comparison scale for $M$ when considering
applicability of the effective field theory.
}

We note that throughout the present study
we have used $\cut$ to denote the cutoff scale,
which is connected to another commonly used notation
$\tilde{\cut}$ via $\,\tilde{\cut}\!=\!|c_j^{}|^{1/4}\cut\,$,\,
as defined around Eq.\eqref{cj}.
The cutoff $\tilde{\cut}$ may be regarded as the true new physics scale.
The operators \eqref{eq:dim8H} invoke both gauge bosons and Higgs doublets.
In a strongly-interacting underlying theory, the coefficient
could be sizeable ($\,c_j^{}\gg 1\,$),\footnote{%
In strongly-coupled theories, one could use the generalized na\"{i}ve dimensional
analysis (GNDA)\,\cite{GNDA} to estimate roughly the size of the coefficient $c_j^{}$,
but should keep in mind that this is by no means a solid derivation, nor is
it really model-independent. For instance, according to the GNDA,
depending on how the $SU(2)$ weak gauge bosons are actually involved in the underlying
strong dynamics, the size of $c_j^{}$ for $\mathcal{O}_{\widetilde{B}W}$ is expected
to vary between $O(1)$ and $O(4\pi)^2$
up to a possible overall gauge coupling factor $g'g$\,.
}
so the actual cutoff scale $\tilde{\cut}$
could be significantly larger than the quantity $\cut$
shown in Fig.\,\ref{fig:88}.
For instance, when $c_j^{}\!\approx 4\pi$,\,
we have  $\,\tilde{\cut}\approx 1.9\cut\,$.\,

\vspace*{1mm}

In passing, it is clear that the same $ZZ\gamma$ nTGC contributes to both the
processes $\,e^+e^-\!\!\to\! Z\gamma\,$ and $\,e^+e^-\!\!\to\! ZZ\,$,
but the $Z\gamma$ final state is much simpler to analyze experimentally
than $ZZ$, since the out-going photon can be detected directly
and its energy and direction can be measured with good precision.
Moreover, in this case the final state
is a 3-body system after $Z$ decay, while the diboson
$ZZ$ decays produce a more complicated 4-body final state.
Besides, if one seeks to analyze leptonic $Z$-decays,
each small leptonic $Z$-decay branching ratio
Br$[Z\!\to\!\ell\bar\ell\,]\simeq 0.1$\,
reduces the signal by an order of magnitude, while
hadronic $Z$-decays are harder to measure and disentangle.
In the future, however, when a specific lepton collider and detector are eventually approved,
it may be beneficial to study systematically
all possible channels, including the $ZZ$ channel, for probing the nTGCs.

\vspace*{1mm}

Before concluding this section, we make a final clarification.
Since some dimension-6 operators contribute to
the $Ze^+e^-$ vertex, they may cause possible deviations from the SM
prediction for $e^+e^-\!\!\to\! Z \gamma$\,
via the $t$-channel exchange diagram.
However, the $Ze^+e^-$ coupling has already been highly constrained
by the current electroweak precision data,
and will be much more severely constrained by $Z$-pole measurements
at the future $e^+e^-$ colliders such as CEPC.
For instance, Ref.\,\cite{GHX-2016} did a systematical study on the
Higgs-related dimension-6 operators at $e^+e^-$ colliders, and derived the
constraints from both the current data and the future $e^+e^-$ Higgs factory.
We find that the following dimension-6 operators will contribute to the
$Ze^+e^-$ vertex\,\cite{GHX-2016},
\beqs
\vspace*{-1mm}
\label{eq:dim6-Zee}
\beqa
O^{(3)}_L &\,=\,& (\ii H^\dagger\sigma^a\!\! \stackrel \leftrightarrow D_\mu^{}\!\! H)
(\overline \Psi_L^{} \gamma^\mu \sigma^a \Psi_L^{})\,,
\\
\mathcal O_L^{} &\,=\,& (\ii H^\dagger\! \stackrel \leftrightarrow D_\mu^{}\! H)
(\overline \Psi_L^{} \gamma^\mu \Psi_L^{})\,,
\\
\mathcal O_R^{}\! &\,=\,& (\ii H^\dagger\!\! \stackrel \leftrightarrow D_\mu^{}\!\! H)
  (\overline \psi_R^{} \gamma^\mu \psi_R^{})\,,
\eeqa
\eeqs
where each operator is associated with the corresponding cutoff suppression factor
$\, c_j^{}/\tilde{\cut}^2\equiv \cut_j^2\,$.\,
From the analysis of Ref.\,\cite{GHX-2016}, we can extract the constraints
from the current electroweak precision data, finding that
cutoff scale of the dimension-6 operator $O^{(3)}_L$ is already bounded by
$\,\cut>7.7\,$TeV ($2\sigma$ level). This is because $O^{(3)}_L$ contributes
to both $Ze\bar{e}$ and $We\bar{\nu}$ vertices at the same time, and thus modifies
$G_F^{}$; while the other two operators $O^{}_L$ and $O^{}_R$ do not contribute to
any of the precision observables among
($M_Z^{}$, $M_W^{}$, $G_F^{}$, $\alpha_{\text{em}}^{}$) \cite{GHX-2016}.\footnote{%
As clarified in Ref.\,\cite{GHX-2016},
although the SM has 3 free input-parameters $(g,\,g',\,v)$,
there are at least 4 most precisely measured observables
($M_Z^{}$, $M_W^{}$, $G_F^{}$, $\alpha_{\text{em}}^{}$)
which were {\it all used} in the fits\,\cite{GHX-2016}
of the current electroweak precision data
and the projected $Z$-pole measurements at CEPC.
As stressed in Sec.\,2 of Ref.\,\cite{GHX-2016},
this is a {\it scheme-independent approach} for fitting the precision data,
and there was no need to choose either the $Z$-scheme
($M_Z^{},\,G_F^{},\,\alpha_{\text{em}}^{}$)
or the $W$-scheme ($M_Z^{},\,G_F^{},\,M_W^{}$).
Hence, it is fully consistent and expected
that strong bounds on these dimension-6 operators can be derived\,\cite{GHX-2016}
by using the current precision data and the projected $Z$-pole measurements at CEPC.
As we have further verified from Ref.\,\cite{GHX-2016},
our fit of the current precision measurements on
($M_Z^{}$, $M_W^{}$, $G_F^{}$, $\alpha_{\text{em}}^{}$) put strong bound on $O^{(3)}_L$
because only $O^{(3)}_L$ can contribute to $G_F^{}$, but $O^{}_L$ and $O^{}_R$ contribute
to none of the 4 observables. The size of this current bound on $O^{(3)}_L$ is mainly
controlled by the $M_W^{}$ measurement since it has the largest uncertainty among
these 4 precision observables.}
After including the projected $Z$-pole measurements
at the CEPC, we find that all the dimension-6 operators \eqref{eq:dim6-Zee} will
be severely constrained with their cutoff scales bounded by
$\,\cut_6^{} > 21.2\!-\!34.5\,$TeV ($13.4\!-\!21.8\,$TeV)
at the $2\sigma$ ($5\sigma$) level,
as summarized in Table\,\ref{tab:4new}.
On the other hand, our present nTGC study of the dimension-8 operator
shows that the nTGC bounds obey $\,\cut_8^{}\! > 2.6$\,TeV (2.1\,TeV)
at $2\sigma$ ($5\sigma$) level for collider energies up to $\sqrt{s}=3$\,TeV
(cf.\ Tables\,\ref{tab:55} and \ref{tab:66}).
Hence, the cutoff scale $\,\cut_6^{}$ of the dimension-6 operators are already
severely constrained to be substantially higher than our present nTGC bounds
on the dimension-8 operator by about an order of magnitude, namely,
$\,\cut_6^{}=O(10)\cut_8^{}\gg \cut_8^{}\,$.\,

\begin{table}[t]
	\begin{center}
		\begin{tabular}{c||c|c|c|c}
			\hline\hline
&&&& \\[-2.6mm]
Dim-6 Operators & $\mathcal{O}_L^{(3)}$ & $\mathcal{O}_L^{(3)}$
& $\mathcal{O}_L^{}$& $\mathcal{O}_R^{}$ \\
& (current bound) &&& \\[-0.3mm]
\hline\hline
&&&& \\[-3.5mm]
$\Lambda_{2\sigma}^{}$\,(TeV) & 7.7 & 34.5 & 19.2 & 21.2 \\
&&&& \\[-2.9mm]
\hline
&&&& \\[-2.75mm]
$\Lambda_{5\sigma}^{}$\,(TeV) &4.8& 21.8 & 12.2 &13.4\\
\hline\hline
\end{tabular}
\end{center}
\vspace*{-3.7mm}
\caption{{\it Constraints on the scales of dimension-6 operators \eqref{eq:dim6-Zee} by the current
electroweak precision data (2nd column) and by further including the $Z$-pole measurements
at the CEPC (3rd--5th columns), at the $2\sigma$ level (2nd row) and the $5\sigma$ level (3rd row).}	
}
\label{tab:4new}
\end{table}

\vspace*{1mm}

Using the relation  $\,\cut_6^{}=O(10)\cut_8^{}$,\,
we can compare the contributions to
$e^-e^+\!\!\to\! Z\gamma$
from a dimension-6 operator $\mathcal{O}_6^{}$ in Eq.\eqref{eq:dim6-Zee}
and from the dimension-8 operator
$\mathcal{O}_8^{}\!=\!\mathcal{O}_{\widetilde{B}W}^{}$.\,
We have presented in Appendix\,\ref{sec:A1} the contributions of the
dimension-8 operator $\mathcal{O}_8^{}$ to the helicity amplitudes
for $ee\!\to\! Z\gamma$\, as in Eq.\eqref{eq:T8};
while for the current purpose of comparison,
we can estimate the contributions of dimension-6 operators
\eqref{eq:dim6-Zee} by power-counting.
With these, we summarize the estimated sizes of the contributions
to the helicity amplitudes by both dimension-8 and dimension-6 operators,
{\small
\beqs
\vspace*{-3mm}
\beqa
&&\hspace*{-15mm}
\mathcal{T}_{(8)}^{}[\pm\pm]\sim \frac{\,M_Z^2E^2\,}{\cut_8^4},~~~~~
\mathcal{T}_{(8)}^{}[\pm\mp]=0\,,~~~~~
\mathcal{T}_{(8)}^{}[0\pm]\sim \frac{\,M_Z^{}E^3\,}{\cut_8^4},~~~~~
\\
&&\hspace*{-15mm}
\mathcal{T}_{(6)}^{}[\pm\pm]\sim \frac{\,M_Z^4\,}{\,E^2\cut_6^2\,},~~~~~
\mathcal{T}_{(6)}^{}[\pm\mp]\sim\frac{M_Z^2}{\,\cut_6^2\,},~~~~
\mathcal{T}_{(6)}^{}[0\pm]\sim \frac{\,M_Z^3\,}{\,E\cut_6^2\,},
\eeqa
\eeqs
}
\hspace*{-3mm}
where the helicity amplitude $\mathcal{T}_{(8)}^{}[\pm\mp]$ vanishes
due to the symmetry structure of $\mathcal{O}_8^{}$.\,
For the $Z\gamma$ helicity combinations $\lambda\lambda'=\pm\pm,\,0\pm$,\,
we find that the ratios of the
$\mathcal{O}_6^{}$ and $\mathcal{O}_8^{}$ contributions are given by
%
\beqa
\label{eq:T6/T8}
\frac{\,\mathcal{T}_{(6)}^{}\,}{\,\mathcal{T}_{(8)}^{}\,}
\sim
\frac{M_Z^2}{\,E^2\,}
\frac{\cut_8^2}{\,E^2\,}\frac{\,\cut_8^2\,}{\cut_6^2}
\sim\frac{M_Z^2}{\,E^2\,}\frac{\,\cut_8^2\,}{\cut_6^2}
\ll O(10^{-2})\,,
\eeqa
%
where $\,E\!=\!\!\sqrt{s}$\, and $M_Z^2/E^2\!\ll\! 1\,$.\,
For the nTGC analysis of $\mathcal{O}_8^{}$, our Tables 6 and 7 show
that the probed cutoff scale $\Lambda_8^{}$ is close to or somewhat below
the collider energy, $\Lambda_8^{}\!\sim\! E$\,,\,
which is used in the second step of the above estimates.
For the $Z\gamma$ helicity combinations $[\pm\mp]$\,,\,
the $\mathcal{O}_8^{}$ contribution vanishes, and
$\mathcal{T}_{(6)}^{}[\pm\mp]$ just interferes with the SM amplitude
$\mathcal{T}_{\text{sm}}^{}[\pm\mp]$.\,
Thus, we can estimate this interference contribution
to the cross section as
$\,\sigma_{\!(6)}^{}[\pm\mp]\sim\frac{M_Z^2}{\,E^2\cut^2_6\,}\,$.
This will be compared to the interference contribution of $\mathcal{O}_8^{}$ to the
total cross section $\,\sigma_{\!(8)}^{}\!\sim\!\frac{M_Z^2}{\cut_8^4}$\,
from Eq.\eqref{Zgamma}, or,
$\sigma_8^{}(\mathbb{O}^1_c)\sim \frac{EM_Z^{}}{\cut_8^4}$ from Eq.\eqref{eq:O1c}.
With these, we estimate the following ratios:
\beqa
\label{eq:T6/T8+-}
\frac{\sigma_{\!(6)}^{}[\pm\mp]}{\sigma_{\!(8)}^{}}
\sim \frac{\,\cut_8^2\,}{\cut_6^2} = O(10^{-2})\,,~~~~~
%
\frac{\sigma_{\!(6)}^{}[\pm\mp]}{\sigma_{\!(8)}^{}(\mathbb{O}^1_c)}
\sim \frac{\,M_Z^{}}{E}\frac{\,\cut_8^2\,}{\cut_6^2} < O(10^{-2})\,.
\hspace*{15mm}
\eeqa
The above analysis demonstrates
that the contributions of such dimension-6 operators \eqref{eq:dim6-Zee}
are always {\it negligible} for energy scales $E\!\lesssim\!\Lambda_8^{}$\,,\,
due to the severe independent constraints on their cutoff $\cut_6^{}$
imposed by the current electroweak precision data and the projected future $Z$-pole
measurements at CEPC, namely,
$\,\cut_6^{}=O(10)\cut_8^{}\gg \cut_8^{}\,$.\,
Moreover, we recall that these dimension-6 operators
do not contribute to the nTGC and thus are irrelevant to our nTGC study.
Hence, it is well justified to drop such dimension-6 operators \eqref{eq:dim6-Zee}
for the present nTGC study, which considers individually the relevant dimension-8 operator.

\section{\hspace{-5.5mm}.\,Analysis of Invisible Decay Channels
	\boldmath{$Z\!\to\!\nu\bar{\nu}$}}
\label{sec:44}

\begin{figure}[t]
	\includegraphics[width=7.8cm,height=5.5cm]{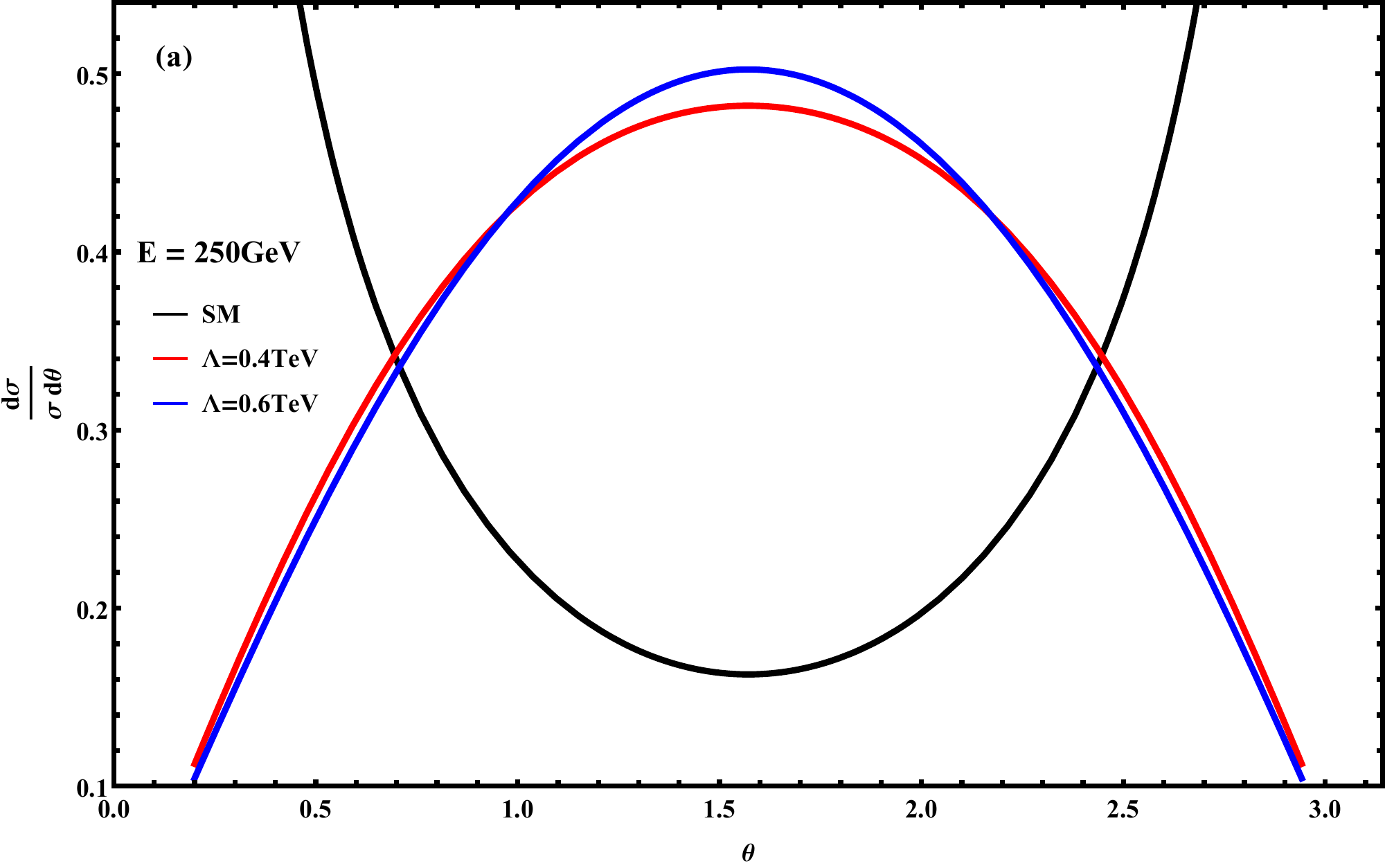}
	\includegraphics[width=7.8cm,height=5.5cm]{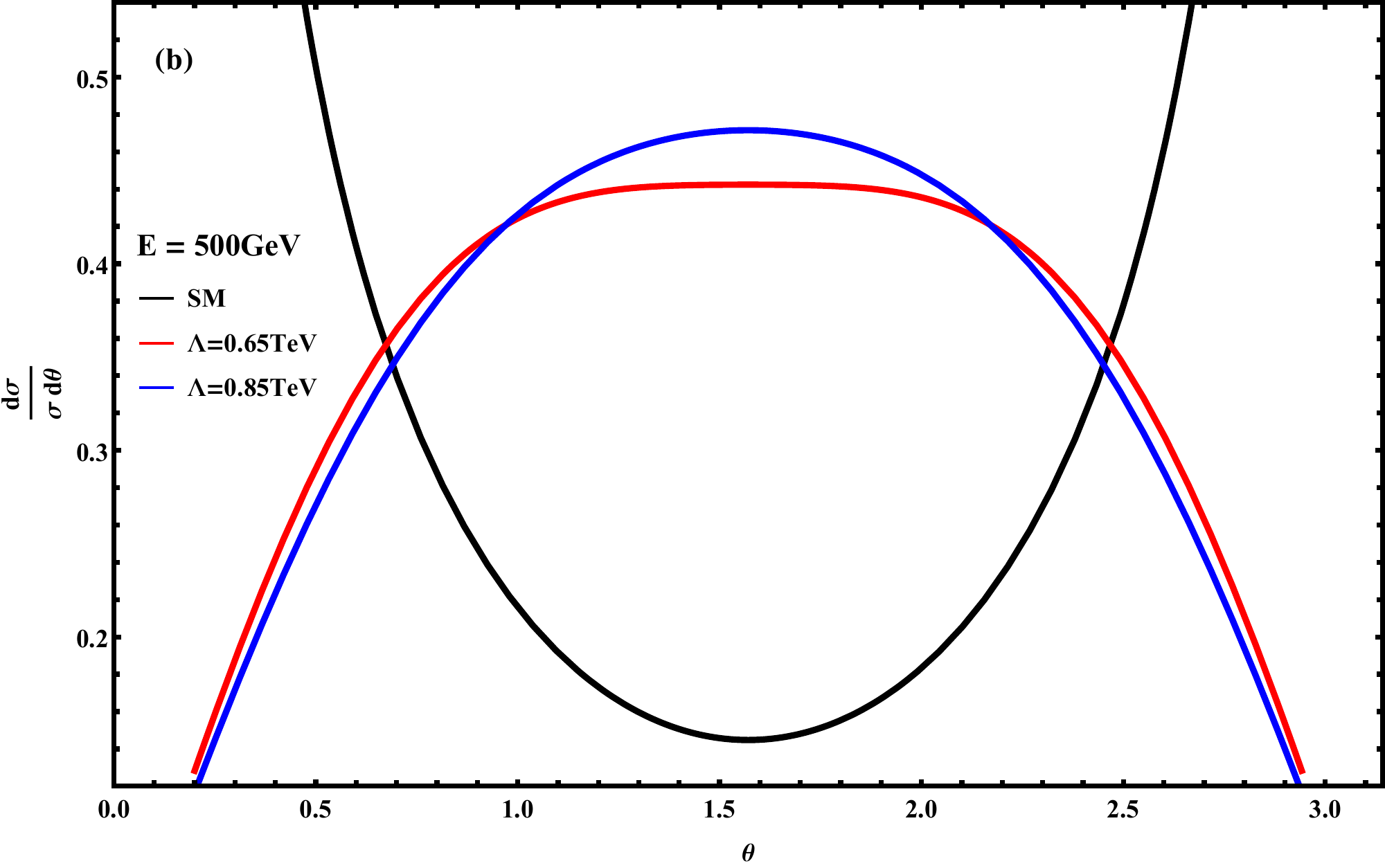}
	\\[3mm]
	\includegraphics[width=7.8cm,height=5.5cm]{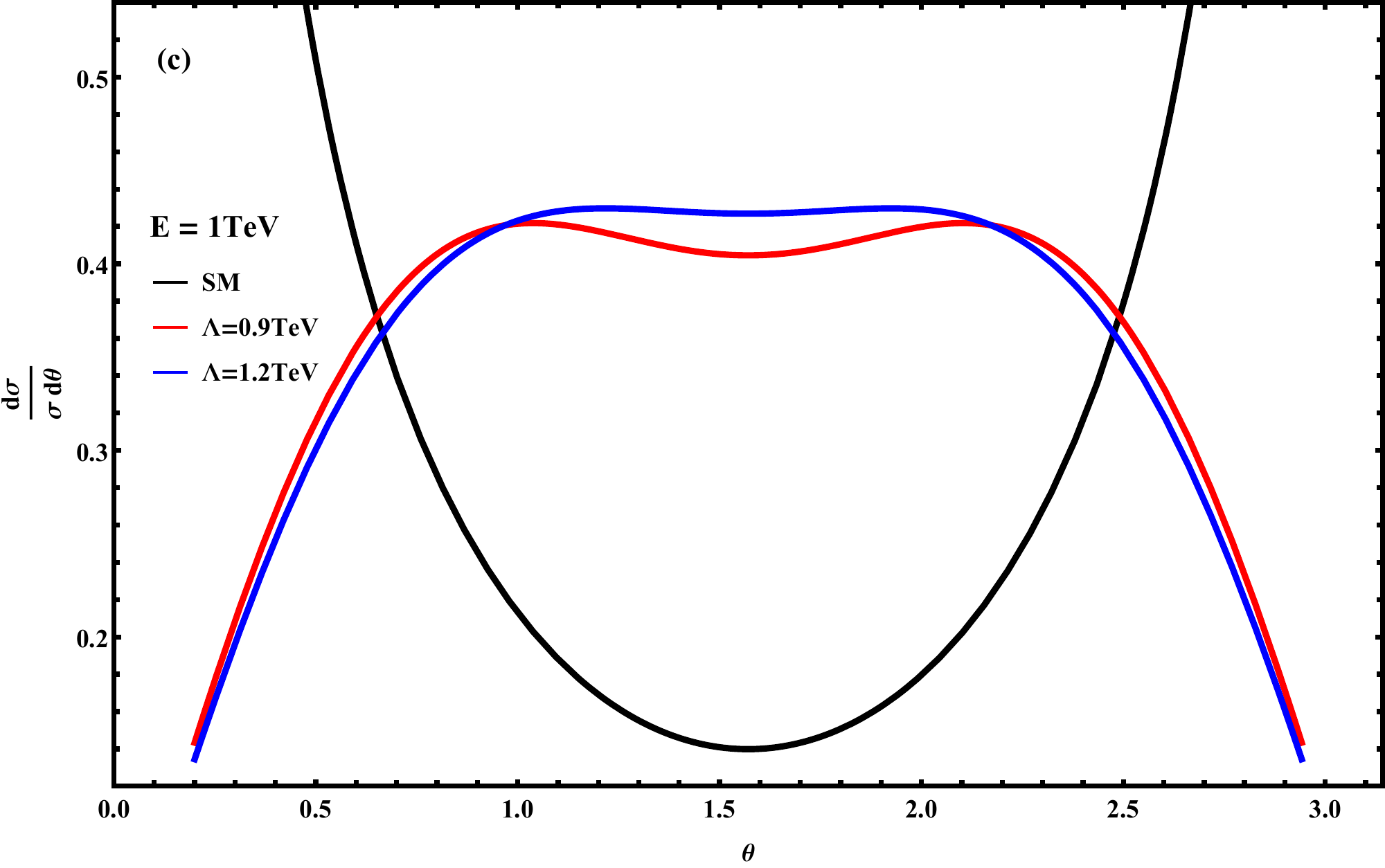}
	\includegraphics[width=7.8cm,height=5.5cm]{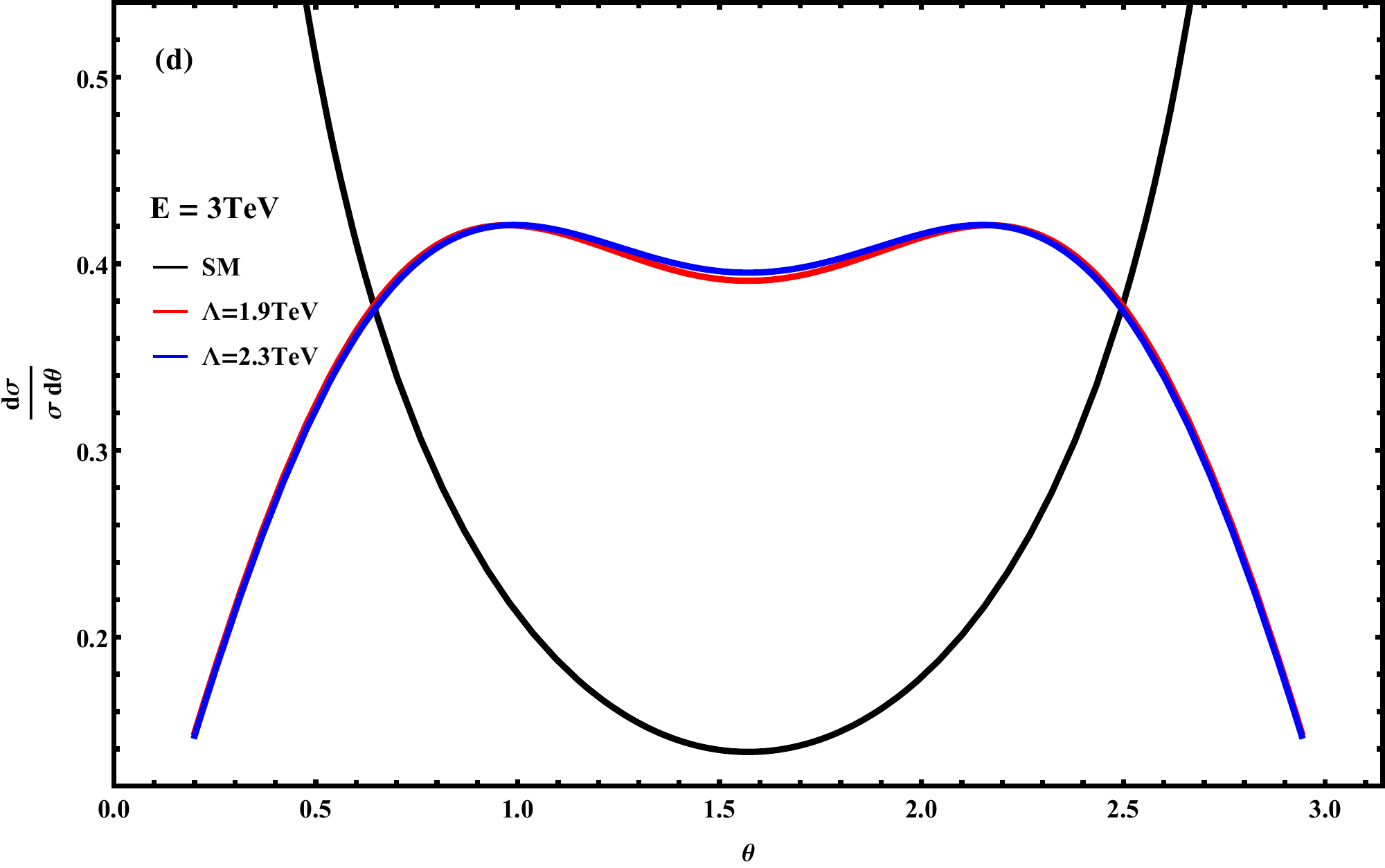}
	\vspace*{-7mm}
\caption{{\it Normalized angular distributions for the scattering angle $\theta$
        of the reaction $e^-e^+\!\!\to\!Z\ga$ (with invisible decays $Z\!\!\to\!\nu\bar{\nu}$)
		in the lab frame and at different collision energies,}
		$\sqrt{s}=(250\,\text{GeV},\, 500\,\text{GeV},\, 1\,\text{TeV},\, 3\,\text{TeV})$.
		{\it In each plot, the black curve denotes the SM contribution and
        the red and blue curves present the contributions from
	    new physics with two sample values of $\,\Lambda$\,.}}
\label{figthetan}
\label{fig:99}
\vspace*{2mm}
\end{figure}

In this section, we analyze $e^-e^+\!\!\!\to\!Z\ga$ production
followed by the invisible decays $Z\!\!\to\!\nu\bar{\nu}$\,.\,
Then, we combine its sensitivity with that of the leptonic channels
$Z\!\!\to\!\ell^-\ell^+$ presented in Section\,\ref{sec:3}.

\vspace*{1mm}

In the case of the invisible decay channels $Z\!\!\to\!\nu\bar{\nu}$,\,
we can apply the angular cut on the scattering angle
of the final state mono-photon,
$\,\delta_m^{}\!\!<\!\theta\!<\pi\!-\delta_m^{}$,\,
which corresponds to a cut on the photon transverse momentum
$P^{\ga}_T=q\sin\!\theta> q\sin\!\delta_m^{}$.\,
The angular distributions of $\,\theta\,$ are presented in Fig.\,\ref{fig:99}.
We estimate the following optimal cuts
$\,\delta_m^{}\!\!<\!\theta\!<\pi\!-\delta_m^{}$\, with
\beqa
\label{eq:theta-cut}
\delta_m^{}=\,(0.633,\,0.626,\,0.624,\,0.623,\,0.623),
\eeqa
for various collider energies
$\sqrt{s\,}\!=(250,\,500,\,1000,\,3000,\,5000)$\,GeV, respectively.
We see that the optimal angular cut $\,\delta_m^{}$\, is not very sensitive
to the variation of collider energy $\sqrt{s\,}$\,.\,
The averaged value of the cuts \eqref{eq:theta-cut} equals
$\,\sin\!\delta_m^{}\!\simeq 0.586$\,,\,
which restricts the scattering angle within the range
\beqa
\label{eq:theta-cutA}
36^\circ \lesssim \,\theta\, \lesssim 144^\circ\,.
\eeqa

\begin{figure}[t]
	\includegraphics[width=7.8cm,height=5.5cm]{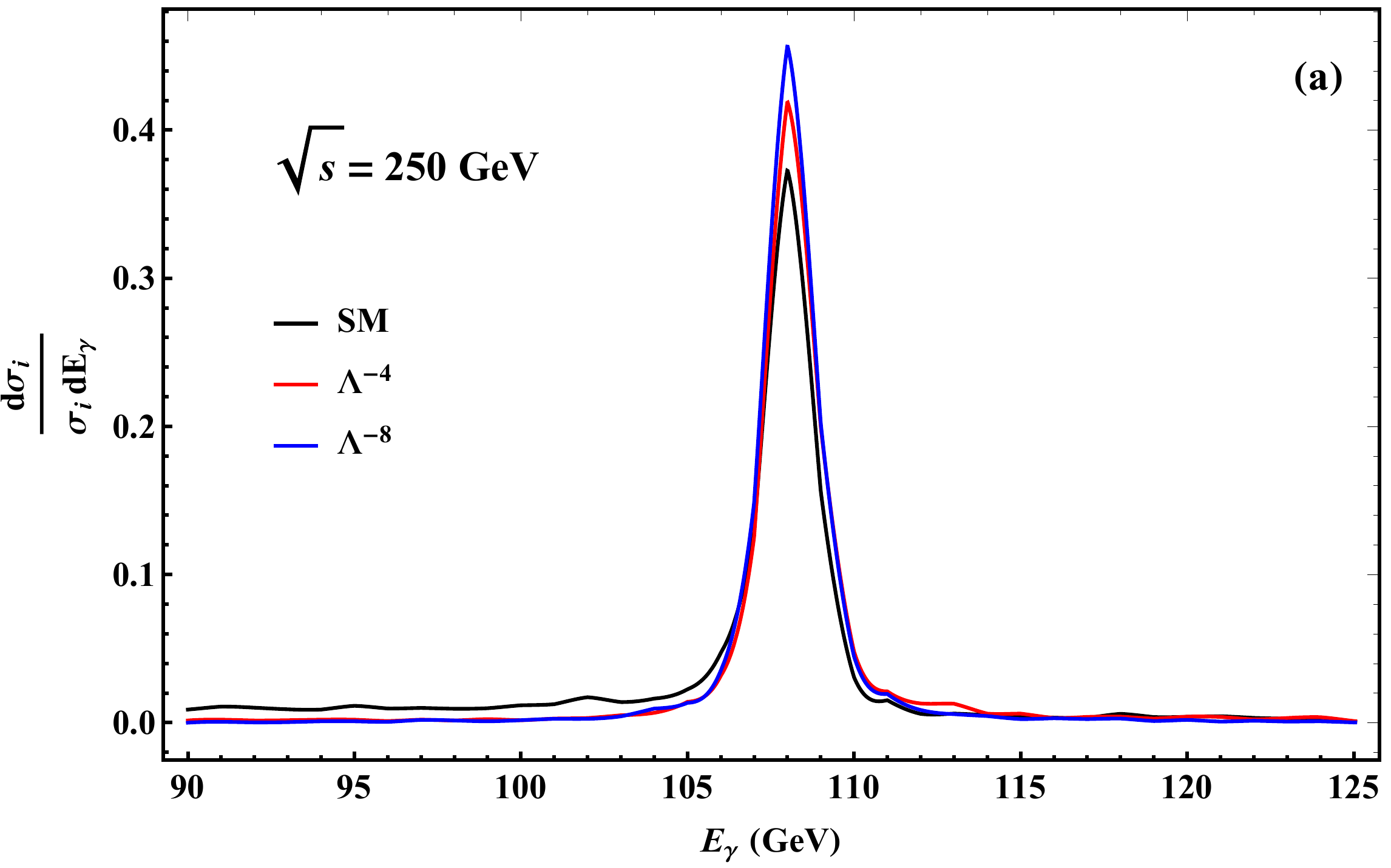}
	\includegraphics[width=7.8cm,height=5.5cm]{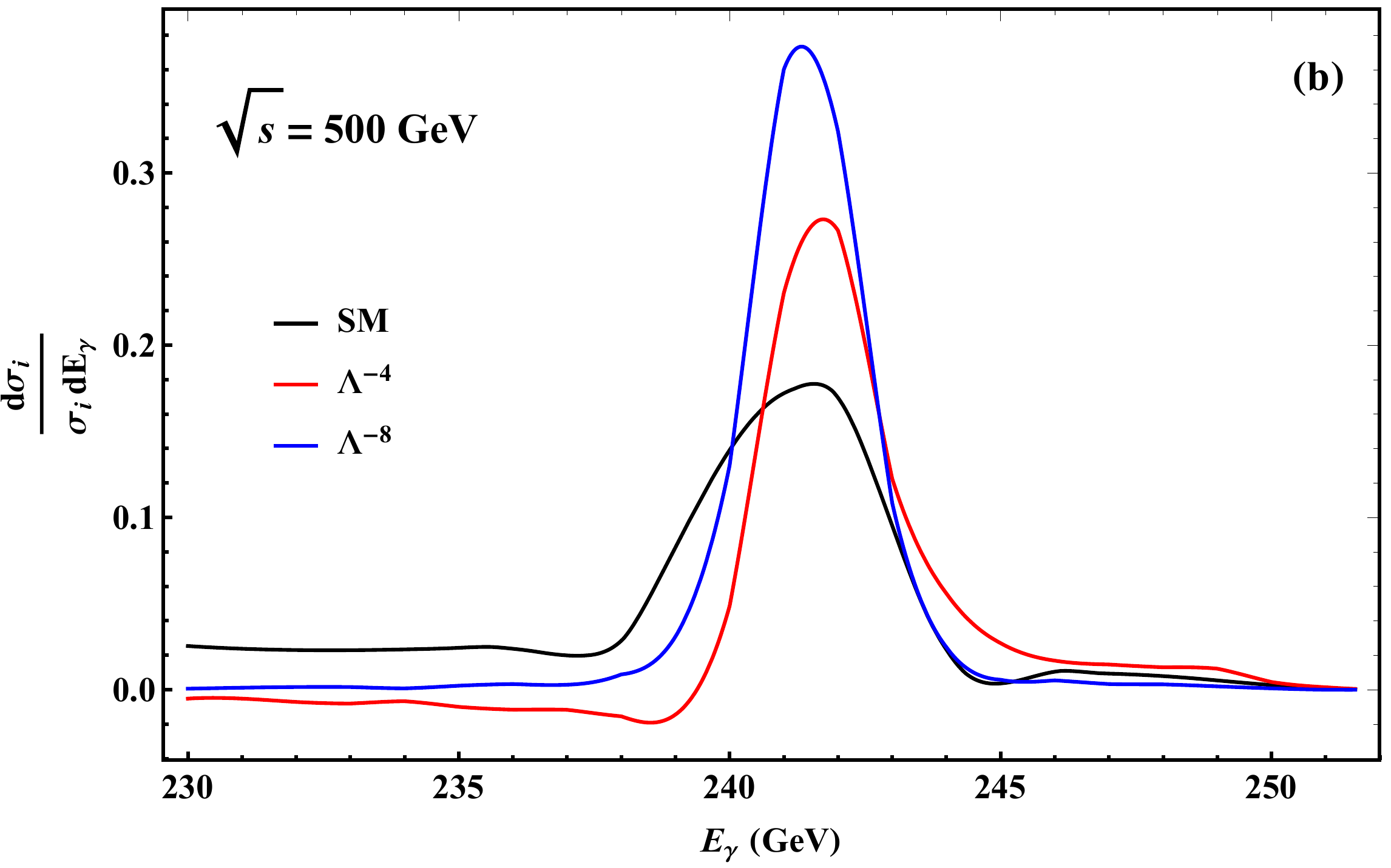}
	\\[3mm]
	\includegraphics[width=7.8cm,height=5.5cm]{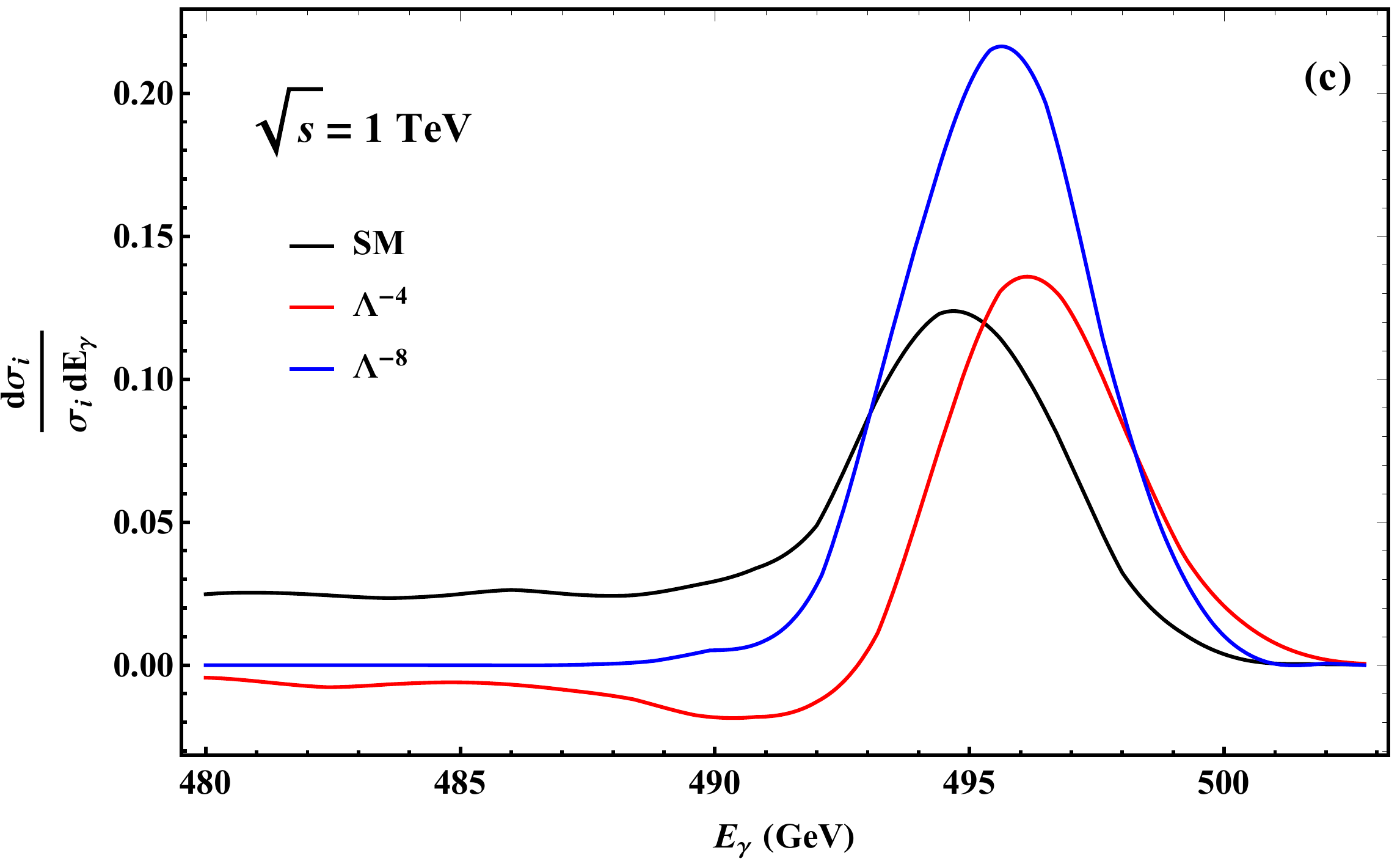}
	\includegraphics[width=7.8cm,height=5.5cm]{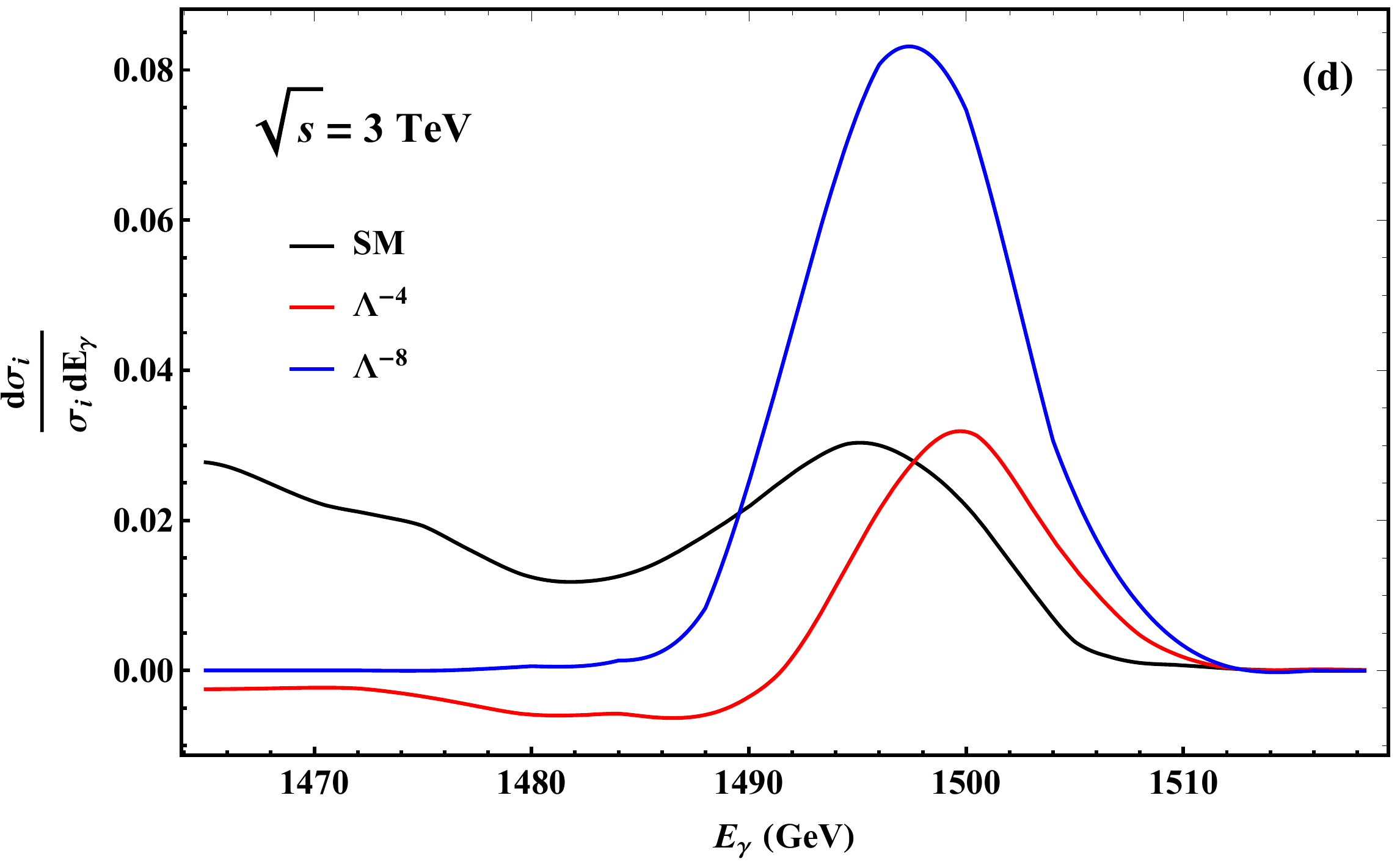}
\vspace*{-7mm}
\caption{{\it Normalized photon energy distributions for different collision energies,}
$\sqrt{s}=(250\,\text{GeV},\, 500\,\text{GeV},\, 1\,\text{TeV},\, 3\,\text{TeV})$,
{\it where the experimental photon energy resolution \eqref{eq:dE-CMS} is included.
In each plot, the (black, red, blue) curves present the SM contributions, the
new physics contributions of $\,O(\cut^{-4})$ and $\,O(\cut^{-8})$, respectively.}}
\label{figE}
\label{fig:10E}
\end{figure}

Next, we study the photon energy distributions of the signals and backgrounds.
According to the signal kinematics, the photon energy
$\,E_\gamma^{}\!=(s\!-\!M_{\nu\bar\nu}^2)/(2\sqrt s)$\,
and the invariant mass
$\,M_{\nu\bar\nu}^{}\!=M_Z^{}$\,.\,
To require $\nu\bar\nu$ mainly from on-shell $Z$-decays, we would normally
set an invariant mass window,
$\,M_{\nu\bar\nu}^{\pm}\!= M_Z\pm \delta$\, with $\delta =10$\,GeV.
This corresponds to a variation in photon energy,
$\,E_\gamma^\pm =(s-{M_{\nu\bar\nu}^{\mp}}^2)/(2\sqrt{s})$,\,
with
\beqa
\label{eq:dE-photon}
\Delta E_\gamma^{} =\, E_\gamma^+ - E_\gamma^-
= \frac{\,2M_Z^{}\delta\,}{\,\sqrt{s\,}\,},
\eeqa
which is suppressed by \,$2M_Z^{}/\!\sqrt{s\,}$\, and
may be comparable to or smaller than
the experimental energy resolution of photons in the detector for high energy
$\,\sqrt{s}\gg 2M_Z^{}$\,.\,
As a result, one cannot impose a naive photon energy cut
to ensure the on-shell decays $Z\!\!\to\!\nu\bar\nu$ via
$\,E_\gamma^- < E_\gamma^{} < E_\gamma^+$\,.\,
To take into account the experimental photon-energy resolution, we may choose
the current $E_\gamma^{}$ resolution ($\sigma_E^{}$) of the CMS
detectors\,\cite{Sirunyan:2017ulk,Chatrchyan:2013dga} as a good estimate,
\begin{equation}
\label{eq:dE-CMS}
\frac{\,\sigma_E^{}\,}{E} \,=\,
\frac{2.8\%}{\,\sqrt{E/\GeV\,}\,} \oplus
\frac{12\%}{E/\GeV\,} \oplus 0.3\% \,.
\end{equation}	
Thus, we set the photon energy cut as follows,
\beqa
\label{eq:Ephoton-cut}
\min (E_\gamma^-,\bar E_\gamma\!-\!2\sigma_E^{})
< E_\gamma^{} <
\max (E_\gamma^+,\bar E_\gamma\!+\!2\sigma_E^{}) \,,
\eeqa
where
$\,\bar{E}_\gamma^{}=\,(s\!-\!M_Z^2)/(2\sqrt{s})$\,.\,
In our event simulations,
$E_\gamma^{}$ is the photon energy convoluted with the experimental resolution
\eqref{eq:dE-CMS}.

\vspace*{1mm}

In Fig.\,\ref{fig:10E}, we present the normalized photon energy distributions
for different collision energies
$\sqrt{s\,}\!=(250\,\GeV,\,500\,\GeV,\,1\,\TeV,\,3\,\TeV )$.
We find that the interference term of $O(\cut^{-4})$ is negligible
for high collision energies $\sqrt{s}\gtrsim 3$\,TeV; while for
$\sqrt{s}\lesssim 1$\,TeV, the interference term becomes important,
in which the contribution of the SM $W$-exchange diagram
[Fig.\,\ref{fig:77}(e)] is negligible.
We inspect the photon energy distributions at different collision energies
in Fig.\,\ref{fig:10E}. We see that the SM background distribution (black color)
and the new physics distribution (red color) have similar shapes for
$\sqrt{s}\lesssim 500$\,GeV,
implying that the photon energy cut is not so useful here.
This is because the SM $W$-exchange contribution [Fig.\,\ref{fig:77}(e)]
is negligible in such cases.
But, for $\sqrt{s}\gtrsim\! 1$\,TeV,
the shape of the new physics distribution differs
from that of the SM backgrounds substantially, since the SM $W$-exchange
contribution [Fig.\,\ref{fig:77}(e)] becomes important.
Hence, the photon energy cut becomes
important for effectively suppressing the SM backgrounds,
especially the SM $W$-exchange diagram in Fig.\,\ref{fig:77}(e).

\vspace*{1mm}

\begin{figure}[t]
\begin{center}
\includegraphics[width=8cm,height=6cm]{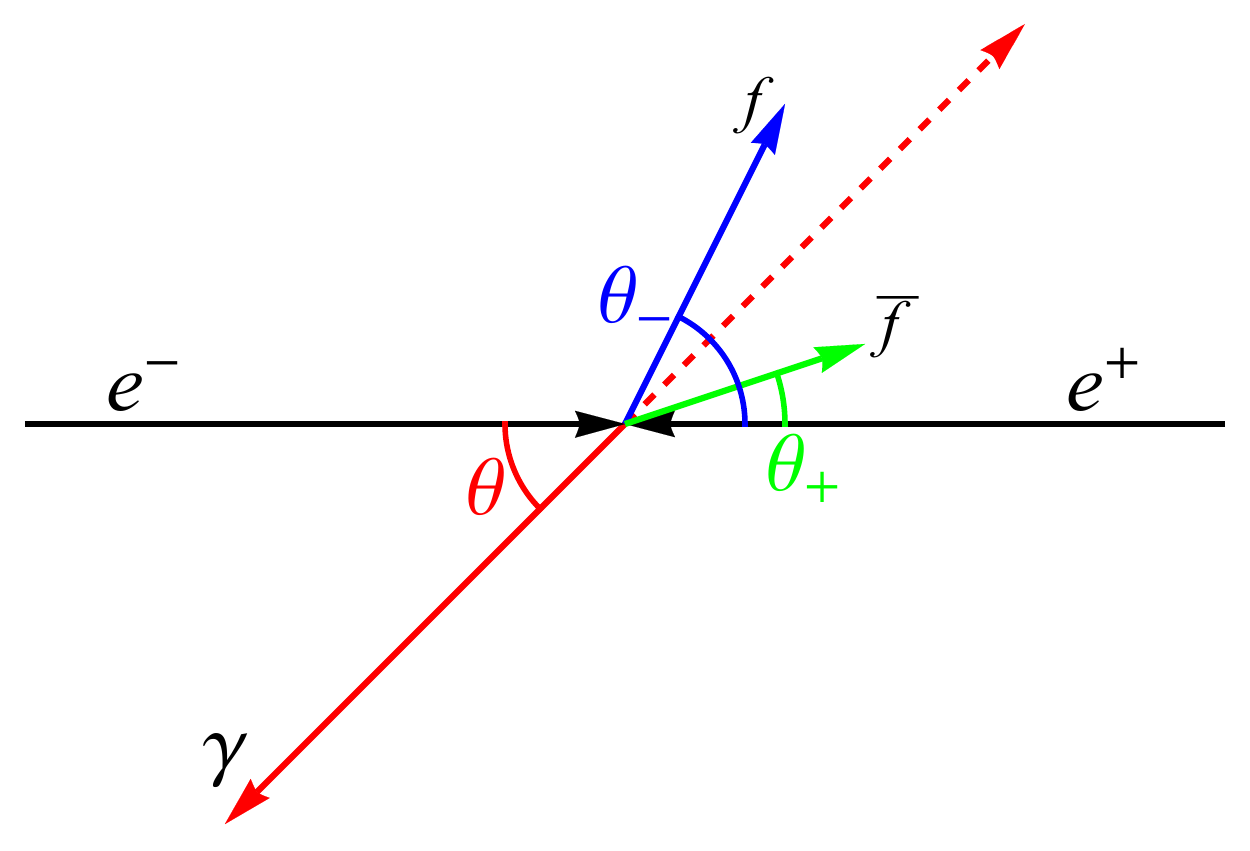}
\end{center}
\vspace*{-8mm}
\caption{{\it Illustration of kinematics of the reaction
$\,e^-e^+\!\to f\bar{f}\gamma\,$, where the vector with red color denotes
the photon 3-momentum $\vec{p}_\gamma^{}$, and the vector with blue (green) color
denotes the 3-momentum $\vec{p}_f^{}$ ($\vec{p}_{\bar{f}}^{}$) of the final state
fermion $f$ ($\bar{f}$). The momentum conservation requires
$(\vec{p}_f^{}\!+\vec{p}_{\bar{f}}^{})+\vec{p}_\gamma^{}=0$\,.\,
From this, we deduce:
(i)~either $\theta_-^{}\!>\theta$\,
or $\,\theta_+^{}\!>\theta$\,
for the case of $\,0^\circ\!<\!\theta\! <\!90^\circ$;\, and
(ii)~either $\theta_-^{}\!<\theta$ or $\,\theta_+^{}\!<\theta$
for the case of $\,90^\circ\!<\!\theta\!<180^\circ$.\,
}}
\label{fig:11new}
\end{figure}

Finally, in addition to the irreducible backgrounds for
$e^-e^+\!\to\nu\bar{\nu}\gamma$ (Fig.\,\ref{fig:77}),
we have also considered the possible reducible background
$e^-e^+\!\!\to\! e^-e^+\gamma$ with both $e^-$ and $e^+$ lost in the beam pipe.
We illustrate the kinematics of this background process as in Fig.\,\ref{fig:11new},
the final state fermions $\,f\bar{f}=e^-e^+$.\,
As defined before (Fig.\,\ref{fig:1}),
the scattering angle $\theta$ is the angle between the moving directions
of the incoming $e^-$ and outgoing $Z$,
which equals the angle between the incoming $e^+$ and outgoing $\gamma$
(Fig.\,\ref{fig:11new}).
For any final state $\,e^-e^+\gamma$\,,\, the sum of the
3-momenta of $e^-$ and $e^+$ must cancel exactly the photon 3-momentum:
$(\vec{p}_{e^-}^{} \!+ \vec{p}_{e^+}^{}\!) +\vec{p}_\gamma^{}=0$\,.\,
In Fig.\,\ref{fig:11new}, we denote the angle between the moving directions of
the final state $e^-$ ($e^+$) and initial state $e^-$ as
$\theta_-^{}$ ($\theta_+^{}$).
Thus, we must have:
(i)~either $\theta_-^{}\!>\theta$\,
or\, $\theta_+^{}\!>\theta$\,
for the case of $\,0^\circ\!<\!\theta\! <\!90^\circ$;\,
and
(ii)~either $\theta_-^{}\!<\theta$\,
or\, $\theta_+^{}\!<\theta$\, for the case of
$\,90^\circ\!<\!\theta\!<180^\circ$.\,
We observe that in both cases (i) and (ii) at least one of the
final state $e^-$ and $e^+$ must have its moving direction deviate
more from the beam pipe than the outgoing photon.
Because we have imposed significant angular cuts
\eqref{eq:theta-cutA} that restrict the scattering angle of the outgoing photon
within the range $\,36^\deg\lesssim\theta\lesssim 144^\deg$,\,
we know that at least one of the outgoing $e^-$ and $e^+$ must have its angle
($\theta_-^{}$ or $\theta_+^{}$) fall into this window
$\,36^\deg\lesssim\theta_\mp^{}\lesssim 144^\deg$,\,
and thus cannot escape the detector.
This shows that there is no chance that both the final-state particles
$e^-$ and $e^+$ could escape detection, so such $e^-e^+\gamma$ events
can be always vetoed.
Therefore, under the cuts \eqref{eq:theta-cutA},
the reducible $e^-e^+\gamma$ background is fully negligible
in our study.

\vspace*{1mm}

With the photon angular cut \eqref{eq:theta-cut} and energy cut
\eqref{eq:Ephoton-cut}, we compute the cross section of
$\,e^-e^+\!\!\to\nu\bar\nu\gamma$\, as a function of the new physics scale
$\cut$\,,
{\small
	\beqs
	\begin{eqnarray}
	\sqrt{s\,}\!=250~\text{GeV}, &~~~&
	\sigma(\nu\bar\nu\gamma) = \left[635
	\pm 1.33\!\left(\!\frac{\,0.5\text{TeV}\,}{\Lambda}\!\right)^{\!\!4}\!
	+ 0.282\!\left(\!\frac{\,0.5\text{TeV}\,}{\Lambda}\!\right)^{\!\!8}
	\right]\!\text{fb}\,,~~~~~~~~
	\\[1mm]
	\sqrt{s\,}\!=500~\text{GeV}, &~~~&
	\sigma(\nu\bar\nu\gamma)
	= \left[126 \pm 0.201\!\(\!\frac{\,0.8\text{TeV}\,}{\Lambda}\!\)^{\!\!4}\!
	+0.124\!\(\!\frac{\,0.8\text{TeV}\,}{\Lambda}\!\)^{\!\!8}\right]\!\text{fb}\,,
	\\[1mm]
	\sqrt{s\,}\!=1~\text{TeV}, &~~~&
	\sigma(\nu\bar\nu\gamma)  = \left[33.8 \pm 0.086\!\(\!\frac{\,\text{TeV}\,}{\Lambda}\!\)^{\!\!4}
	+0.370\!\(\!\frac{\,\text{TeV}\,}{\Lambda}\!\)^{\!\!8}\right]\!\text{fb}\,,
	\\[1mm]
	\sqrt{s\,}\!=3~\text{TeV}, &~~~&
	\sigma(\nu\bar\nu\gamma)
	= \left[4.28 \pm 0.004\!\(\!\frac{\,2\text{TeV}\,}{\Lambda}\!\)^{\!\!4}
	+0.124\!\(\!\frac{\,2\text{TeV}\,}{\Lambda}\!\)^{\!\!8}\right]
	\!\text{fb}\,,
	\\[1mm]
	\sqrt{s\,}\!=5~\text{TeV}, &~~~&
	\sigma(\nu\bar\nu\gamma) = \left[1.78 \pm 0.001\!\(\!\frac{\,2.5\text{TeV}\,}{\Lambda}\!\)^{\!\!4}
	+0.160\!\(\!\frac{\,2.5\text{TeV}\,}{\Lambda}\!\)^{\!\!8}\right]\!\text{fb}\,.
	\hspace*{15mm}
	\end{eqnarray}
	\eeqs
}

With these, we derive the signal significance $\SZZ_8^{}$ as follows:
{\small
\beqs
\label{eq:Z8-Znunu}
\begin{eqnarray}
	\sqrt{s}=250\,\text{GeV}, &~~~&
	\SZZ_{8,\vvb}^{}
	= \left|\pm
	2.36\!\(\!\frac{\,0.5\text{TeV}\,}{\Lambda}\!\)^{\!\!4}\! +
	0.501\!\(\!\frac{\,0.5\text{TeV}\,}{\Lambda}\!\)^{\!\!8}\right|
	\!\times\! \sqrt{\ep\,}\,, \hspace*{25mm}
	\\[0mm]
	\sqrt{s}=500\,\text{GeV}, &~~~&
	\SZZ_{8,\vvb}^{} =
	\left|\pm
	0.801\(\!\frac{\,0.8\text{TeV}\,}{\Lambda}\!\)^{\!\!4}\! +
	0.493\!\(\!\frac{\,0.8\text{TeV}\,}{\Lambda}\!\)^{\!\!8}\right|
	\!\times\!\sqrt{\ep\,}\,,
	\\[0mm]
	\sqrt{s}=1\,\text{TeV}, &~~~&
	\SZZ_{8,\vvb}^{} =
	\left|\pm
	0.650\!\(\!\frac{\,\text{TeV}\,}{\Lambda}\!\)^{\!\!4}\! +
	2.85\!\(\!\frac{\,\text{TeV}\,}{\Lambda}\!\)^{\!\!8}\right|
	\!\times\!\sqrt{\ep\,}\,,
	\\[0mm]
	\sqrt{s}=3\,\text{TeV}, &~~~&
	\SZZ_{8,\vvb}^{} =
	\left|\pm
	0.09\!\(\!\frac{\,2\,\text{TeV}\,}{\Lambda}\!\)^{\!\!4}\! +
	2.69\!\left(\!\frac{\,2\,\text{TeV}\,}{\Lambda}\!\right)^{\!\!8} \right|
	\!\times\!\sqrt{\ep\,}\,,
	\\[0mm]
	\sqrt{s}=5\,\text{TeV}, &~~~&
	\SZZ_{8,\vvb}^{} =
	\left|\pm
	0.04\!\(\!\frac{\,2.5\text{TeV}\,}{\Lambda}\!\)^{\!\!4}\! +
	5.38\!\(\!\frac{\,2.5\text{TeV}\,}{\Lambda}\!\)^{\!\!8}\right|
	\!\times\! \sqrt{\ep\,} \,.
\end{eqnarray}
\eeqs
}

\begin{table}[t]
\begin{center}
\begin{tabular}{c|ccccc}
\hline\hline
			$\sqrt{s\,}$ (GeV) & 250 & 500 & 1000 & 3000 & 5000 \\
			\hline
			$\Lambda$~(TeV) & 0.5 & 0.7 & 1.0 & 1.9 & 2.6
			\\
			\hline\hline
			$\SZZ_{\ell\bar\ell}^{}$ & 3.6(3.2) & 4.1(3.4) & 4.5(3.9) & 4.4(4.2) & 4.2(4.1)
			\\
			\hline
			$\SZZ_{8,\nu\bar\nu}^{}$ & 2.9(2.3) & 2.8(0.07) & 3.4(2.2) & 4.1(3.9) & 4.0(3.9)
            \\
			\hline
			\\[-4.4mm]
			$\SZZ$(combined) & 4.6(3.9) & 5.0(3.4) & 5.6(4.5) & 6.0(5.8) & 5.8(5.6)
			\\
			\hline\hline
\end{tabular}
\end{center}
\vspace*{-4.2mm}
\caption{{\it Signal significances for the dilepton channels (3rd row) and invisible channels
(4th row) at different collider energies (shown in the 1st row).
For each collider energy $\sqrt{s\,}$,
we assume a representative new physics scale $\cut$ (shown in the 2nd row).
The combined signal significance}
$\SZZ$({combined}) {\it for each collision energy is presented in the last row.
Here the numbers in the parentheses correspond to the case of the
dimension-8 operator whose coefficient is negative.
For illustration, we have assumed a fixed representative integrated luminosity}
$\,\LL=2\,\text{ab}^{-1}\!$ {\it and an ideal detection efficiency $\,\ep=100\%$.}
}
\vspace*{7mm}
\label{tab:4}
\label{tab:44}
\end{table}
%

In Table\,\ref{tab:44}, we present the signal significances
at different collider energies (shown in the first row),
for the dilepton channels $Z\!\!\to\!\ell^-\ell^+$ (third row)\footnote{%
The dilepton channel results shown here are based on the analysis
of Section\,\ref{sec:3.3}.}
and for the invisible channels $Z\!\!\to\!\vvb$ (fourth row).
For each collider energy $\sqrt{s\,}$,
we input the relevant sample new physics scale $\cut$\,,\,
as shown in the second row.
We see that the signal significances of these two types of channels are
{\it comparable,} with the dilepton channels being more sensitive for
$\sqrt{s\,}\lesssim 2.5$\,TeV,
whereas their sensitivities become fairly close for
$\sqrt{s\,}\gtrsim 3$\,TeV.
We present the combined signal significance
$\,\SZZ(\text{combined})\!=\!\sqrt{\!\SZZ_{\ell\bar\ell}^2+\!\SZZ_{8,\vvb}^2}$\,,\,
in the last row for each given collision energy.
This shows that in each case the combined sensitivity
is enhanced over the individual channels
by a sizeable factor of about $1.2-\!1.4$.\,
As previously, the numbers in the parentheses of Table\,\ref{tab:44}
correspond to the case of the dimension-8 operator with a negative coefficient.
For illustration, we have assumed a fixed representative integrated luminosity
$\,\LL \!=\!2\,\text{ab}^{-1}\!$  and an ideal detection efficiency $\,\ep=100\%$\,.

\vspace*{1mm}

By requiring $\SZZ_{8}^{}\!=2$ and $\SZZ_{8}^{}\!=5$\, in Eq.\eqref{eq:Z8-Znunu},
we derive the reaches for the new physics scale $\,\cut\,$ at the $2\sigma$ and $5\sigma$ levels,
denoted as $\Lambda^{2\sigma}_{\nu\bar\nu}$ and $\Lambda^{5\sigma}_{\nu\bar\nu}$\,,\,
respectively.
We summarize the findings in Table\,\ref{tab:55}, as shown in the fourth and fifth rows.
For comparison, we also list the new physics reaches
$\Lambda^{2\sigma}_{\ell\bar\ell}$  and $\Lambda^{5\sigma}_{\ell\bar\ell}$
(second and third rows of Table\,\ref{tab:55})
from the dilepton channels of Table\,\ref{tab:3}.
Then, we derive the combined sensitivity reaches
of the new physics scale $\cut$ from both the dilepton channels and invisible channels,
which are presented in the sixth and seventh rows of the current Table\,\ref{tab:55},
denoted as
$\Lambda^{2\sigma}_{\ell\nu,\text{comb}}$ and $\Lambda^{5\sigma}_{\ell\nu,\text{comb}}$.
We see that the combined bounds
$\Lambda^{2\sigma}_{\ell\nu,\text{comb}}$ and $\Lambda^{5\sigma}_{\ell\nu,\text{comb}}$
are only slightly enhanced compared to the analysis using the
$Z\!\!\to\! \ell^+ \ell^-$\, channel alone.
This can be understood by noting that the new physics scale $\cut$ is
rather insensitive to the significance $\SZZ$,\,
because Eqs.\eqref{lam4} and \eqref{lam8} show,
$\cut\!\propto\! \SZZ_4^{-\frac{1}{4}}$
(when the $\cut^{\!-4}$ contribution dominates the signal) and
$\cut\!\propto\!\SZZ_8^{-\frac{1}{8}}$
(when the $\cut^{\!-8}$ contribution dominates the signal).

\begin{table}[t]
\begin{center}
\begin{tabular}{c|ccccc}
\hline\hline
\\[-4.4mm]
$\sqrt{s\,}$~(GeV) & 250 & 500 & 1000~~ & 3000~ & 5000 \\[0.4mm]
\hline\hline
\\[-4.4mm]
$\Lambda^{2\sigma}_{\ell\bar\ell}$\,(TeV)& 0.57(0.56) & 0.82(0.80) & 1.2~ & 2.1~ & 2.9
\\[-4.4mm]
\\
\hline
\\[-4.4mm]
$\Lambda^{5\sigma}_{\ell\bar\ell}$\,(TeV) & 0.46(0.44) & 0.67(0.64) & 0.98(0.95) & 1.9~ & 2.5
			\\[0.48mm]
			\hline\hline
			\\[-4.4mm]
$\Lambda^{2\sigma}_{\nu\bar\nu}$\,(TeV)& 0.54(0.32) & 0.74(0.61) & 1.1(1.0) & 2.1 & 2.8
\\[-4.4mm]
\\
\hline
\\[-4.4mm]
$\Lambda^{5\sigma}_{\nu\bar\nu}$\,(TeV) & 0.44(0.32) & 0.64(0.56) & 0.96(0.91) & 1.9(1.8) & 2.5
			\\[0.48mm]
			\hline\hline
			\\[-4.4mm]
$\Lambda^{2\sigma}_{\ell\nu,\text{comb}}$\,(TeV)& 0.61(0.59) & 0.85(0.80) & 1.2 & 2.2~ & 3.0
			\\[-4.4mm]
			\\
\hline
\\[-4.4mm]
$\Lambda^{5\sigma}_{\ell\nu,\text{comb}}$\,(TeV) & 0.49(0.46) & 0.70(0.64) & 1.0(0.91) & 2.0(1.9) & 2.6
\\[0.7mm]
\hline\hline
\end{tabular}
\end{center}
\vspace*{-4.2mm}
\caption{{\it %
			Sensitivity reaches for the new physics scale $\cut$
			from the $\,e^-e^+\!\!\to\!\nu\bar{\nu}\ga$\, channel,
			and from combining both $\ell^-\ell^+\gamma$ and $\nu\bar{\nu}\gamma$ channels,
			at the $2\sigma$ and $5\sigma$ levels, for different collider energies.
			Here again the numbers in parentheses correspond to the case of the
			dimension-8 operator whose coefficient has a minus sign, while in all other
			entries the differences between the two signs of the coefficient are negligible.
			For illustration, we have assumed a fixed representative integrated luminosity}
$\,\LL=2\,\text{ab}^{-1}\!$ {\it and an ideal detection efficiency $\,\ep=100\%$\,.}
}
\vspace*{1mm}
\label{tab:3B}
\label{tab:55}
\end{table}
%

\vspace*{2mm}
\section{\hspace{-5mm}.\,Improvements from $\bd{e^\mp}$ Beam Polarizations}
\label{sec:55}
\label{sec:5.1}

In this section, we extend our analysis to include the effects
of initial-state electron/positron polarizations, and demonstrate how
the sensitivity reaches of new physics scale $\cut$ can be improved.

\vspace*{1mm}

Using the helicity amplitude \eqref{eq:T-llgamma}, we derive the 
leading contribution to the differential cross section at $\mO(\Lambda^{\!-4})$,\,
which is proportional to the interference term,
\vspace*{-2mm}
{\small
\beqa
\label{eq:dSigmaZA}
&\hspace*{-5mm}&
\Re\texttt{e}\!\!
\left[\mathcal{T}^L_{(8)}(0\pm)\mathcal{T}^{T*}_{\sm}(\mp\pm)\right]
\!\sin\theta\sin\theta_*^{}
\\ 
&\hspace*{-5mm}&
\propto
\dis\frac{v^2\!\sqrt{s\,}}{\,\Lambda^4 M_Z^{}\,}\!
  \!\left[({e_L^2}\!+\!{e_R^2})(f_L^2\!-\!f_R^2)(1\!+\!\cos^2\theta)
  +2({e_L^2}\!-\!{e_R^2})(f_L^2\!+\!f_R^2)\cos\!\theta\cos\!\theta_*^{}
\right]\sin^2\!\theta_*^{}\cos\phi_*^{},~~~
\nn
\eeqa
}
\hspace*{-3mm}
where
$\,(e_L^{},\,e_R^{})=(c_L^{}\delta_{\!s,-\frac{1}{2}}^{},\,
   c_R^{}\delta_{\!s,\frac{1}{2}}^{})$\,
are the $Z$ gauge couplings to the (left, right)-handed electrons, 
with the index
$\,s =\mp\fr{1}{2}$\, denoting the initial-state electron helicities.
The final-state $Z$ boson decays into leptons $\ell^-\ell^+$ with couplings
$\,(f_L^{},\,f_R^{})= (c_L^{}\delta_{\!\si,-\frac{1}{2}}^{},\,
    c_R^{}\delta_{\!\si,\frac{1}{2}}^{})$,\,
where $\si$ denotes the helicity of the massless lepton $\ell^-$ and
\,$(c_L^{},\,c_R^{})=(s_W^2\!-\!\fr{1}{2},\, s_W^2)$\, give the
$Z$ gauge couplings to the (left,\,right)-handed leptons.
We note from the right-hand-side (RHS) of Eq.\eqref{eq:dSigmaZA}
that for unpolarized initial states $e^\mp$
the observable $\mathbb{O}^1_c$ is suppressed by the coupling factor
$\,f_L^2\!-\!f_R^2\propto \frac{1}{4}\!-\!\sin^2\!\theta_W^{}$\,
in the first term,
and the second term is suppressed by the coupling factor $\,e_L^2\!-\!e_R^2$
plus the factors $\cos\!\theta\cos\!\theta_*^{}$
which can be either positive or negative.
If the initial-state $e^\mp$ are polarized,
we can largely remove the suppressions in the second term of the RHS of
Eq.\eqref{eq:dSigmaZA}, since the coupling factor $\,e_L^2\!-\!e_R^2$
is replaced by $e_L^2$ (or $e_R^2$) in the fully-polarized case,
and the factors $\cos\!\theta\cos\!\theta_*^{}$ can be made positive
by defining $\mathbb{O}^1_c$ appropriately.\,

\vspace*{1mm}

In the ideal case of a fully left-polarized $e^-$ beam, 
we can redefine $\mathbb{O}^1_c$ as follows:
{\small
\beqs
\begin{eqnarray}
\mathbb{O}^1_c &=&
\left|\sigma_1^{}\!\int\!\! \di\theta \di\theta_*^{}\di\phi_*^{} \di M_*^{}\,
f^1
\text{sign}(\cos\theta)\,\text{sign}(\cos\theta_*^{})\,
\text{sign}(\cos\phi_*^{})
\right| ,
\hspace*{12mm}
\\
f_j^{} &=& \dis
\frac{\di^4\sigma_j^{}}
     {\,\sigma_j^{}\di\theta\, \di\theta_*^{} \di\phi_*^{}\di M_*^{}\,} \,,
\end{eqnarray}
\eeqs
}
\hspace*{-3.5mm}
where $\,M_*^{}\,$ is the invariant-mass of the final state leptons $\ell^-\ell^+$.\,
As can be readily seen,
in this case the first term on the RHS of Eq.\eqref{eq:dSigmaZA} gives
zero contribution to the observable $\,\mathbb{O}^1_c$\,.\,
Thus, $\,\mathbb{O}^1_c$\, is dominated
by the leading contributions of the second term on the RHS of Eq.\eqref{eq:dSigmaZA},
and is proportional to $\,c_L^2(c_L^2\!+\!c_R^2)$\, 
rather than $\,(c_L^2\!-\!c_R^2)(c_L^2\!+\!c_R^2)$\,.
Thus, at different collider energies,
we can derive the following signal significance $\SZZ_4^{}$ for
the final state $e^-e^+\ga$\,,\,
{\small
	\beqs
	\vspace*{-1mm}
	\label{eq:LZ4-ee}
	\begin{eqnarray}
	\sqrt{s}=250\,\text{GeV}, &~~~&
	\mathcal{Z}_{4}^e =\,
	4.46\!\(\!\frac{\,0.5\text{TeV}\,}{\Lambda}\!\)^{\!\!\!4}\!\times
	\!\sqrt{\epsilon\,}\,,
	\hspace*{12mm}
	\\
	\sqrt{s}=500\,\text{GeV}, &~~~&
	\mathcal{Z}_{4}^{e} =\,
	3.64\!\left(\!\frac{\,0.8\text{TeV}\,}{\Lambda}\!\)^{\!\!\!4}\! \times
	\!\sqrt{\epsilon\,}\,,
	\\
	\sqrt{s}=1\,\text{TeV}, &~~~&
	\mathcal{Z}_{4}^{e} =\,
	6.40\!\(\!\frac{\,\text{TeV}\,}{\Lambda}\!\)^{\!\!\!4}\!\times
	\!\sqrt{\epsilon\,}\,,
	\\
	\sqrt{s}=3\,\text{TeV}, &~~~&
	\mathcal{Z}_{4}^{e} \,=\,
	3.80\!\left(\!\frac{\,2\text{TeV}\,}{\Lambda}\!\right)^{\!\!\!4}\!\times
	\!\sqrt{\epsilon\,}\,,
	\\
	\sqrt{s}=5\,\text{TeV}, &~~~&
	\mathcal{Z}_{4}^{e} =\,
	4.32\!\(\!\frac{\,2.5\text{TeV}\,}{\Lambda}\!\)^{\!\!\!4}\! \times
	\!\sqrt{\epsilon\,}\,;
	\end{eqnarray}
	\vspace*{-3mm}
	\eeqs}
\hspace*{-3.5mm}
for the final state $\mu^-\mu^+\ga$\,,\,
{\small
\beqs
\vspace*{1mm}
\label{eq:LZ4-mumu}
\begin{eqnarray}
	\sqrt{s}=250\,\text{GeV}, &~~~&
	\mathcal{Z}_{4}^\mu =\,
	4.86\!\(\!\frac{\,0.5\text{TeV}\,}{\Lambda}\!\)^{\!\!\!4}\!\times
	\!\sqrt{\epsilon\,}\,,
	\hspace*{12mm}
	\\
	\sqrt{s}=500\,\text{GeV}, &~~~&
	\mathcal{Z}_{4}^{\mu} =\,
	3.78\!\left(\!\frac{\,0.8\text{TeV}\,}{\Lambda}\!\)^{\!\!\!4}\! \times
	\!\sqrt{\epsilon\,}\,,
	\\
	\sqrt{s}=1\,\text{TeV}, &~~~&
	\mathcal{Z}_{4}^{\mu} =\,
	6.47\!\(\!\frac{\,\text{TeV}\,}{\Lambda}\!\)^{\!\!\!4}\!\times
	\!\sqrt{\epsilon\,}\,,
	\\
	\sqrt{s}=3\,\text{TeV}, &~~~&
	\mathcal{Z}_{4}^{\mu} \,=\,
	3.80\!\left(\!\frac{\,2\text{TeV}\,}{\Lambda}\!\right)^{\!\!\!4}\!\times
	\!\sqrt{\epsilon\,}\,,
	\\
	\sqrt{s}=5\,\text{TeV}, &~~~&
	\mathcal{Z}_{4}^{\mu} =\,
	4.33\!\(\!\frac{\,2.5\text{TeV}\,}{\Lambda}\!\)^{\!\!\!4}\! \times
	\!\sqrt{\epsilon\,}\,;
\end{eqnarray}
\eeqs
}
and for the $\tau^-\tau^+\ga$ final state we have
$\,\SZZ_{4}^\tau\simeq \SZZ_{4}^\mu$.\,

\vspace*{1mm}

In reality, the $e^\mp$ beams could only be partially polarized.
Let $P_L^e$ ($P_R^{\bar e})$ denote the left (right) polarization of the
electrons (positrons) in the beam. We then have the following relations between the
observable $\mathbb{O}^1_c$ with partial and full polarizations:
\beqs
\begin{eqnarray}
\mathbb{O}^1_c(P_L^e,P_R^{\bar e})
&\simeq&
\left|
\frac{\,c_L^2P_L^eP_R^{\bar e}\!-\!c_R^2(\!1\!-\!P_L^e)(\!1\!-\!P_R^{\bar e})\,}
     {c_L^2}\right|
\mathbb{O}^1_c(1,1)\,,
\\
\sigma^0_c(P_L^e,P_R^{\bar e})
&\simeq&
\frac{\,c_L^2P_L^eP_R^{\bar e}\!+\!c_R^2(\!1\!-\!P_L^e)(\!1\!-\!P_R^{\bar e})\,}
{c_L^2}\sigma^0_c(1,1) \,,
\\
\SZZ_4^{}(P_L^e,P_R^{\bar e})
&\simeq&
\frac{\,\left|c_L^2P_L^eP_R^{\bar e}\!-\!c_R^2(\!1\!-\!P_L^e)(\!1\!-\!P_R^{\bar e})\right|\,}
{\,|c_L^{}|\sqrt{c_L^2P_L^eP_R^{\bar e}\!+\!c_R^2(\!1\!-\!P_L^e)(1\!-\!P_R^{\bar e})}\,}
\SZZ_4^{}(1,1)\,,
\label{eq:Z4-pol}
\end{eqnarray}
\eeqs
where the signal significance of the fully polarized case,
$\,\SZZ_4^{}(1,1)\!=\!
 \sqrt{(\SZZ_4^e)^2\!+\!(\SZZ_4^{\mu})^2\!+\!(\SZZ_4^{\tau})^2}$,\,
with $(\SZZ_4^e,\SZZ_4^{\mu},\SZZ_4^{\tau})$
given by Eqs.\eqref{eq:LZ4-ee}-\eqref{eq:LZ4-mumu}.
For instance, assuming the polarizations
$(P_L^e,\,P_R^{\bar e})=(0.9,\,0.65)$, we derive
\beqa
\label{eq:Z4-pol(0.9,0.65)}
\SZZ_4^{}(P_L^e,P_R^{\bar e}) \,\simeq\, 0.715\;\SZZ_4^{}(1,1) \,.
\eeqa
We note that the $e^\mp$ polarization possibilities have been well studied for the linear
colliders ILC\,\cite{ILC} and CLIC\,\cite{CLIC}, whereas the longitudinal polarization is
harder to realize at the circular colliders CEPC\,\cite{CEPC}
and FCC-$ee$\,\cite{FCCee}.

\vspace*{1mm}

We can derive the following relations between the
the cross sections $(\sigma_0^{},\,\sigma_1^{},\,\sigma_2^{})$ for the partially polarized
and unpolarized cases:
\beqs
\begin{eqnarray}
\sigma_0^{}(P_L^e,P_R^{\bar e})
&=&
\frac{\,c_L^2P_L^eP_R^{\bar e}\!+\!c_R^2(\!1\!-\!P_L^e)(\!1\!-\!P_R^{\bar e})}
{\,0.5^2(c_L^2\!+\!c_R^2)\,}\sigma_0^{}(0.5,0.5)\,,
\\[1mm]
\sigma_1^{}(P_L^e,P_R^{\bar e})
&=&
\frac{\,c_L^2P_L^eP_R^{\bar e}\!-\!c_R^2(\!1\!-\!P_L^e)(\!1\!-\!P_R^{\bar e})}
{\,0.5^2(c_L^2\!-\!c_R^2)\,}
\sigma_1^{}(0.5,0.5)\,,
\\[1mm]
\sigma_2^{}(P_L^e,P_R^{\bar e})
&=&
\frac{\,c_L^2P_L^eP_R^{\bar e}\!+\!c_R^2(\!1\!-\!P_L^e)(\!1\!-\!P_R^{\bar e})}
{\,0.5^2(c_L^2\!+\!c_R^2)\,}
\sigma_2^{}(0.5,0.5) \,,
\end{eqnarray}
\eeqs
where $\,(P_L^e,\, P_R^{\bar e})\!=\!(0.5,\,0.5)$\, corresponds to the unpolarized case.

\begin{table}[t]
	\begin{center}
	\begin{tabular}{c|ccccc}
		\hline\hline
		\\[-4.4mm]
		$\sqrt{s\,}$~(GeV) & 250 & 500 & 1000~~ & 3000~ & 5000 \\[0.4mm]
		\hline\hline
		\\[-4.4mm]
		$\Lambda^{2\sigma}_{\ell\bar\ell}$\,(TeV)& 0.81 & 1.1 &1.5(1.4)~ & 2.5~ & 3.2
		\\[-4.4mm]
		\\
		\hline
		\\[-4.4mm]
		$\Lambda^{5\sigma}_{\ell\bar\ell}$\,(TeV) & 0.64 & 0.87(0.85) & 1.2(1.1) & 2.1(2.0)~ & 2.7
		\\[0.48mm]
		\hline\hline
		\\[-4.4mm]
		$\Lambda^{2\sigma}_{\nu\bar\nu}$\,(TeV)& 0.85 & 1.0 & 1.3(0.86)~ & 2.1~ & 2.9(2.8)
		\\[-4.4mm]
		\\
		\hline
		\\[-4.4mm]
		$\Lambda^{5\sigma}_{\nu\bar\nu}$\,(TeV) & 0.67 & 0.84(0.80) & 1.1(0.82) & 1.9~ & 2.6(2.5)
		\\[0.48mm]
		\hline\hline
		\\[-4.4mm]
		$\Lambda^{2\sigma}_{\ell\nu,\text{Comb}}$\,(TeV)
        & 0.90 & 1.2(1.0)& 1.5 & 2.6(2.5)~ & 3.3
		\\[-4.4mm]
		\\
		\hline
		\\[-4.4mm]
		$\Lambda^{5\sigma}_{\ell\nu,\text{Comb}}$\,(TeV) & 0.72 & 0.93(0.91) & 1.2 & 2.1(2.0) & 2.8(2.7)
		\\[0.7mm]
		\hline\hline
	\end{tabular}
\end{center}
\vspace*{-3.7mm}
\caption{{\it %
Sensitivity reaches of the new physics scale $\cut$ via
$\,e^-e^+\!\!\to\!\ell^-{\ell}^+\ga$\, and
$\,e^-e^+\!\!\to\!\nu\bar{\nu}\ga$\, channels and their combinations,
for polarized
$e^\mp$ beams with $(P_L^e,\, P_R^{\bar e})=(0.9,\,0.65)$.\,
The bounds are shown at the $2\sigma$ and $5\sigma$ levels,
and for different collider energies.
As previously, the numbers in parentheses correspond to the case of the
dimension-8 operator with negative coefficient, while in all other
entries the effects of the sign of the coefficient are negligible.
For illustration, we assume a representative integrated luminosity}
$\,\LL=2\,\text{ab}^{-1}\!$ {\it and an ideal detection efficiency $\,\ep=100\%$.}
}
\vspace*{3mm}
\label{tab:pol}
\label{tab:66}
\end{table}

\vspace*{1mm}

For the signal significance $\SZZ_8^{}$,
we denote its $\mO(\Lambda^{-4})$ contribution as
$\SZZ_8^{(4)}$ and its $\mO(\Lambda^{-8})$ contribution as $\SZZ_8^{(8)}$,\,
with $\SZZ_8^{}=\SZZ_8^{(4)}\!\!+\!\SZZ_8^{(8)}$.\,
We obtain the following signal significances $\SZZ_8^{}$ with partial polarizations
$(P_L^e,P_R^{\bar e})$:
\beqs
\label{eq:Z8-PL-PR}
\begin{eqnarray}
\SZZ_8^{}(P_L^e,P_R^{\bar e}) &=&
\SZZ_8^{(4)}\!(P_L^e,P_R^{\bar e})+\SZZ_8^{(8)}\!(P_L^e,P_R^{\bar e}),
\\[2mm]
\label{eq:Z8(4)-PL-PR}
\SZZ_8^{(4)}\!(P_L^e,P_R^{\bar e}) &=&
\frac{\,2\!\sqrt{c_L^2\!+\!c_R^2}\,}{\,c_L^2\!-\!c_R^2\,}
\frac{\,c_L^2P_L^eP_R^{\bar e}\!-\!c_R^2(\!1\!-\!P_L^e)(\!1\!-\!P_R^{\bar e})\,}
{\,\sqrt{c_L^2P_L^eP_R^{\bar e}\!+\!c_R^2(\!1\!-\!P_L^e)(\!1\!-\!P_R^{\bar e})}\,}
\SZZ_8^{(4)}\!(0.5,0.5),
\hspace*{16mm}
\\[2mm]
\label{eq:Z8(8)-PL-PR}
\SZZ_8^{(8)}\!(P_L^e,P_R^{\bar e}) &=&
2\sqrt{\frac{c_L^2P_L^eP_R^{\bar e} \!+\! c_R^2(\!1\!-\!P_L^e)(\!1\!-\!P_R^{\bar e})}
{c_L^2\!+\!c_R^2}}\SZZ_8^{(8)}\!(0.5,0.5).
\end{eqnarray}
\eeqs
For instance, with $e^\mp$ beam polarizations
$(P_L^e,\, P_R^{\bar e})=(0.9,\,0.65)$,\, we find the following relation,
\beqa
\label{eq:Z8-pol(0.9,0.65)}
\SZZ_8^{}(P_L^e,P_R^{\bar e}) \,\simeq\,
7.18\SZZ_8^{(4)}\!(0.5,0.5)+1.19\SZZ_8^{(8)}\!(0.5,0.5) \,,
\eeqa
where the contributions $\SZZ_8^{(4)}\!(0.5,0.5)$ and $\SZZ_8^{(8)}\!(0.5,0.5)$
correspond to the unpolarized case, as computed
in Section\,\ref{sec:3}-\ref{sec:44}.

\vspace*{1mm}

From Eqs.\eqref{eq:Z4-pol(0.9,0.65)}\eqref{eq:LZ4-ee}\eqref{eq:LZ4-mumu}
and Eqs.\eqref{eq:Z8-pol(0.9,0.65)}, we compute the sensitivity reaches
of the new physics scale $\cut$ via
$\,e^-e^+\!\!\to\!\ell^-{\ell}^+\ga$\, and
$\,e^-e^+\!\!\to\!\nu\bar{\nu}\ga$\, channels,
for polarized electron/positron beams with
$(P_L^e,\, P_R^{\bar e})=(0.9,\,0.65)$.\,
These results are summarized in Table\,\ref{tab:66}.
Here the numbers in the parentheses correspond to the case of the
dimension-8 operator with negative coefficient, while in all other
entries the differences for opposite signs of the coefficient are negligible.
For illustration, we have assumed a representative integrated luminosity
$\,\LL=2\,\text{ab}^{-1}\!$ and an ideal detection efficiency $\,\ep=100\%$\,.

\vspace*{1mm}

We present in Table\,\ref{tab:66} the $2\sigma$ and $5\sigma$ bounds on $\cut$
for different collider energies. The limits from the
$\,\ell^-{\ell}^+\ga$\, channel are shown in the 2nd and 3rd rows,
while the 4th and 5th rows give the limits in the $\nu\bar{\nu}\ga$
channel. Finally, we derive the combined limits of
$\,\ell^-{\ell}^+\ga$ and $\nu\bar{\nu}\ga$ channels, as shown in
the 6th and 7th rows. Comparing the results in Table\,\ref{tab:66} with
those in the previous Table\,\ref{tab:55}, we see that for collider energies
$\sqrt{s}=(250\!-\!1000)$\,GeV, the $e^\mp$ beam polarizations can
enhance the sensitivity reaches of $\cut$ significantly,
by about $(47\!-\!25)\%$ [$(47\!-20\!)\%$] for the $2\sigma$ [$5\sigma$] limits;
whereas for $\sqrt{s}=(3\!-\!5)$\,TeV,
the polarization effects are much milder,
yielding an enhancement around $(18\!-\!10)\%$ [$(5\!-\!8)\%$]
for the $2\sigma$ [$5\sigma$] limits.
This feature is expected because the polarized $e^\mp$ beams help to remove
the large cancellation in the coupling factor
$\,c_L^2\!-c_R^2\!\propto\! \frac{1}{4}\!-\sin^2\!\theta_W^{}$
of the $O(\cut^{-4})$ interference cross section \eqref{Zgamma0}
[or the coupling factor $\,e_L^2\!-e_R^2=c_L^2\!-c_R^2\,$ in Eq.\eqref{eq:dSigmaZA}],
and thus will mainly enhance the sensitivity $\mathcal{Z}_4^{}$\,.
But, the polarized $e^\mp$ have rather minor effect on the $O(\cut^{-8})$
cross section because the coupling factor $\,c_L^2\!+c_R^2$\, in Eq.\eqref{Zgamma0}
has no special cancellation.
This is clear from Eqs.\eqref{eq:Z8-pol(0.9,0.65)} and \eqref{eq:Z8-PL-PR}
which show that adding the $e^\mp$ polarizations enhances the
$\mathcal{Z}_8^{(4)}$ contribution substantially over the
$\mathcal{Z}_8^{(8)}$ contribution.
Fig.\,\ref{fig:88}(a) shows that $\mathcal{Z}_8^{}$ dominates over $\mathcal{Z}_4^{}$
for the collider energy $\sqrt{s\,}\!\gtrsim\! (1\!-\!2)$\,TeV, while it is
the other way around for $\sqrt{s\,}\!\lesssim\!(1\!-\!2)$\,TeV.
This explains why the $e^\mp$ polarizations only improve slightly the
sensitivity reaches for large collider energies $\sqrt{s}=(3-5)$\,TeV,
whereas they play a significant role for $\sqrt{s\,}\lesssim 1$\,TeV.

\begin{figure}[t]
\centering
\hspace*{-5mm}
\includegraphics[width=7.7cm,height=6.8cm]{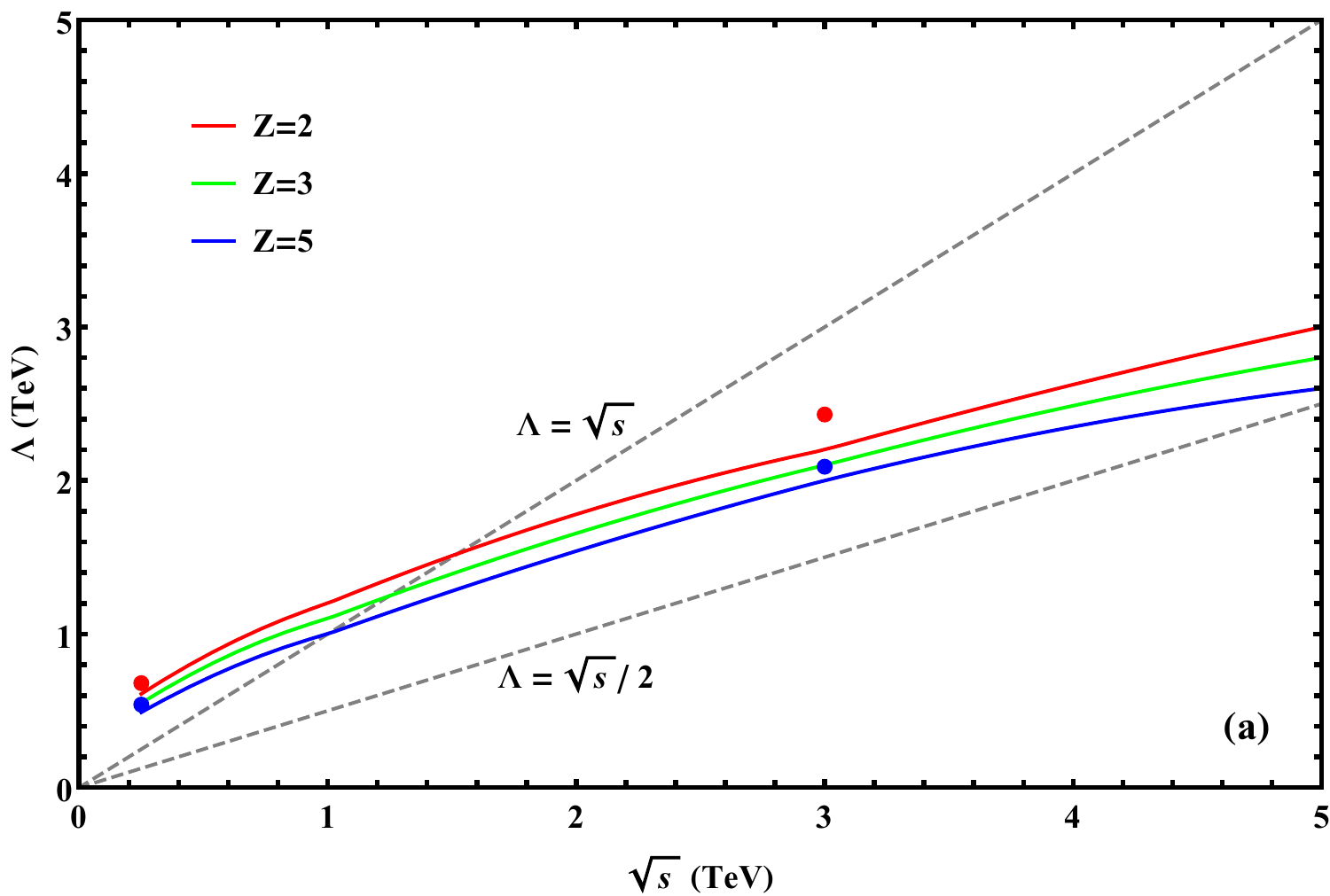}
\includegraphics[width=7.7cm,height=6.8cm]{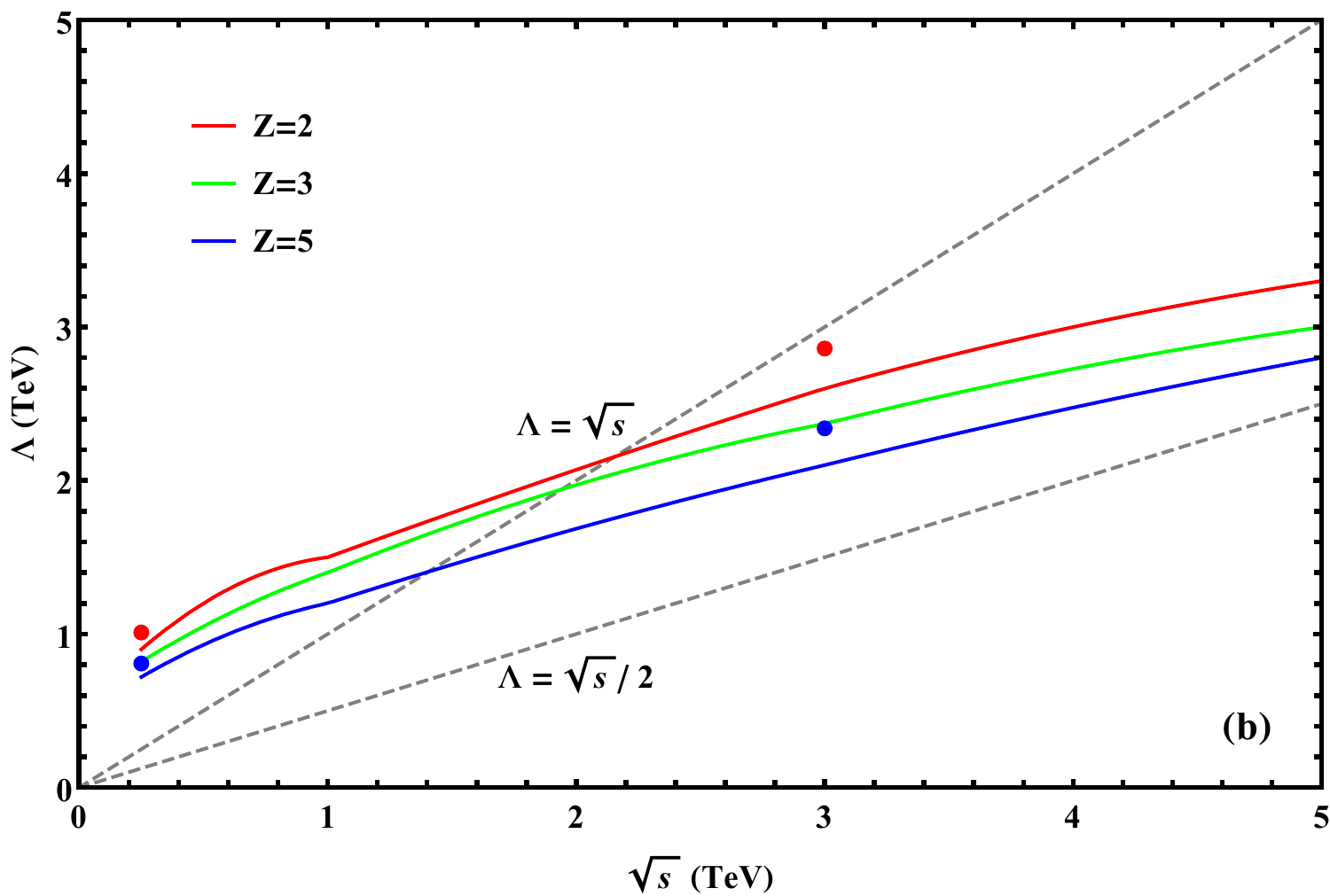}
\vspace*{-3mm}
\caption{\it Reaches for the new physics scale $\cut$ as functions
of the $e^+ e^-$ collision energy $\sqrt{s\,}$
with an integrated luminosity $\,\mL \!=\! 2\,\text{ab}^{-1}$.\,
Plot\,(a) depicts the combined sensitivity
$\,\SZZ \!=\!\sqrt{\SZZ_{\ell\bar\ell}^{2}\!+\!\SZZ_{\nu\bar \nu}^{2}\,}\!
  =(2,\,3,\,5)\sigma$\,
for unpolarized $e^\mp$ beams,
as shown by the (red, green, blue) curves.
Plot\,(b) depicts the combined sensitivity
$\,\SZZ \!=\!\!\sqrt{\SZZ_{\ell\bar\ell}^{2}\!+\!\SZZ_{\nu\bar\nu}^{2}\,}
        \!=\!(2,\,3,\,5)\sigma$,\, with $e^\mp$ beam polarizations
$(P_L^e,\,P_R^{\bar{e}})=(90\%,\,65\%)$,\,
as shown by the (red, green, blue) curves.
In each plot, the (red, blue) dots show the $(2\si ,\,5\si )$ reaches
with a projected integrated luminosity $\,\mL =5\,\text{ab}^{-1}$
at $\sqrt{s\,}=\rm{250GeV}$ (CEPC and FCC-ee) and $\sqrt{s\,}=\rm{3TeV}$ (CLIC).
For reference, we also show the lines
$\,\cut\! =\! \sqrt{s\,}$\, and $\,\cut\! = \!\sqrt{s\,}/2$\, in each plot.}
\label{fig:10}
\label{fig:11}
\label{fig:12}
\end{figure}

\vspace*{1mm}

We present in Fig.\,\ref{fig:10} the sensitivity reaches for the new physics scale
$\cut$ as functions of the collision energy $\sqrt{s\,}$,\,
with a universal integrated luminosity $\,\mathcal{L} =2\,\text{ab}^{-1}$.\,
It compares our results for the
unpolarized and polarized cases in plots (a) and (b), respectively,
assuming $(P_L^e,\,P_R^{\bar{e}})=(90\%,\,65\%)$
for the polarized electron and positron beams in the plot\,(b).
In each plot, we show the limits  $\SZZ=(2,\,3,\,5)\sigma$
by the (red,\,green,\,blue) curves, where the signal significance
$\,\SZZ \!=\!\!\sqrt{\SZZ_{\ell\bar\ell}^{2}\!+\!\SZZ_{\nu\bar\nu}^{2}\,}$\,
combines both $\,\ell^-{\ell}^+\ga$\, and $\,\nu\bar{\nu}\ga$\, channels.
For comparison, we further present the sensitivity reaches with
a projected integrated luminosity $\,\mathcal{L} =5\,\text{ab}^{-1}$
at $\sqrt{s\,}=\rm{250\,GeV}$ (CEPC and FCC-ee) and $\sqrt{s\,}=\rm{3\,TeV}$ (CLIC),
shown as (red, blue) dots at $(2\sigma ,\,5\sigma )$ level.
We see that electron/positron beam polarizations can
improve significantly the sensitivity reaches for the new physics scale.
For reference, we also draw the dashed lines of
$\,\cut\! =\! \sqrt{s\,}$\, and $\,\cut\! = \!\sqrt{s\,}/2$\, in each plot.
%

\section{\hspace{-5.5mm}.~Conclusions}
\label{sec:4}	
\label{sec:66}

As we have discussed in this work, the reaction
$\,e^+ e^- \!\!\to\! Z \gamma$\, provides a rare
opportunity to probe an effective dimension-8 operator in the SMEFT.
The $Z V \gamma$ vertices ($V\!=\! Z,\gamma$)
have no tree-level SM contributions, and nor do they receive any
contributions from dimension-6 operators,
opening up the possibility of probing the new physics scale associated with
the dimension-8 operator.
Such dimension-8 operators invoke the Higgs doublets and are connected to the
Higgs boson and the spontaneous electroweak symmetry breaking.
We have presented a general analysis of the angular distributions for $Z \gamma$
production in the lab frame and for $\,Z\!\!\to\!\ell^+ \ell^-$\,
in the $Z$ rest frame to identify distinctive angular distributions
and cuts that maximize the statistical sensitivity to the possible
new physics scale $\Lambda$, either including only the
${\cal O}(\Lambda^{\!-4})$ contributions that interfere with the SM contributions,
or including together the ${\cal O}(\Lambda^{\!-8})$ term
of the dimension-8 operator for its full contribution.

\vspace*{1mm}

As seen in Fig.\,\ref{Z4L}(b) and Tables\,\ref{tab:1}-\ref{tab:3},
the prospective sensitivities to
$\,\Lambda\,$ extend into the multi-TeV range.
As expected from the energy dependences of the dimension-8 contributions to the
cross section for $\,e^+ e^-\!\!\to\! Z \gamma$,\, Eqs.\eqref{lam4} and \eqref{lam8}
show that the prospective sensitivities increase with the collision energy
$\!\sqrt{s\,}$, but much slower with the integrated luminosity $\mathcal{L}$\,.
So taking a constant integrated luminosity in our current tables and figures
is a fairly sensible assumption for the simplicity of our presentation.
Rescaling the sensitivity reaches to a different $\mathcal{L}$ is straightforward
as shown by the (blue,\,red) dots in Fig.\,\ref{fig:88}(b) and Fig.\,\ref{fig:12}.
The sensitivities at the $(2\sigma,\,3\sigma,\,5\sigma)$ levels of significances
are not greatly different, as discussed in the text and seen by comparing the
(red,\,green,\,blue) curves in Fig.\,\ref{Z4L}(b) and Fig.\,\ref{fig:12}.

\vspace*{1mm}

We have also drawn in Fig.\,\ref{Z4L} the two reference lines
$\,\Lambda \!=\!\sqrt{s\,}$\, and $\,\cut \!=\!\sqrt{s\,}/2$\,.\,
In general, one would expect the SMEFT approach to be suitable only for energy scales
that are small compared to $\cut\,$.\,
However, the way that we have defined $\cut$ in Eq.(\ref{cj}) of this paper corresponds
to the true new physics scale $\,\tilde{\Lambda}\,$ only if the unknown coefficient $c_j^{}$
has a magnitude of unity. If, on the other hand, the true magnitude of
$\,c_j^{} \!\gg\! 1$ \cite{GNDA}, the true new physics scale $\,\tilde{\Lambda}$\,
could be sizably larger than the value of $\,\cut\,$ extracted from our analysis,
and the SMEFT approach would have broader applicability.

\vspace*{1mm}

We have studied the effect of including the reaction
$e^-e^+\!\!\to\! Z\ga$ with invisible decay channel
$Z\!\to\!\nu\bar{\nu}$ in Section\,\ref{sec:44}.
We have presented the sensitivities of this channel in Table\,\ref{tab:44} and
Table\,\ref{tab:55}, and derived the combined new physics reaches for
both the leptonic and invisible channels $\ell^-\ell^+\ga$
and $\nu\bar{\nu}\ga$\,.\,
We found that the sensitivity of the invisible channel is
comparable to that of the lepton channel (cf.\ Table\,\ref{tab:44}).
Then, we demonstrated in Section\,\ref{sec:55} that
including the electron/prostrion beam polarizations
can further improve significantly the signal sensitivities.
We have presented our findings for the
polarized case in Fig.\ref{fig:10}(b), to be compared with the unpolarized case
in Fig.\ref{fig:10}(a).
We have summarized the $2\si$ and $5\si$ bounds on the new physics scale $\,\cut\,$
in Table\,\ref{tab:66},
including the combined reaches of both leptonic and invisible channels.

\vspace*{1mm}

In summary, we have presented here the sensitivity reaches for
the new physics scale $\,\cut$\, in nTGCs using the reaction $e^-e^+\!\!\to\! Z\ga$
with both leptonic and invisible $Z$ decays, and including their combinations
in Table\,\ref{tab:55} and Fig.\ref{fig:12}(a).
In addition, we have presented in Table\,\ref{tab:66} and Fig.\ref{fig:12}(b)
the improved sensitivities obtainable with polarized $e^-$ and $e^+$ beams.
Comparing Tables\,\ref{tab:66} and \ref{tab:55},
we see that for collider energies
$\sqrt{s}=(250\!-\!1000)$\,GeV, using polarized $e^\mp$ beams can
enhance the sensitivity reaches of $\cut$ significantly,
by about $(47\!-\!25)\%$ [$(47\!-20\!)\%$] for the $2\sigma$ [$5\sigma$] limits;
while for $\sqrt{s}=(3\!-\!5)$\,TeV,
the polarization effects are much smaller,
giving an enhancement around $(18\!-\!10)\%$ [$(5\!-\!8)\%$]
for the $2\sigma$ [$5\sigma$] limits.

\vspace*{1mm}

Finally, it is interesting to compare the sensitivity to the dimension-8 coefficient
found here with that found previously in studies of the dimension-8 operator
contributions to light-by-light scattering and the process $\,gg\!\to\!\gamma\gamma$\,
at the LHC. The former is sensitive to a dimension-8 scale that is
${\cal O}(100)$\,GeV\,\cite{gaga-gaga}, whereas the latter is sensitive to a dimension-8
scale that is ${\cal O}(1)$\,TeV\,\cite{gg-gaga}. The dimension-8 operators studied
in those analyses contain gauge fields only and differ from what we studied here,
hence they probe very different aspects of dimension-8 physics.
However, it is encouraging that we have found in this work that
future $e^+e^-$ colliders (such as the CEPC, FCC-ee, ILC, and CLIC) may be able
to provide very competitive sensitive probes of the scale of new physics.
We therefore encourage further detailed studies of the reaction
$\,e^+ e^-\!\!\!\to\!\! Z \gamma$\,
by our experimental colleagues.

\vspace*{8mm}
\appendix

\noindent	
{\bf\large Appendix}
\vspace*{-3mm}

\section{\hspace{-5mm}.\,Helicity Amplitudes for \boldmath{$Z\gamma$} Production
         with $\mathbf{Z}$ Decays}
\label{sec:A}

In this Appendix we present the helicity amplitudes for the production process
$e^-e^+\!\to Z\ga$, and then we include leptonic $Z$ decays. These results
are used in the analyses of Sections\,\ref{sec:2} and \ref{sec:3} of the main text.

\vspace*{1mm}
\subsection{\hspace{-5mm}.~Helicity Amplitudes for \boldmath{$Z\gamma$} Production}
\label{sec:A1}
\vspace*{1.5mm}

The helicity amplitudes for
$\,e^-(p_1^{}) e^+(p_2^{}) \to Z(q_1^{},\epsilon_\lambda^{})
  \gamma(q_2^{},\epsilon_{\lambda'}^{})$\,
can be written as
{\small
\begin{eqnarray}
\mathcal{T}^{ss'}_{\lambda\lambda'} \!&=&
\bar v^{s'}\!(p_2^{})\!\left[\!\frac{e^2}{\,s_W^{}c_W^{}\,}\!\!\(\!\!
\frac{\,\slashed \ep_{\lambda'}^*(q_2^{})(\slashed q_1^{}\!\!-\!\slashed p_1^{})\slashed
       \ep_{\lambda}^*(q_1^{})\,}{t}
+\frac{\,\slashed \ep_{\lambda}^*(q_1^{})(\slashed q_2^{}\!\!-\!\slashed p_1^{})
         \slashed \ep_{\lambda'}^*(q_2^{})\,}{u} \!\)
\right.
\nonumber
\\[1mm]
&& \left.\hspace*{14mm}
- \frac{\,\ii 2M_Z^2\,}{\,\cut^4\,}
\ep^{\mu\nu\al\be}\gamma_\mu^{}
\ep_{\lambda,\nu}^*(q_1^{})\ep_{\lambda'\!,\al}^*(q_2^{})q_{2\be}^{}\!\right]\!
\!\(c_L^{}P_L^{} \!\!+\! c_R^{}P_R^{}\)\!u^s(p_1^{})\,,
\hspace*{13mm}
\label{eq:amp}
\end{eqnarray}
}
%

\noindent
where we have used the standard spinor notations $u^s(p_1^{})$ and $\bar{v}^{s'}\!(p_2^{})$
\cite{PeskinBook} for the initial-state $\,e^-$ and $e^+$,\,
and $(\ep_\lambda^{},\,\ep_{\lambda'}^{})$ denote the polarization vectors
of the final-state gauge bosons $(Z,\,\ga)$.\,
In the above, $\,P_{L,R}^{}=\fr{1}{2}(1\mp\ga_5^{})\,$ are
the chirality projection operators, and the coefficients
\,$(c_L^{},\,c_R^{})=(s_W^2\!-\!\fr{1}{2},\, s_W^2)$\,
arise from the (left-,\,right)-handed gauge couplings of electrons to the $Z$ boson.
In the above, we have used the Mandelstam variables
$\,t=(p_1^{}\!-q_1^{})^2=-\fr{1}{2}(s\!-\!M_Z^2)(1\!-\cos\!\theta)\,$ and
$\,u=(p_1^{}\!-q_2^{})^2=-\fr{1}{2}(s\!-\!M_Z^2)(1\!+\cos\!\theta)\,$.\,

\vspace*{1mm}

In Eq.\eqref{eq:p1p2q1q2} we defined the momenta of the final-state particles
$Z(q_1^{})\ga (q_2^{})$ as
$\,q_1^{} = (E_Z^{},\, q\sin\!\theta,\, 0,\, q\cos\!\theta)$\, and
$\,q_2^{} = q(1,\, -\sin\!\theta,\, 0,\, -\cos\!\theta)$.\,
Then, we can express the three polarization vectors $\ep_\lambda^{}(q_1^{},\theta)$
of the $Z$ boson as follows:
\beqs
\begin{eqnarray}
\ep_\pm^{Z}(\theta) &=&
\fr{1}{\sqrt{2\,}\,}\!\(0,\mp{\cos\!\theta},\,-{\ii},\,\pm{\sin\!\theta}\)\!,
\\[1mm]
\ep_0^{Z}(q_1^{},\theta) &=&
\fr{1}{M_Z^{}}\!
\(q_1^{},\,E_Z^{}{\sin\!\theta},\,0,\,E_Z^{}{\cos\!\theta}\)\!,
\end{eqnarray}
\eeqs
where $E_Z^{}=\sqrt{q_1^2\!+\!M_Z^2\,}$.\,
The final-state photon has two transverse polarization vectors
that are similar to those of the $Z$ boson,
\beqa
\label{eq:pol-A-Z}
\ep_\pm^{\ga}(\theta)=\ep_\pm^Z(\theta\!+\!\pi )=\ep_\mp^Z(\theta )\,.\,
\eeqa
The first two terms in Eq.\eqref{eq:amp} arise from the SM contributions via
the $t$- and $u$-channel exchanges, respectively, while the third term is contributed
by the dimension-8 operator.
For the final-state $Z(\lam)\ga(\lam')$ helicity combinations
$\lam\lam'\!=\!(--,-+,+-,++)$ and $\lam\lam'\!=\!(0-,0+)$,\,
we compute the SM contributions to the scattering amplitudes as follows:
{\small
\beqs
\label{eq:Tsm-T+L}
\begin{eqnarray}
\hspace*{-2mm}
\mathcal{T}_{\text{sm}}^{ss'\!,\text{T}}\!\!\left\lgroup\!\!
\begin{array}{cc}
-- & -+ \\
+- & ++\\
\end{array}\!\!\right\rgroup
\!\!&=& \frac{2e^2}{\,s_W^{}c_W^{}(s\!-\!M_Z^2)\,}\!\!
\left\lgroup\!\!
\begin{array}{ll}
\(e_L^{}\!\cot\!\frac{\theta}{2}\!-\!e_R^{}\!\tan\!\frac{\theta}{2}\)\!M_Z^2~
&
\(-e_L^{}\!\cot\!\frac{\theta}{2}\!+\!e_R^{}\!\tan\!\frac{\theta}{2}\)\!s
\\[2mm]
\(e_L^{}\!\tan\!\frac{\theta}{2}\!-\!e_R^{}\!\cot\!\frac{\theta}{2}\)\!s
&
\(-e_L^{}\!\tan\!\frac{\theta}{2}\!+\!e_R^{}\!\cot\!\frac{\theta}{2}\)\!M_Z^2
\end{array}
\!\!\right\rgroup\!\!,\hspace*{13mm}
\label{msmT}
\label{eq:Tsm-T}
\\[2mm]
\hspace*{-2mm}
\mathcal{T}_{\text{sm}}^{ss'\!,\text{L}}(0-,0+)
&=& \frac{\,2\sqrt{2}(e_L^{}\!\!+\!e_R^{})e^2M_Z^{}\sqrt{s\,}\,}
      {\,s_W^{}c_W^{}(s\!-\!M_Z^2)\,}\(1,\,-1\),
\label{msm}
\label{eq:Tsm-L}
\end{eqnarray}
\eeqs
}
\hspace*{-3mm}
where $\,(e_L^{},\,e_R^{}) = (c_L^{}\delta_{\!s,-\frac{1}{2}}^{},\,
   c_R^{}\delta_{\!s,\frac{1}{2}}^{})$,\, with the subscript index
$\,s =\mp\fr{1}{2}$\, denoting the initial-state electron helicities.
For the massless initial-state $e^-$ and $e^+$, we have $s=-s'$.\,
We note that, in Eq.\eqref{eq:Tsm-T}, the identical-helicity amplitudes
$\,\mathcal{T}_{\text{sm}}^{ss'\!,\text{T}}(\pm\pm)$\, in the diagonal entries
are proportional to $M_Z^2$ (unlike the off-diagonal entries, which are
$\propto s$\,). This is expected because
the identical-helicity amplitudes should vanish exactly in the massless limit $M_Z^{}\!\to\!0$,\,
after ignoring the tiny electron mass, as in the pair-annihilation process
$\,e^-e^+\!\to\!\ga\ga$\, in QED\,\cite{PeskinBook}. Hence the non-zero
amplitudes have asymptotic behaviors\,
$\mathcal{T}_{\text{sm}}^{ss'\!,\text{T}}(\pm\pm)\propto M_Z^2/s$\,.\,

\vspace*{1mm}

Next, we compute the corresponding helicity amplitudes from the new physics
contributions of the dimension-8 operator, which are as follows:
{\small
\beqs
\label{eq:T8}
\begin{eqnarray}
\label{eq:T8-T}
\mathcal{T}_{(8)}^{ss'\!,\text{T}}
\!\!\left\lgroup\!\!
\begin{array}{cc}
-- & -+ \\
+- & ++\\
\end{array}\!\!\right\rgroup
\!&=&
\frac{\,(e_L^{}\!\!+\!e_R^{})\sin\!\theta M_Z^2(s\!-\!M_Z^2)\,}{\cut^4}\!\!
\left\lgroup\!
\begin{array}{cr}
1 ~& 0
\\[1mm]
0 ~& -1
\end{array}
\!\right\rgroup \!\!,
\label{m8T}
\\[2mm]
\mathcal{T}_{(8)}^{ss'\!,\text{L}} (0-,0+)
&=& \frac{\,\sqrt{2}M_Z^{}(s\!-\!M_Z^2)\sqrt{s}\,}{\cut^4}\!
\(\!e_L^{}\!\sin^2\!\frac{\theta}{2}\!-e_R^{}\!\cos^2\!\frac{\theta}{2},~
e_R^{}\!\sin^2\!\frac{\theta}{2}\!-e_L^{}\!\cos^2\!\frac{\theta}{2}
\)\!.
\hspace*{15mm}
\label{eq:T8-L}
\label{m8}
\end{eqnarray}
\eeqs
}
We note that in Eq.\eqref{eq:T8-T} the off-diagonal amplitudes
$\mathcal{T}_{(8)}^{ss'\!,\text{T}}(-+)$ and $\mathcal{T}_{(8)}^{ss'\!,\text{T}}(-+)$
vanish exactly.
This can be understood by noting
that the $Z\ga Z^*$ vertex [cf.\ Eqs.\eqref{eq:ZAZ*-vertex} and \eqref{eq:amp}]
contains the rank-4 antisymmetric tensor $\epsilon^{\mu\nu\al\be}$,
which contracts with the $Z$ and $\gamma$ polarization vectors
$\ep_{\lam ,\nu}^Z(\theta)$ and $\ep_{\lam'\! ,\al}^\ga (\theta)$\,
as in Eq.\eqref{eq:amp}.
For the off-diagonal $Z\ga$ helicity combinations
$\,\lam\lam '=+-,-+$,\, we deduce
\beqa
\label{eq:T8T-off=0}
\ep^{\mu\nu\al\be}\gamma_\mu^{}
\ep_{\pm ,\nu}^{Z*}(\theta)\ep_{\mp\! ,\al}^{\ga *} (\theta)
\,=\,
\ep^{\mu\nu\al\be}\gamma_\mu^{}
\ep_{\pm ,\nu}^{Z*}(\theta)\ep_{\pm\! ,\al}^{Z*} (\theta)
\,=\,0 \,,
\eeqa
according to Eq.\eqref{eq:pol-A-Z}.

\vspace*{1mm}
\subsection{\hspace{-5mm}.~Helicity Amplitudes Including $\mathbf{Z}$ Decays}
\label{sec:A2}
\vspace*{1mm}

In this Appendix we incorporate the leptonic decays of the $Z$ boson.
We first consider $Z$ decay in its rest frame,
$Z\!\to\!\ell^-(k_1^{})\ell^+(k_2^{})$,
where the final-state leptons have momenta:
\\[-8mm]
\begin{eqnarray}
k_1^{} &=&
k(1,\,\sin\!\theta_*^{}\cos\phi_*^{},\,
\sin\!\theta_*^{}\sin\!\phi_*^{},\,\cos\!\theta_*^{})\,,
\nn\\[-3mm]
\label{eq:k1k2}\\[-3mm]
k_2^{} &=&
k(1,-\sin\theta_*^{}\cos\phi_*^{},-\sin\theta_*^{}\sin\phi_*^{},-\cos\theta_*^{})\,,
\nn
\end{eqnarray}
where the leptons are treated as effectively massless and $\,k=|\vec{k}_1^{}\!|\simeq\fr{1}{2}M_Z^{}$\,.
In the $Z$ rest frame, the massless lepton spinors are defined as follows,
\begin{eqnarray}
u_+^{}(k_1^{}) &=& \sqrt{2k}\!
\(0,\, 0,\, e^{-\frac{\ii\phi_*^{}}{2}}\!\cos\!\frac{\theta_*^{}}{2},\,
  e^{\frac{\ii\phi_*^{}}{2}}\! \sin\!\frac{\theta_*^{}}{2}\)\!,
\nn\\
u_-^{}(k_1^{}) &=&
\sqrt{2k}\!\(-e^{-\frac{\ii\phi_*^{}}{2}}\!\sin\!\frac{\theta_*^{}}{2},\,
e^{\frac{\ii\phi_*^{}}{2}}\! \cos\!\frac{\theta_*^{}}{2},\, 0,\, 0\)\!,
\nn\\[-2mm]
\\[-3mm]
v_+^{}(k_2^{}) &=&
\sqrt{2k}\!\(e^{-\frac{\ii\phi_*^{}}{2}}\! \cos\!\frac{\theta_*^{}}{2},\,
e^{\frac{\ii\phi_*^{}}{2}}\!\sin\!\frac{\theta_*^{}}{2},\, 0,\, 0\)\!,
\nonumber\\
v_-^{}(k_2^{}) &=&
\sqrt{2k}\!\(0,\, 0,\, e^{-\frac{\ii\phi_*^{}}{2}}\!\sin\!\frac{\theta_*^{}}{2},\,
-e^{\frac{\ii\phi_*^{}}{2}}\!\cos\!\frac{\theta_*^{}}{2}\!\),
\nn
\end{eqnarray}
where $u_+^{}$ ($u_-^{}$) correspond to spin-up (-down) and
$v_+^{}$ ($v_-^{}$) correspond to spin-down (-up) along their directions of motion
in Eq.\eqref{eq:k1k2}.

\vspace*{1mm}

Then, we write down the left-handed and right-handed spinor currents
in the $Z$ boson rest frame,
\beqs
\begin{eqnarray}
\widetilde{C}_L^{\mu} &=&
\bar{v}_L^{} \gamma^\mu u_L^{}
= M_Z^{}\!\(0,\, -\cos\!\theta_*^{}\!\cos\!\phi_*^{}\!\!-\! \ii\sin\!\phi_*^{},\,
-\cos\!\theta_*^{}\!\sin\!\phi_*^{}\!\!+\!\ii\cos\!\phi_*^{},\, \sin\!\theta_*^{}\)\!,
\hspace*{16mm}
\\[1mm]
\widetilde{C}_R^{\mu} &=&
\bar v_R^{} \gamma^\mu u_R^{}
= M_Z^{}\!\(0,\,-\cos\!\theta_*^{}\!\cos\!\phi_*^{}\!\!+ \ii \sin\!\phi_*^{},\,
-\cos\!\theta_*^{}\!\sin\!\phi_*^{}\!\!- \ii \cos\!\phi_*^{},\, \sin\!\theta_*^{} \)\!,
\end{eqnarray}
\eeqs
where $\,(u_L^{},\,u_R^{})=(u_-^{},\,u_+^{})\,$ and
$\,(v_L^{},\,v_R^{})=(v_+^{},\,v_-^{})\,$.\,
After making a Lorentz boost $\hat{L}$ back to the laboratory frame
(i.e., the c.m.\ frame of the $Z\gamma$ pair)
and rotating the axis $z^*$ back to the axis $z$ by the rotation $\hat{R}$,\,
we have new currents $\,C_{L,R}^\mu =\hat{R}\hat{L}\widetilde{C}_{L,R}^\mu$\,
in the lab frame.
The Lorentz boost $\hat{L}$ acts on the (0,\,3) components, with
$\,\hat{L}_{00}^{}\!=\! \hat{L}_{33}^{}\!=\ga$\, and
$\,\hat{L}_{03}^{}\!=\! \hat{L}_{30}^{}\!=\ga\be$\,,\,
where  $(\beta ,\,\ga)=(p_Z^{}/E_Z^{},\,E_Z^{}/M_Z^{})$.\,
The rotation matrix $\hat{R}$ acts on the (1,\,3) components, with elements
$\,\hat{R}_{11}^{}\!=\!\hat{R}_{33}^{}\!=\cos\!\theta$\, and
$\,\hat{R}_{13}^{}\!=\!-\hat{R}_{31}^{}\!=\sin\!\theta$.\,
Thus, we can derive the currents $\,C_{L,R}^\mu\,$ and express them in terms of
$Z$ boson polarization vectors,
{\small
\beqs
\vspace*{-3mm}
\label{eq:CL-CR}
\beqa
\label{eq:CL}
C_L^{\mu} &=&
M_Z^{}\!\(\!\sin\!\theta_*^{}\ep_{0}^Z
-\sqrt{2}\sin^2\!\frac{\,\theta_*^{}}{2}e^{-\ii\phi_*^{}}\ep_+^Z
-\sqrt{2}\cos^2\!\frac{\,\theta_*^{}}{2}e^{\ii\phi_*^{}}\ep_{-}^Z\)\!,
\hspace*{16mm}
\\[0mm]
\label{eq:CR}
C_R^{\mu} &=&
M_Z^{}\!\(\!\sin\!\theta_*^{}\ep_{0}^Z
+\sqrt{2}\cos^2\!\frac{\,\theta_*^{}}{2}e^{-\ii\phi_*^{}}\ep_{+}^Z
+\sqrt{2}\sin^2\!\frac{\,\theta_*^{}}{2}e^{\ii\phi_*^{}}\ep_{-}^Z\)\!.
\eeqa
\eeqs
}
Next, we can obtain the amplitude for $\,e^-e^+\!\!\to\!\ell^-\ell^+\ga$\,
by replacing the $Z$ polarization vector $\ep_{\lam}^{*\mu}(q_1^{})$ in
Eq.\eqref{eq:amp} with $\,C_{L,R}^{\mu}(q_1^{})\mathcal{D}_Z^{}$,
where $\mathcal{D}_Z^{}=1/(q_1^2\!-\!M_Z^2\!+\!\ii M_Z^{}\Gamma)$\,
is from the $Z$ propagator.
Since $\,q_{1\mu}^{}C_{L,R}^\mu(q_1^{})\!=0$, we can drop the
$\,q_1^\mu q_1^\nu\,$ term in the $Z$-propagator.
Then, we derive the $\ell^-\ell^+\ga$ amplitude as follows,
{\small
\begin{eqnarray}
\mathcal{T}^{ss'}_{\si\si'\!\lambda}(\ell\bar\ell\ga)
\!&=&
\frac{\,ef_{L,R}^{}\mathcal{D}_{\!Z}^{}\,}{s_W^{}c_W^{}}
\bar v^{s'}\!(p_2^{})\!\!\left[\!\frac{e^2}{\,s_W^{}c_W^{}\,}\!\!\(\!\!
\frac{\,\slashed \ep_{\lambda}^*(q_2^{})(\slashed q_1^{}\!\!-\!\slashed p_1^{})
      \slashed C_{L,R}^*(q_1^{})\,}{t}
+\frac{\,\slashed C_{L,R}^*(q_1^{})(\slashed q_2^{}\!\!-\!\slashed p_1^{})
         \slashed \ep_{\lambda}^*(q_2^{})\,}{u} \!\)
\right.
\hspace*{5mm}
\nonumber
\\[1mm]
&& \left.\hspace*{14mm}
- \frac{\,\ii 2M_Z^2\,}{\,\cut^4\,}
\ep^{\mu\nu\al\be}\gamma_\mu^{}
C_{L,R;\nu}^*(q_1^{})\ep_{\lambda\!,\al}^*(q_2^{})q_{2\be}^{}\!\right]\!
\!\(c_L^{}P_L^{} \!\!+\! c_R^{}P_R^{}\)\!u^s(p_1^{})\,,
\hspace*{10mm}
\label{eq:amp-LLA}
\end{eqnarray}
}
where $\,(\si ,\, \si'\!,\,\lambda)$\,
denote the helicities of the final-state particles $(\ell^-,\,\ell^+,\,\ga)$
with $\,\si = -\si'\,$ for massless leptons, and
we have defined the coefficients
$\,(f_L^{},\,f_R^{})= (c_L^{}\delta_{\!\si,-\frac{1}{2}}^{},\,
   c_R^{}\delta_{\!\si,\frac{1}{2}}^{})$.\,

\vspace*{1mm}

Substituting Eq.\eqref{eq:CL-CR} into Eq.\eqref{eq:amp-LLA},
we can express the amplitude \eqref{eq:amp}
in terms of the helicity amplitudes \eqref{eq:Tsm-T+L}-\eqref{eq:T8}
of Appendix\,\ref{sec:A1},
{\small
\begin{eqnarray}
\label{eq:T-llgamma}
\mathcal{T}^{ss'}_{\si\si'\!\lambda}\!(\ell\bar\ell\ga) &=&
\frac{\,eM_Z^{}\mathcal{D}_Z^{}\,}{s_W^{}c_W^{}}
\left[\sqrt{2\,}e^{\ii\phi_*^{}}\!\!
\(\!f_R^{\si}\cos^2\!\frac{\,\theta_*^{}}{2}\!-f_L^{\si}\!\sin^2\!\frac{\,\theta_*^{}}{2}\!\)\!
\mathcal{T}_{ss'}^T(+\lam )
\right.
\\
&&
\left.
+\sqrt{2\,}e^{-i\phi_*^{}}\!\!
\(\!f_R^{\si}\sin^2\!\frac{\,\theta_*^{}}{2}\!
-f_L^{\si}\!\cos^2\!\frac{\,\theta_*^{}}{2}\!\)\!
\mathcal{T}_{ss'}^T(-\lam )
+(f_R^{\si}\!+\!f_L^{\si})\sin\!\theta_* \mathcal{T}_{ss'}^L(0\lam )\right]\!,
\hspace*{13mm}
\nn
\end{eqnarray}
}
where $\mathcal{T}_{ss'}^T(\pm\lam )$ and $\mathcal{T}_{ss'}^L(0\lam )$
are the on-shell helicity amplitudes of $\,e^-e^+\!\!\to Z\ga$\,,
\beqa
\label{eq:T-llgamma-sum}
\mathcal{T}_{ss'}^T(\pm\lam ) &=&
\mathcal{T}^{ss'\!,T}_{\text{sm}}(\pm\lam )
+ \mathcal{T}^{ss'\!,T}_{(8)}(\pm\lam )\,,
\nn\\[-3mm]
\\[-2mm]
\mathcal{T}_{ss'}^L(0\lam ) &=&
\mathcal{T}^{ss'\!,L}_{\text{sm}}(0\lam )
+ \mathcal{T}^{ss'\!,L}_{(8)}(0\lam )\,,
\nn
\eeqa
which sum up the contributions from both the SM and the dimension-8 operator
as derived in Eqs.\eqref{eq:Tsm-T+L}-\eqref{eq:T8} of Appendix~\ref{sec:A1}.
From Eq.\eqref{eq:T-llgamma}, we see that the full cross section for
$\,e^-e^+\!\!\to\! \ell^-\ell^+\gamma$\, depends on the angle $\,\phi_*^{}$,\,
due to the interference between the terms with different $Z$ boson helicities
$\lam'\!=+,-,0$\,.\, Eq.\eqref{eq:T-llgamma} also exhibits
the $\,\theta_*^{}$ dependence associated with each $Z$ boson helicity.
We have used Eqs.\eqref{eq:T-llgamma}-\eqref{eq:T-llgamma-sum} in
the analysis of angular observables in Section\,\ref{sec:3}.

\section{\hspace{-5mm}.\,Sensitivities $\mathcal{Z}_4^{}$ and $\mathcal{Z}_8^{}$
have Negligible Correlation}
\label{sec:B}

In this Appendix, we prove that the two statistical significances
denoted by $\mathcal{Z}_4^{}$ and $\mathcal{Z}_8^{}$
in Sections~\ref{sec:3.2.1} and \ref{sec:3.2.2}
have negligible correlation. Hence, their combination
$\,\mathcal{Z}\!=\!\!\sqrt{\mathcal{Z}_4^2\!+\!\mathcal{Z}_8^2\,}$\,
is well justified. In the following, we will first demonstrate that
there is no overlap or double-counting between the two signals
$S_I^{}$ (for $\mathcal{Z}_4^{}$) and $S_{II}^{}$ (for $\mathcal{Z}_8^{}$).
Then, we further prove that there is negligible correlation
between the corresponding background events $B_I^{}$ and $B_{II}^{}$.

\vspace*{1mm}

We first explain why there is no correlation between the two signals
$S_I^{}$ and $S_{II}^{}$, as analyzed in Sections~\ref{sec:3.2.1} and \ref{sec:3.2.2}.
We presented the angular distributions of $\phi_*^{}$ in
Eq.\eqref{eq:f-phi*} and Fig.\,\ref{fig:44}.
It is clear that for the distributions $\,f_{\phi_*^{}}^{0,2}$, the constant terms
mainly dominate $\,f_{\phi_*^{}}^{0,2}$,
and all the $\phi_*$ terms are suppressed by $M_Z/\!\sqrt{s}$ and have no significant effect.
But in the distribution $\,f_{\phi_*^{}}^{1}$\, the $\cos\phi_*^{}$ term
is enhanced by $\sqrt{s}/M_Z^{}$ and dominates $\,f_{\phi_*^{}}^{1}$,
while all other terms are negligible.
We note that for computing the significance $\mathcal{Z}_{4}^{}$ in Eq.\eqref{eq:Z4},\,
the signal $S_{I}^{}$ and the SM background error
$\,\Delta_{B_{I}}^{}$ [in Eqs.\eqref{eq:SI-BI},\eqref{eq:BI-DeltaBI}]
originate from the $\cos\!\phi_*^{}$ term of $\,f_{\phi_*^{}}^{1}$ and
the constant term of $\,f_{\phi_*^{}}^{0}$ [in Eq.\eqref{eq:f-phi*}],
respectively.
From the above explanation and the discussion around Eq.\eqref{eq:SI-cos(phi)},
we see that the constant terms in
$\,f_{\phi_*^{}}^{1,2}$ are strongly cancelled in the signal $S_{I}^{}$,
so $S_{I}^{}$ originates from the $\cos\!\phi_*^{}$ term of $\,f_{\phi_*^{}}^{1}$
under the asymmetric integration \eqref{eq:Oj} [and thus \eqref{eq:SI-cos(phi)}].
Next, we inspect the significance $\mathcal{Z}_{8}^{}$ defined in Eq.\eqref{eq:Z48}.
It contains the signal $S_{II}^{}$ and the SM background error
$\,\Delta_{B_{II}}^{}$ in Eq.\eqref{eq:SII-BII-DBII},
which originate from the constant terms of
$\,f_{\phi_*^{}}^{1,2}$ and $\,f_{\phi_*^{}}^{0}$ in Eq.\eqref{eq:f-phi*},
respectively.
For $\mathcal{Z}_{8}^{}$, our integrations over the full range of $\phi_*^{}$
are uniform and receive no cancellation for the constant terms in $\,f_{\phi_*^{}}^{1,2}$
and $\,f_{\phi_*^{}}^{0}$,\, but integrations over the terms of
$\cos\phi_*^{}$ and $\cos2\phi_*^{}$ vanish identically.
Hence, the significance $\mathcal{Z}_{8}^{}$ only extracts the constant terms
of $\,f_{\phi_*^{}}^{1,2}$ and $\,f_{\phi_*^{}}^{0}$ for
$S_{II}^{}$ and $\,\Delta_{B_{II}}^{}$.\,
This is contrary to the case of $\mathcal{Z}_{4}^{}$,
where the signal $S_I^{}$ extracts the $\cos\phi_*^{}$ term alone.
The above analysis proves that
for the significances $\mathcal{Z}_{4}^{}$ and $\mathcal{Z}_{8}^{}$,
their corresponding signals {$S_{I}^{}$ and $S_{II}^{}$
originate from totally different parts of the $\phi_*^{}$ distributions.}
This means that {the two subgroups of signal events included in $S_{I}^{}$ and $S_{II}^{}$
are different and do not overlap.}
Hence, the signal samples $S_{I}^{}$ and $S_{II}^{}$ have no
double-counting and are not correlated.

\vspace*{1mm}

Because of the fact that $S_{I}^{}$ and $S_{II}^{}$ contain two different subgroups
of signal events without overlap,
it is clear that any other cuts (such as the $\theta$ cut we used in
Table\,\ref{tab:11}) cannot cause extra correlation between $S_{I}^{}$ and $S_{II}^{}$.
To make this fully obvious, we consider the angular distributions respect to
both $\,\phi_*^{}\,$ and $\,\theta\,$ for instance, namely,
the differential cross sections $\di^2\sigma_j^{} /\di\phi_*^{}\di\theta$ (with $j=0,1,2$),
where $(\sigma_0^{},\,\sigma_1^{},\,\sigma_2^{})$ correspond to the contributions
from the SM, the $O(\cut^{-4})$ term, and the $O(\cut^{-8})$ term, respectively.
Expanding in the small parameter $M_Z/\!\sqrt{s}$,\,
we can formally write down the structure of $\di^2\sigma_j^{} /\di\phi_*^{}\di\theta$
as follows,
{\small
\beqs
\label{eq:dsigma/dphi/dtheta}
\beqa
\label{eq:dsigma/dphi/dtheta-s0}
\frac{1}{\sigma_0^{}}
\frac{\di^2\sigma_0^{}}{\,\di\phi_*^{}\di\theta\,} &=&
h_{00}^{}(\theta) + h_{01}^{}(\theta)\cos\!\phi_*^{}\frac{M_Z^{}}{E\,}
+O\!\!\(\!\frac{M_Z^2}{E^2}\!\)\!,
\\[1mm]
\label{eq:dsigma/dphi/dtheta-s1}
\frac{1}{\sigma_1^{}}
\frac{\di^2\sigma_1^{}}{\di\phi_*^{}\di\theta} &=&
h_{10}^{}(\theta) + h_{11}^{}(\theta)\cos\!\phi_*^{}\frac{E}{\,M_Z^{}}
+ h_{12}^{}(\theta)\cos\! 2\phi_*^{}
+O\!\!\(\!\frac{M_Z^2}{E^2}\!\)\!,~~~~~
\\[1mm]
\label{eq:dsigma/dphi/dtheta-s2}
\frac{1}{\sigma_2^{}}
\frac{\di^2\sigma_2^{}}{\di\phi_*^{}\di\theta} &=&
h_{20}^{}(\theta) + h_{21}^{}(\theta)\cos\!\phi_*^{}\frac{M_Z^{}}{E}
+O\!\!\(\!\frac{M_Z^3}{E^3}\!\)\!,~~~~~
\eeqa
\eeqs
}
\hspace*{-3mm}
where $E=\!\sqrt{s\,}$.\,
It is clear that $\,{\di^2\sigma_1^{}}/{\di\phi_*^{}\di\theta}$\, is dominated
by the $\,\cos\!\phi_*^{}$ term with coefficient $h_{11}^{}(\theta)$,
while ${\di^2\sigma_{0}^{}}/{\di\phi_*^{}\di\theta}$
and ${\di^2\sigma_2^{}}/{\di\phi_*^{}\di\theta}$ are dominated by the
$E$-independent terms $h_{00}^{}(\theta)$ and $h_{20}^{}(\theta)$.\,
Integrating Eq.\eqref{eq:dsigma/dphi/dtheta} over
$\,\theta\,$, we obtain the angular distributions $f_{\phi_*^{}}^{0,1,2}$ in
Eq.\eqref{eq:f-phi*}. We note that the signal $S_I^{}$ (for $\mathcal{Z}_4^{}$)
is given by the term $h_{11}^{}(\theta)\cos\!\phi_*^{}$ after the integration
over $\,\theta$\, and the asymmetric integration over $\,\phi_*^{}\,$,\,
while the signal $S_{II}^{}$
(for $\mathcal{Z}_8^{}$) comes from the $E$-independent terms
$h_{10}^{}(\theta)$ and $h_{20}^{}(\theta)$ in
${\di^2\sigma_1^{}}/{\di\phi_*^{}\di\theta}$\,,\,
after integrating over $\,\theta$\, and $\,\phi_*^{}\,$.\,
For the $\,\theta$\, integration, we can add different $\,\theta$\, cuts
on the integrations of $h_{11}^{}(\theta)$ and $h_{10,20}^{}(\theta)$,
respectively, as in Table\,\ref{tab:11}.
The key point is that the term $h_{11}^{}(\theta)$ and the terms $h_{10,20}^{}(\theta)$
contain {two different subgroups of events without overlap.}
Hence, the $\,\theta$\, cuts (Table\,\ref{tab:11}) {cannot} cause any correlation
between $S_I^{}$ and $S_{II}^{}$.\footnote{This is contrary to a hypothetical situation
where the two subgroups of events $S_{I}^{}$ and $S_{II}^{}$ would have an overlap,
in which case adding $\theta$ cuts on the two signals clearly could cause nonzero correlation.}

\vspace*{1.5mm}


In the following, we further prove
that there is {negligible statistical correlation} between
the backgrounds $B_I^{}=N_0^b-N_0^a$ (for $\mathcal{Z}_4^{}$)
and $B_{II}^{}=\over{N}_0^b+\over{N}_0^a$  (for $\mathcal{Z}_8^{}$),
where $B_{I}^{}$ and $B_{II}^{}$ are defined in
Eqs.\eqref{eq:BI-DeltaBI} and \eqref{eq:SII-BII-DBII}.
We follow the concepts and definitions of
statistical errors and correlations given by a standard
textbook such as Ref.\,\cite{book}.
We stress that the following proof is based on pure statistical reasoning only
and is completely general, {independent of any additional
kinematical cuts (such as the cuts on $\theta$ and $\theta_*^{}$ distributions).}

\vspace*{1mm}

We denote by $x$ and $y$ two {independent} observables (such as the
event numbers $N_a^{}$ and $N_b^{}$
in regions (a) and (b) of the $\phi_*^{}$ distribution discussed
in Section\,\ref{sec:3.2}),
\begin{eqnarray}
x = \overline x+\delta_x^{}\,, ~~~~~~
y = \overline y+\delta_y^{}\,,
\end{eqnarray}
where $(\bar{x},\,\bar y)$ are the statistical means (expectation values)
of $(x,\, y)$,  and $(\delta_x^{},\,\delta_y^{})$ denote
the corresponding statistical errors. 
Since $x$ and $y$ are independent of each other by definition, we have
$\overline{xy\,} =\overline{x\,}\,\overline{y\,}$.

\vspace*{1mm}

By definition, we have $\overline {\delta_x^{}}=\overline{\delta_y^{}}=0$
and
$(\,\overline{\delta_x^2},\,\overline{\delta_y^2}\,)
 =(\sigma_x^2,\,\sigma_y^2)$,\,
where $(\sigma_x^2,\,\sigma_y^2)$ are the corresponding variances
whose square-roots $(\sigma_x^{},\,\sigma_y^{})$
are the so-called standard deviations.
The covariance of $(x,\,y)$ is given by
\begin{eqnarray}
\text{Cov}(x,y) \,=\, \overline{(x-\bar x)(y-\bar y)}
\,=\,\overline{\delta_x\delta_y}
\,=\,\overline{xy\,}-\overline{x\,}\,\overline{y\,}
\,=\, 0 \,,
\end{eqnarray}
since $\,\overline{xy\,} =\overline{x\,}\,\overline{y\,}$.

\vspace*{1mm}

Any observable $\mathcal{O}(x,y)$
can be expanded up to linear order
in $(\delta_x^{},\,\delta_y^{})$:
\begin{eqnarray}
\mathcal{O}(x,y) \,=\,\overline{\mathcal{O}}
+\frac{\partial\mathcal{O}}{\partial x} \delta_x^{} \!
+\frac{\partial\mathcal{O}}{\partial y} \delta_y^{}\,.
\end{eqnarray}
For the current purpose, we may define
the two observables $\mathcal{O}_+^{}$ and $\mathcal{O}_-^{}$ as follows,
\begin{eqnarray}
\mathcal{O}_+^{} &\!=& x+y
\,=\, (\overline x +\overline y)+(\delta_x^{}\!+\delta_y^{}),
\\ \nonumber
\mathcal{O}_-^{} &\!=& x-y
\,=\, (\overline x-\overline y)+(\delta_x^{}\!-\delta_y^{}).
\end{eqnarray}
Then, we can deduce the variances and covariances of
$(\mathcal{O}_+^{},\mathcal{O}_-^{})$:
\beqs
\begin{eqnarray}
\sigma_+^2 = \sigma_-^2 &=& \sigma_x^2+\sigma_y^2,
\\[2mm]
\text{Cov}(\mathcal{O}_+^{},\mathcal{O}_-^{})
&=& \overline{(\mathcal{O}_+\!-\!\overline{\mathcal{O}}_+)
           (\mathcal{O}_-\!-\!\overline{\mathcal{O}}_-)}
=\overline{(\delta_x^{}\!+\!\delta_y^{})(\delta_x^{}\!-\!\delta_y^{})}
\hspace*{15mm}
\nn\\
&=& \overline{ \delta_x^2}-\overline{\delta_y^2} \,=\, \sigma_x^2\!-\sigma_y^2\,.
\end{eqnarray}
\eeqs
Thus, we can derive the statistical correlation $\rho$
between $\mathcal{O}_+^{}$ and $\mathcal{O}_-^{}$ as follows,
\begin{eqnarray}
\rho \,=\, \frac{\,\text{Cov}(\mathcal{O}_+^{},\mathcal{O}_-^{})\,}{\sigma_+ \sigma_-}
\,=\, \frac{~\sigma_x^2\!-\!\sigma_y^2~}{\sigma_x^2\!+\!\sigma_y^2} \,.
\end{eqnarray}
For our current study, we choose the two independent observables $(x,\,y)$
to be the SM background event numbers $(N_0^b,\,N_0^a)$
in regions (b) and (a) of the $\phi_*^{}$ distribution, i.e.,
$(x,\,y)=(N_0^b,\,N_0^a)$.\,
Thus, $(\mathcal{O}_-^{},\mathcal{O}_+^{})
=(B_{I}^{},\,B_{II}^{})=(N_0^b\!-\!N_0^a,\,N_0^b\!+\!N_0^a)$.\footnote{%
Since in both regions (a) and (b) the numbers of signal events are
much smaller than the corresponding backgrounds, we have
$(N_b^{},\,N_a^{})\approx (N_0^b,\,N_0^a)$.\,}\,
So, we have
$(\sigma_x^{},\,\sigma_y^{}) 
 = (\sqrt{N_0^b},\,\sqrt{N_0^a})$.\,
Because our analysis chooses the regions (a) and (b) to have the same size
in the $\phi_*^{}$ range and the background distribution $f^0_{\phi_*^{}}$
is nearly flat, we always have $N_0^a\simeq \!N_0^b$\,.
Thus, we can deduce the correlation
\begin{eqnarray}
\label{eq:rho<<1}
|\rho | \,=\,
\frac{\,|N_0^b\!-\!N_0^a|\,}{N_0^b\!+\!N_0^a}\ll 1 \,.
\end{eqnarray}
This means that as long as our choices of regions (a) and (b) obeys
the relation  $N_0^a\approx\! N_0^b\gg |N_0^b-N_0^a|$,\,
the correlation $\,\rho\,$ between the
$B_I^{}$ and $B_{II}^{}$
can be safely ignored. Hence, $B_I^{}$ and $B_{II}^{}$
are independent of each other.
Because we already showed in Sec.\,\ref{sec:3.2.2} that the signals
$S_I^{}$ and $S_{II}^{}$ in the two analyses are also uncorrelated,
the significances $\mathcal{Z}_4^{}$ and $\mathcal{Z}_8^{}$
are independent of each other.

\vspace*{1mm}

Furthermore, we can apply additional different cuts on $B_I^{}$ and $B_{II}^{}$
(such as the angular cuts on $\theta$ and $\theta_*^{}$ distributions)
to optimize the significanes $\mathcal{Z}_4^{}$ and $\mathcal{Z}_8^{}$,
respectively. We denote
\begin{eqnarray}
B_I^{} = N_{0bI}^{}\!-\!N_{0aI}^{},\, ~~~~~
B_{II}^{} = N_{0bII}^{}\!+\!N_{0aII}^{},
\end{eqnarray}
where the regions $aI,aII\in a$ and $bI,bII\in b$\,.\footnote{%
Here for the convenience of discussion, we have used slightly different notations
of the background event numbers from $N_0^{a,b}$ and $\over{N}_0^{a,b}$
in Sec.\,\ref{sec:3.2}. They are connected by
$(N_{0aI}^{},\,N_{0bI}^{})=(N_0^{a},\,N_0^b)$\, and
$(N_{0aII}^{},\,N_{0bII}^{})=(\over{N}_0^{a},\,\over{N}_0^b)$.}\,
For any $\,aX\in a\,$ and $\,bY\in b\,$ (with $X,Y=I,II$), we have
$\text{Cov}(aX,bY)=\overline{\delta_{aX}\delta_{bY}}=0$\,
because $a\bigcap b=0$ and thus $aX\bigcap bY\!=0$\,.\,
Thus, we can derive the covariance between the backgrounds $B_I^{}$ and $B_{II}^{}$,
\begin{eqnarray}
\text{Cov}(I,II)
&=& \overline{(\delta_{bI}^{}\!-\!\delta_{aI}^{})(\delta_{bII}^{}\!+\!\delta_{aII}^{})}
\,=\, \overline{\delta_{bI}^{}\delta_{bII}^{}}\!-\overline{\delta_{aI}^{}\delta_{aII}^{}}
\nn\\[1.5mm]
&=& \text{Cov}(bI,bII)\!-\!\text{Cov}(aI,aII)
\nn\\
&=& N_{0(bI\bigcap bII)}^{}-N_{0(aI\bigcap aII)}^{} \,,~~~~
\end{eqnarray}
where $N_{0(aI\bigcap aII)}^{}$ denotes the SM background event number in the overlapping region
$\,aI\bigcap aII$,\, and
$N_{0(bI\bigcap bII)}^{}$ gives the background event number in the overlapping region
$\,bI\bigcap bII$.\,
Finally, we derive the correlation between the backgrounds $B_I^{}$ and $B_{II}^{}$,
\begin{eqnarray}
\rho \,=\frac{\text{Cov}(I,II)}{\sigma_I^{} \sigma_{II}^{}}
=\frac{\,N_{0(bI\bigcap bII)}^{}-N_{0(aI\bigcap aII)}^{}\,}
      {\sqrt{(N_{0aI}^{}\!+\!N_{0bI}^{})(N_{0aII}^{}\!+\!N_{0bII}^{})\,}}\,,
\end{eqnarray}
where the variances
\beqa
\sigma_I^2 = \over{\delta_{aI}^2}+\over{\delta_{bI}^2}=N_{0aI}^{}\!+\!N_{0bI}^{}\,,
~~~~
\sigma_{II}^2 = \over{\delta_{aII}^2}+\over{\delta_{bII}^2}=N_{0aII}^{}\!+\!N_{0bII}^{}\,.\,
\eeqa
Since the cuts imposed in our current analysis are symmetric, we have
\beqa
N_{0(aI\bigcap aII)}^{} \,\approx\, N_{0(bI\bigcap bII)}^{} \,,
\eeqa
and thus
$\,|N_{0(bI\bigcap bII)}^{}-N_{0(aI\bigcap aII)}^{}|\ll
 \sqrt{(N_{0aI}^{}\!+\!N_{0bI}^{})(N_{0aII}^{}\!+\!N_{0bII}^{})\,}\,$.\,
This leads to the correlation between the backgrounds $B_I^{}$ and $B_{II}^{}$,
\begin{eqnarray}
\label{eq:rho-cuts}
|\rho |
\,=\, \frac{\,|N_{0(bI\bigcap bII)}^{}-N_{0(aI\bigcap aII)}^{}|\,}
      {\sqrt{(N_{0aI}^{}\!+\!N_{0bI}^{})(N_{0aII}^{}\!+\!N_{0bII}^{})\,}}
\ll\, 1\,.
\end{eqnarray}
which is negligible.
Here we note that for deriving Eq.\eqref{eq:rho-cuts}, we already take into account
the effects of additional kinematical cuts (such as the angular cuts on
$\theta$ and $\theta_*^{}$, etc), which allow the unequal parameter spaces
of $\,aI\!\neq\! aII$ and $\,bI\!\neq\! bII$.\,
This is an extension of our result \eqref{eq:rho<<1}
(with $aI \!=\! aII$ and $\,bI\!=\! bII$) to the general case.

\vspace*{1mm}

We stress that the above conclusion is based on pure statistical reasoning only.
Hence,  {\it the fact of $\,|\rho |\!\ll\! 1$  is independent of any additional
kinematical cuts (such as the cuts on $\theta$ and $\theta_*^{}$ distributions).}

\vspace*{1mm}

With the above, we conclude that the significanes $\mathcal{Z}_4^{}$ and $\mathcal{Z}_8^{}$
are independent of each other. Hence, the combination
$\,\mathcal{Z}\!=\!\!\sqrt{\mathcal{Z}_4^2\!+\!\mathcal{Z}_8^2\,}$\,
is well justified.

\vspace*{7mm}
\noindent
{\large {\bf Acknowledgments}}
\\[1mm]
We thank Manqi Ruan for discussing the electron/positron beam polarizations
at the linear and circular colliders, and thank him and Kun Liu for discussing
the photon energy resolution at the lepton colliders and the LHC.
The work of JE was supported in part by United Kingdom STFC Grant ST/P000258/1,
in part by the Estonian Research Council via a Mobilitas Pluss grant, and in part by the
TDLI distinguished visiting fellow programme.
The work of HJH and RQX was supported in part
by the National NSF of China (under grants 11675086 and 11835005).
HJH is also supported in part by the CAS Center for Excellence in Particle Physics (CCEPP),
by the National Key R\,\&\,D Program of China (No.\,2017YFA0402204),
by the Shanghai Laboratory for Particle Physics and Cosmology (No.\,11DZ2260700),
and by the Office of Science and Technology, Shanghai Municipal Government (No.\,16DZ2260200).



\end{document}